\documentclass[aps,pre,showpacs,twocolumn,superscriptaddress,groupedaddress]{revtex4}
\usepackage{appendix}
\usepackage{graphicx}  
\usepackage{bm}        
\usepackage{amssymb}   
\usepackage{amsmath}
\usepackage{color}
\usepackage{pgfplots}
\usepackage{tikz}
 \usepackage{epstopdf}
\usetikzlibrary{calc,decorations.markings}

\begin{document}

\title{Stochastic thermodynamics of Langevin systems under time-delayed feedback control: I. Second-law-like inequalities }
\author{ M.L. Rosinberg}
\affiliation{Laboratoire de Physique Th\'eorique de la Mati\`ere Condens\'ee, Universit\'e Pierre et Marie Curie,CNRS UMR 7600,\\ 4 place Jussieu, 75252 Paris Cedex 05, France}
\email{mlr@lptmc.jussieu.fr}
\author{T. Munakata}
\affiliation{Department of Applied Mathematics and Physics, 
Graduate School of Informatics, Kyoto University, Kyoto 606-8501, Japan
} 
\author{G. Tarjus}
\affiliation{Laboratoire de Physique Th\'eorique de la Mati\`ere Condens\'ee, Universit\'e Pierre et Marie Curie,CNRS UMR 7600,\\ 4 place Jussieu, 75252 Paris Cedex 05, France}
\email{tarjus@lptmc.jussieu.fr}
\begin{abstract}
Response lags are generic to almost any physical system and often play a crucial role in the feedback loops present in artificial nanodevices and biological molecular machines. In this paper, we perform a comprehensive study of small stochastic systems governed by an underdamped Langevin equation and driven out of equilibrium by a time-delayed continuous feedback control. In their normal operating regime, these systems settle in a nonequilibrium steady state in which work is permanently extracted from the surrounding heat bath. By using the Fokker-Planck representation of the dynamics, we derive a set of second-law-like inequalities that provide bounds to the rate of extracted work. These inequalities involve additional contributions characterizing the reduction of entropy production due to the continuous measurement process. We also show that the non-Markovian nature of the  dynamics requires a modification of the basic relation linking dissipation to the breaking of time-reversal symmetry at the level of trajectories. The new relation includes a contribution arising from the acausal character of the reverse process. This in turn leads to another second-law-like inequality. We illustrate the general formalism by a detailed analytical and numerical study of a harmonic oscillator driven by a linear feedback, which describes actual experimental setups.
\end{abstract} 


\maketitle

\section{Introduction}

There is currently a lot of interest in exploring the role of feedback control in the thermodynamic exchanges of heat, work, and entropy between small  stochastic  systems such as molecular motors and their environment\cite{BS2014}. How is the information about the microscopic state of the system processed so as to provide energy beyond the bounds set by the standard second law ?  This fundamental  question, which has been actively debated since the birth of thermodynamics\cite{LR2002,MNV2009,SU2011}, is now becoming of practical interest owing to the progress in the fabrication of micro- and nanomechanical devices\cite{C2009} and to the increasing ability to manipulate single molecules\cite{ARR2012}. In this direction,  a major achievement of the last decade has been the derivation of various second-law-like inequalities and fluctuation relations\cite{KQ2004,TL2004,SU2008,CF2009,SU2010,HV2010,P2010,FS2010,TSUMS2010,AS2012,SU2012,LRJ2012} that extend our understanding of far-from-equilibrium thermodynamics\cite{J2011,S2012} to feedback driven systems

The majority of these recent  developments have been concerned with systems governed by a Markovian dynamics, which allows a consistent theoretical description within the context of stochastic thermodynamics\cite{S2012}. However, in many  processes encountered in engineering and physics in which feedback loops operate continuously,  non-Markovian effects occur due to the finite time interval needed by the feedback mechanism to measure, transmit, or process the information\cite{B2005}. As is well documented, such response lags are also ubiquitous in biological processes\cite{BR2000}, including gene regulation\cite{R2010}, neural networks\cite{A2002},  or human postural sway\cite{M2009}, to name just a few.
Depending on the situation, time delays may have a significant (positive or negative) impact on the control performance (see \textit{e.g.} \cite{FC2007,CLPL2008,LKLMLL2008,LGK2014} for applications to Brownian ratchets). In particular, they may stabilize (resp. destabilize) otherwise unstable (resp. stable) dynamical systems\cite{SHFD2010}. How can one integrate such non-Markovian features into a  thermodynamic description of feedback control ?  This is the main issue explored in this paper and its sequel\cite{RTM2015}, where we  focus on feedback processes governed by time-delayed Langevin equations (or stochastic delay-differential equations, SDDEs, as they are usually called in the mathematical literature). These equations, which describe the combined effects of noise and delay in dynamical systems, are the subject of extensive  investigations  due to their wide range of applications and their interesting dynamical properties, such as multi-stability, noise-induced bifurcations, and stochastic resonance (for some general background on SDDEs, see \cite{M1984,M2008,L2010}). Still, to the best of our knowledge, there exists at present no general analysis of SDDEs  as regards the  exchanges of energy or information between the feedback driven system and its environment.

In this perspective,  it is worth noting that artificial or biological nanomachines working under isothermal conditions and subjected to a continuous time-delayed feedback control  often settle in a time-independent nonequilibrium steady-state (NESS) in which work is permanently extracted from the surrounding heat bath. They can thus be viewed  as ``autonomous" machines (although a steady supply of energy - which is not specified at this level of description - is of course needed to implement the feedback mechanism). This  contrasts with the situation that has been predominantly explored in the recent literature on Maxwell's demons, in which measurements and actions are performed step by step at regular time intervals, as illustrated by Szilard-type engines or feedback-controlled ratchets\cite{TSUMS2010,SU2012}. The motivation behind our work is more in line with the studies of \cite{KQ2004, MR2012,MR2013,DS2014,SDNM2014,HS2014} that consider Langevin processes and continuous-time feedback.

Needless to say, the non-Markovian character of the feedback control  complicates the mathematical analysis.  One issue is that a complete characterization of the time-evolving state of the system requires the knowledge of the whole Kolmogorov hierarchy,  {\it i.e.},  the $n$-time joint probability distributions for all $n$.  A second issue is that the relation linking dissipation to time-asymmetry is somewhat unusual, as pointed out  in \cite{MR2014}. Since this connection is at the core of the stochastic thermodynamic description of Markov processes \cite{K1998,C1999,LS1999,M2003,G2004} and provides an effective route to fluctuation relations, this issue needs to be explored in  detail.  In order to keep the present paper to a reasonable size, the discussion will be split into two parts.  In what follows, we will mainly concentrate on the derivation of second-law-like inequalities on the ensemble level. In a forthcoming paper\cite{RTM2015} (hereafter referred to as paper II),  we will then study the probability distributions of the fluctuating thermodynamic quantities, focusing on their long-time behavior.

This present paper is organized as follows. In  Sec. II, we introduce the general second-order Langevin equation to be studied. It describes a Brownian system subjected to a position-dependent, deterministic feedback. We then discuss the non-Markovian features that characterize the Fokker-Planck and path integral representations of the dynamics.  The first law (or energy balance equation) is briefly stated in Sec. III and the main topic of this work is addressed in  Sec. IV. We first use the Fokker-Planck formalism to derive entropy balance equations that correspond to different coarse-grained descriptions of the dynamics and that contain a specific ``entropy pumping'' contribution due to the feedback. In the steady state,  these equations lead to second-law-like inequalities that provide bounds on the maximum amount of extracted work. We also define information-theoretic measures that give insight into the functioning of the feedback loop. We then use the path integral formalism to analyze the time asymmetry of the stochastic trajectories. This leads to a fluctuation theorem which in turn yields another second-law-like inequality. This inequality involves  a new and nontrivial contribution arising from the acausal character of the backward dynamics. This also leads to a possible definition of the entropy production in the feedback-controlled system.
Sec. V, which is the longest part of the paper, is devoted to a detailed study of a harmonic oscillator driven by a linear feedback. Since the stochastic process is Gaussian, it is analytically tractable. This case study illustrates the previous abstract analysis through explicit calculations and is  relevant to several practical applications. In particular, we will discuss  the feedback cooling of nanomechanical resonators which is an important experimental realization of the model  (see \cite{PZ2012} for a review). Summary and conclusions are presented in Section VI and some additional pieces of  information are given in the appendices.

 \section{Langevin equation and other representations of the dynamics}

As is well known, there are three equivalent descriptions of a Markovian diffusion process: the stochastic differential equation, the Fokker-Planck equation, and the path integral.  These three descriptions are complementary and the most convenient one may be chosen according to the needs. In particular, the path integral formalism is well suited for deriving fluctuation relations by comparing the probabilities of a trajectory and of its time-reversed image. The situation is however more complicated if time-delay effects come into play, as shown in this Section where we describe our basic model within the three frameworks successively. All of them will play a role in the following.

 \subsection{Langevin equation}

For simplicity, we consider a one-dimensional dynamical system (\textit{e.g.}, a particle or a nanomechanical resonator) in contact with a heat bath in equilibrium at temperature $T$. (Extension to several dimensions is straightforward.) The stochastic dynamics is assumed to be governed by the second-order Langevin equation 
\begin{align}
\label{EqL1}
m\dot v&=F(x)+F_{fb}(t)-\gamma v+\sqrt{2\gamma T}\xi(t)
\end{align}
where $v=\dot x$, $m$ is an effective mass, $\gamma$ is a damping constant, $F(x)=-dU(x)/dx$ is  a conservative force, and  $\xi(t)$ is a Gaussian white noise with zero mean and variance $1$ (throughout this paper temperatures and entropies are measured in units of the Boltzmann constant $k_B$). 
$F_{fb}(t)$ is the stochastic force associated with the feedback control, which we assume to only depend  on the position of the system at time $t-\tau$, \textit{i.e.},
\begin{align}
F_{fb}(t):= F_{fb}(x_{t-\tau})  
\end{align}
where $\tau>0$ is the time delay. This corresponds to the situation that is often encountered in experimental setups involving nanomechanical resonators (\textit{e.g.}, the cantilever of an AFM or a suspended mirror in an optical cavity) where the information obtained about the instantaneous displacement of the resonator is used to apply a force that damps its thermal motion\cite{PZ2012}. As will be recalled below in Section V, this is done in practice by adjusting the delay in the feedback loop\cite{PCBH2000,LRHKB2010, Mont2012}. 

Eq. (\ref{EqL1}) is a SDDE, so its solution, say for $t>t_i$, depends on the trajectory of the system in the  initial time interval $[t_i-\tau,t_i]$. It thus depends on the dynamics for $t<t_i$ (preparation effects are crucial in non-Markovian processes due to the memory of the dynamics). For instance, if the feedback control is switched on at $t=t_i$, with the measurement beginning at $t=t_i-\tau$, the initial trajectory must be sampled from the equilibrium path probability at temperature $T$.  On the other hand, if the system has already evolved to a stationary state (if it exists), the dynamics in the initial time interval is  governed by Eq. (\ref{EqL1}).

Before proceeding further, a few comments are in order.

\begin{itemize}

\item To clearly identify the effect of the delay on the thermodynamics of feedback, we neglect the influence of measurement errors (usually due to the sensor noise). In other words,  we assume that the feedback is deterministic, whereas most recent studies of feedback control have focused on the role of the mutual information  (or the transfer entropy) between the system and the controller\cite{SU2010,P2010,HV2010,FS2010,SU2012,SDNM2014}. As is well known, the noise degrades the performance of the control loop\cite{B2005}, and affects the entropy reduction, as discussed in a previous work\cite{MR2013}.  As we will see, the delay plays a somewhat similar, though more complicated, role (see also the discussion in \cite{KMSP2014}).
The absence of measurement noise  implies that the stochastic variable that represents the measurement outcome  coincides with $x_{t-\tau}$  and therefore obeys the same Langevin dynamics as $x_t$.  This also contrasts with recent investigations based on specific realizations of autonomous Maxwell's demons\cite{MJ2012,MQJ2013,SSBE2013,HBS2014,HE2014,SDNM2014} or simple models of the measurement dynamics\cite{MR2013,HS2014}.

\item Although inertial effects are often negligible at the molecular scale (see \textit{e.g.} \cite{KLZ2007}), we take them into account in Eq. (\ref{EqL1})  because they play a significant role in experimental setups that involve mechanical or electromechanical systems\cite{B2005} and also in human motor control (see \textit{e.g.} Ref.\cite{PFFBT2006} and references therein). However, the overdamped limit $m=0$ will be also considered below in order to make contact with previous investigations\cite{MIK2009,JXH2011}. It is also more suitable for analytical investigations (see Appendix A and paper II). 

\item For small delays (typically, $\tau$ much shorter than the relaxation time of the system), one can perform a Taylor expansion of $F_{fb}(x_{t-\tau})$ around $F_{fb}(x_t)$.  The system's dynamics then becomes Markovian, and in the case of a linear and positive feedback, the first-order term in $\tau$ yields an additional viscous force that increases the damping rate. One then recovers the so-called `molecular refrigerator' model studied in \cite{KQ2004,MR2012}. (Besides, the effect of the second-order term is to reduce the effective mass, which increases the resonant frequency of a mechanical oscillator\cite{LMCSYS2000}.)

\item Finally, we wish to emphasize that the non-Markovian character of Eq. (\ref{EqL1}) is not a consequence  of coarse-graining, \textit{i.e.},  the fact that one only observes the degrees of freedom of the Brownian system. In particular, it does not arise from the coupling to the thermal bath, as is the case for generalized Langevin equations with a retarded friction kernel. Such equations have been studied in the literature within the standard framework of stochastic thermodynamics\cite{ZBCK2005,OO2007,MD2007,SS2007,CK2009,ABC2010} and the main result is that the various fluctuation relations keep their conventional form. In contrast, the occurrence of a time delay in the feedback loop alters the behavior of the system under time reversal (the local detailed balance condition), which in turn modifies the fluctuation relations and the second law\cite{MR2014} (for other recent studies of non-Markovian stochastic processes, see also \cite{EL2008,AG2008,MNW2009,HT2009,GG2012}).
\end{itemize}

\subsection{Fokker-Planck description}

Due to non-Markovian character of Eq. ({\ref{EqL1}), the  Fokker-Planck description in terms of probability densities consists of an infinite hierarchy of integro-differential equations that has no closed solution in general.  A noticeable exception, which will be investigated in detail in Sec. V is when Eq. ({\ref{EqL1}) is linear, which makes the process Gaussian. For nonlinear equations, one has to rely on various approximation schemes, depending on whether the delay\cite{GLL1999}, the nonlinearity\cite{F2005a}, or the noise and the magnitude of the feedback\cite{TP2001} are small. Some examples will be discussed in paper II.

Let us first consider  the Fokker-Planck (FP) equation  for the probability distribution function (pdf) of the process ({\ref{EqL1}) in  phase space,  $p(x,v,t):= \langle\delta (x-x_t)\delta(v-v_t)\rangle $. Here, $\langle ..\rangle$ denotes an average over all realizations of the Gaussian noise $\xi(t)$  for $t>t_i$ {\it and} over all stochastic trajectories in the previous time interval $[t_i-\tau,t_i]$. The pdf $p(x,v,t)$  thus depends on the initial path probability ${\cal P}_{init}$ associated with the stochastic dynamics for $t<t_i$ (see next subsection). However, to avoid cumbersome expressions,  this dependence will not be kept explicit hereafter; nor will be explicit the time-dependence of the various probability densities and currents when there is no ambiguity (for instance,  $p(x,v)$ will be the short-hand notation for $p(x,v,t\vert {\cal P}_{init})$). On the other hand, we will keep the time-dependence of the various effective forces.

The FP equation for $p(x,v)$ is obtained as usual by differentiating  $\langle\delta (x-x_t)\delta(v-v_t)\rangle$ with respect to $t$ and using Novikov's theorem\cite{N1964} (see \textit{e.g.} \cite{GLL1999,F2005b}). This yields
\begin{align}
\label{EqKramers1}
\partial_t p(x,v)=-\partial_xJ^x(x,v)-\partial_vJ^v(x,v)\ ,
\end{align}
where 
\begin{align}
\label{Eqflux1}
J^x(x,v)&=vp(x,v)\nonumber\\
J^v(x,v)&=\frac{1}{m}[-\gamma v+F(x)+ {\overline F}_{fb}(x,v,t)]p(x,v)\nonumber\\
&- \frac{\gamma T}{m^2}\partial_v p(x,v) 
 \end{align}
are the probability currents in the directions of the $x$ and $v$ coordinates, respectively, and ${\overline F}_{fb}(x,v,t)$ is an effective time-dependent force obtained by formally integrating  out the dependence on the variable $y_t:= x_{t-\tau}$, 
\begin{align}
\label{EqFfb0}
{\overline F}_{fb}(x,v,t):= \int_{-\infty}^{\infty}dy F_{fb}(y) p(y\vert x,v)
\end{align}
where  $p(y\vert x,v):= p(x,v;y)/p(x,v)$ and $p(x,v;y):= \langle\delta (x-x_t)\delta(v-v_t)\delta(y-x_{t-\tau})\rangle$. 

Eq. (\ref{EqKramers1}) looks very much like a usual Fokker-Planck-Kramers equation\cite{R1989}, but the dependence of the  effective force  on the two-time probability density $p(x,v;y)$  clearly shows that this is not a closed equation (see \cite{PFFBT2006} for a similar equation in the case of a velocity-dependent time-delayed feedback). 
A Markovian equation, however, can be recovered in the small-$\tau$ limit as ${\overline F}_{fb}(x_t,v_t,t)\approx F_{fb}(x_t)-\tau v_t F_{fb}'(x_t)$ at first order in $\tau$. 
It is also worth mentioning that the nondelayed Langevin equation 
\begin{align}
\label{EqLeff}
m\dot v=F(x)+{\overline F}_{fb}(x,v,t)-\gamma v +\sqrt{2\gamma T}\xi(t)
\end{align}
would lead to the same FP equation [Eq. (\ref{EqKramers1})] as Eq.(\ref{EqL1}). This may be a useful approximation in some cases,  providing  the effective force   ${\overline F}_{fb}(x,v,t)$ is successfully determined by other means\cite{GLL1999}. However, one must keep in mind that although this equation exhibits exactly the same one-time pdf $p(x,v)$ as Eq. (\ref{EqKramers1}),  the stochastic trajectories and  thus the $n$-time probability distributions for $n>1$ are different. 

Similarly, we may consider the  FP equation for the  pdf in velocity space only, $p(v)=\int dx \: p(x,v)$. Integrating Eq. (\ref{EqKramers1}) over $x$, we obtain
\begin{align}
\label{EqKramers0}
\partial_t p(v)=-\partial_vJ(v)\ ,
\end{align}
where 
\begin{align}
\label{Eqflux0}
J(v)&=\int dx\: J^v(x,v)\nonumber\\
&=\frac{1}{m}[-\gamma v+{\overline F}(v,t)+ {\overline F}_{fb}(v,t)]p(v)- \frac{\gamma T}{m^2}\partial_v p(v) 
\end{align}
and  
\begin{align}
\label{EqbarFv}
{\overline F}(v,t):= \int_{-\infty}^{\infty}dx \: F(x) p(x\vert v)\ ,
\end{align}
\begin{align}
\label{EqbarFfbv}
{\overline F}_{fb}(v,t)&:= \int_{-\infty}^{\infty}dy \: F_{fb}(y) p(y\vert v)\nonumber\\
&=  \int_{-\infty}^{\infty}dx \: {\overline F}_{fb}(x,v,t) p(x\vert v) ,
\end{align}
with $p(x\vert v)= p(x,v)/p(v)$, $p(y\vert v)= p(v;y)/p(v)$. Note that the  effective force ${\overline F}(v,t)$ is  a consequence of the coarse graining over $x$ and is not due to the feedback.

Finally, since Eq. (\ref{EqKramers1}) involves the two-time pdf $p(x,v;y)$, it is  also useful to write down the FP equation for this quantity:
\begin{align}
\label{EqKramers2}
\partial_t p(x,v;y)=&-\partial_xJ^x(x,v;y)-\partial_vJ^v(x,v;y)\nonumber\\
&-\partial_yJ^v(x,v;y)\ ,
\end{align}
where 
\begin{align}
\label{Eqflux2}
J^x(x,v;y)&=vp(x,v;y)\nonumber\\
J^v(x,v;y)&=\frac{1}{m}[-\gamma v+F(x)+F_{fb}(y)]p(x,v;y)\nonumber\\
&- \frac{\gamma T}{m^2}\partial_v p(x,v;y) \nonumber\\
J^y(x,v;y)&={\overline F}_2(x,v,y,t)p(x,v;y) \ ,
\end{align}
and 
\begin{align}
\label{EqFeff2}
{\overline F}_2(x,v,y,t):= \int_{-\infty}^{\infty}dw \:w p(w\vert x,v;y) \ ,
\end{align}
with  $p(w\vert x,v;y)= p(x,v;y,w)/p(x,v;y)$ and  $p(x,v;y,w):=  \langle\delta (x-x_t)\delta(v-v_t)\delta(y-x_{t-\tau})\delta(w-v_{t-\tau})\rangle$. Here, the effective force ${\overline F}_2(x,v,y,t)$ appears because the operator $\partial_t$ also acts on the dynamical variable $x(t-\tau)$. 

One could proceed further and derive the FP equation for $p(x,v;y,w)$ (which involves the three-time probability density $p(x,v;y,w;z):= \langle\delta (x-x_t)\delta(v-v_t)\delta(y-x_{t-\tau})\delta(w-v_{t-\tau})\delta (z-x_{t-2\tau})\rangle$), etc., but the complexity of the equations is rapidly increasing. In particular,  if the system obeys a different dynamics for $t<t_i$, one must distinguish whether $t-\tau$ is smaller or larger that $t_i$. For our present purpose, it is not necessary to go into these complications.
 
\subsection{Path integral formalism}

As a third representation of the dynamics, we   assign a probability weight  to each trajectory  generated by Eq. (\ref{EqL1}). Similar to the Markovian  case, this weight is obtained, at least formally, by inserting the Langevin equation into the probability density functional of the noise realizations.  

In the following, to help readability, an arbitrary trajectory observed during the time interval $[t_i,t_f]$ is  simply denoted by ${\bf {\bf X}}$ whereas the previous trajectory in the time interval $[t_i-\tau, t_f-\tau]$ (if $t_f-t_i\le\tau$), or in the time interval $[t_i-\tau, t_i]$ (if $t_f-t_i\ge \tau$), is denoted by ${\bf Y}$ (the length of this path is thus  $t_f-t_i$ in the first case and $\tau$ in the second one). Since the noise $\xi_t$ is a stationary Gaussian process, with weight
\begin{align}
{\cal P}_n[\{\xi_t\}]\propto  e^{-\frac{1}{2} \int_{t_i}^{t_f}dt \:  \xi_t^2} \ ,
\end{align}
the probability of observing ${\bf X}$, conditioned on ${\bf Y}$ and on the initial state ${\bf x}_i= (x_{t_i},v_{t_i})$, is  given by
\begin{align}
\label{Eqpath0}
{\cal P}[{\bf X}\vert {\bf x}_i;{\bf Y}]\propto \big \vert {\cal J} \big \vert \: e^{-\beta  {\cal S}[{\bf X},{\bf Y}]} \ ,
\end{align}
where ${\cal J}$ is the Jacobian of the transformation $\xi_t\rightarrow x_t$ for $t\in [t_i,t_f]$,  $\beta=1/T$, and 
\begin{align}
\label{Eqaction0}
 {\cal S}[{\bf X},{\bf Y}]:=\frac{1}{4\gamma} \int_{t_i}^{t_f}dt \:\Big[m\ddot x_t+\gamma \dot x_t-F(x_t)-F_{fb}(x_{t-\tau})\Big]^2 
\end{align}
 may be viewed as a  generalized Onsager-Machlup action functional\cite{OM1953}.
As usual,  Eq. (\ref{Eqpath0}) can be made rigorous by discretizing the Langevin dynamics, as done for instance in the Appendix B of \cite{ABC2010}. Note that  there is no need to specify the interpretation (Ito versus Stratonovitch) of the stochastic calculus as long as $m\ne 0$. (On the other hand, there is an additional path-dependent contribution if one sets $m=0$ from the outset, see \textit{e.g.} \cite{ZJ2002,CCJ2006}.)  An important feature of Eq. (\ref{Eqpath0}) is that  ${\cal J}$ is  a path-independent positive quantity, the (discrete) Jacobian matrix being lower triangular due to causality\cite{ABC2010}. 

From  Eq. (\ref{Eqpath0}), we can obtain the  path probability $ {\cal P}[{\bf X}]$  by integrating over all possible trajectories  ${\bf Y}$  sampled from some initial probability ${\cal P}_{init}[{\bf x}_i;{\bf Y}]$
\begin{align}
\label{Eqpath1}
{\cal P}[{\bf X}]=\int d{\bf y}_i d{\bf y}_f\int_{{\bf y}_i}^{{\bf y}_f} {\cal D}[{\bf Y}]{\cal P}[{\bf X}\vert {\bf x}_i;{\bf Y}]{\cal P}_{init}[{\bf x}_i;{\bf Y}]
\end{align}
where ${\bf y}_i= (x_{t_i-\tau},v_{t_i-\tau})$ and ${\bf y}_f= (x_{t_f-\tau},v_{t_f-\tau})$ (if $t_f-t_i\le\tau$), or ${\bf y}_f= (x_{t_i},v_{t_i})$ (if $t_f-t_i\ge \tau$). At this point, some clarification is needed. It is indeed redundant  to condition the probability of observing ${\bf X}$ on both ${\bf Y}$ and ${\bf x}_i$  when $t_f-t_i\ge \tau$ since the two trajectories are then contiguous in the continuous-time  limit. However, it is still convenient to treat ${\bf y}_f$, the end point of ${\bf {\bf Y}}$, and ${\bf x}_i$, the initial point of ${\bf X}$, as two distinct microstates. This allows us to write equations that are valid for both $t_f-t_i\le \tau$ and $t_f-t_i\ge \tau$. Specifically, in the first case, one has 
\begin{align}
{\cal P}_{init}[{\bf x}_i;{\bf Y}]=\int_{{\bf y}_f}^{{\bf x}_i} {\cal D}[{\bf Z}]{\cal P}_{init}[{\bf Y}_{\cup}{\bf Z}]\ ,
\end{align}
 where ${\bf Y}_{\cup}{\bf Z}$ denotes the whole trajectory between the time $t_i-\tau$ and the time $t_i$. In the second case, one simply has 
${\cal P}_{init}[{\bf x}_i;{\bf Y}]={\cal P}_{init}[{\bf Y}]\delta ({\bf y}_f-{\bf x}_i)$, where $\delta$ is the Dirac distribution.

In general, ${\cal P}[{\bf X}]$  depends on the choice of the initial path probability ${\cal P}_{init}$ (which thus implies that $p({\bf x}_f,t_f)=\int d{\bf x}_i \int_{{\bf x}_i}^{{\bf x}_f} {\cal D}[{\bf X}] {\cal P}[{\bf X}]$ also depends on ${\cal P}_{init}$, as already pointed out). On the other hand, in the stationary regime,  Eq. (\ref{Eqpath1}) yields a closed functional integral equation for the  probability ${\cal P}_{st}^{(1)}[{\bf X}]$ of observing a trajectory of duration $t_f-t_i\le \tau$,
\begin{align}
\label{Eqpathst}
{\cal P}_{st}^{(1)}[{\bf X}]=\int d{\bf y}_i d{\bf y}_f\int_{{\bf y}_i}^{{\bf y}_f} {\cal D}[{\bf Y}]{\cal P}[{\bf X}\vert{\bf x}_i; {\bf Y}]{\cal P}_{st}^{(1)}[{\bf x}_i;{\bf Y}]\ .
\end{align}
 This is rather formal, however, because the solution of this equation  is in general out of reach (see below). If it were available, one could obtain the stationary probabilities  ${\cal P}_{st}^{(2)},{\cal P}_{st}^{(3)}$, ..., ${\cal P}_{st}^{(n)}$ of  trajectories of lengths $t_f-t_i\le 2\tau$, $t_f-t_i\le 3\tau$, ..., $t_f-t_i\le n\tau$ by  repeatedly applying the non-integrated version of Eq. (\ref{Eqpath1}), for instance
  \begin{align}
\label{EqCK}
{\cal P}_{st}^{(2)}[{\bf X}_{\cup}{\bf Y}]= {\cal P}[{\bf X}\vert {\bf x}_i;{\bf Y}]{\cal P}_{st}^{(1)}[{\bf Y}]\delta ({\bf y}_f-{\bf x}_i)
\end{align}
for $n=2$. This is nothing else than a generalization of the so-called ``step-method" that is used to  solve linear SDDEs  by decomposing the evolution into ``slices" or time intervals of length $\tau$, using the solution to the previous interval as an inhomogeneous term (see \textit{e.g.} \cite{KM1992}). In other words, Eqs. (\ref{Eqpathst}) and (\ref{EqCK}) illustrate the well-known fact that a non-Markovian process with time delay can be described within each slice as a Markovian process\cite{M1984,M2008,F2002}.

In passing, we wish to mention that finding a solution to Eq. (\ref{Eqpathst}) seems to be a challenging task even in the case of a linear Langevin equation. Although the process is  Gaussian and all $n$-time distribution functions are thus completely determined by the two-time autocorrelation matrix, we have only been able to find the explicit expression of  ${\cal P}_{st}^{(1)}[{\bf X}]$  in the overdamped limit $m=0$.  We refer the interested reader to Appendix  A where the method of solution is described. This also makes the formalism presented in this section less abstract.

Finally, we stress that replacing the original Langevin Eq. (\ref{EqL1}) by Eq. (\ref{EqLeff}) in the  action functional [Eq. (\ref{Eqaction0})]  (for instance, by using the steady-state expression of  the effective force ${\overline F}_{fb}(x,v,t)$) is only an approximation. As we already pointed out,  Eq. (\ref{EqLeff}) leads to the same one-time pdf $p(x,v)$ as Eq. (\ref{EqL1}), but it generates different  trajectories. In this respect, the discussion of the stochastic thermodynamics associated with  the linear overdamped Langevin equation presented in \cite{JXH2011} is inaccurate. We shall come back to this issue later (see also Appendix A).

\section{Energy balance equation}

An important assumption underlying Eq. (\ref{EqL1}) is that the feedback control does not affect the environment which always stays at equilibrium. This allows us to use the stochastic energetics approach of \cite{S1997,Sbook2010} to define the energy balance equation  at the level of   an individual fluctuating trajectory. Of course, because of the delay, there is also a dependence on the past trajectory. 

As is now standard, we thus identify the heat dissipated into the medium during the time interval $dt$ as the work done by the system on the environment,  \textit{i.e.}, the work done by the ``reaction'' force $-(-\gamma v_t+\sqrt{2\gamma T}\xi_t)$ \cite{Sbook2010}
\begin{align}
\label{Eqheat0}
dq&=(\gamma v_t-\sqrt{2\gamma T}\xi_t)\:   dx_t \nonumber\\
&=-[m \dot v_t -F(x_t)-F_{fb}(x_{t-\tau})]\: dx_t  \ .
\end{align}
Next, we identify the work done on the system by the feedback force $F_{fb}(x_{t-\tau})$ as
\begin{align}
dw= F_{fb}(x_{t-\tau})\: dx_t \ .
\end{align}
Note that we use  the same sign convention as in \cite{S2005,S2012}: the heat is counted as positive when given to the environment and the work is counted as positive when done on the system  (the heat is defined with the opposite sign in \cite{Sbook2010}). The energy balance equation or first law is then expressed as
\begin{align}
\label{Eq1law}
du=dw-dq \ ,
\end{align}
where 
\begin{align}
du&=[m \dot v_t -F(x_t)] dx_t \nonumber\\
&=d\big[\frac{1}{2}mv_t^2+V(x_t) \big]
\end{align}
is the change in internal energy of the system.  (Throughout this paper, all  fluctuating thermodynamic quantities are defined within the  Stratonovich prescription which allows us to use the standard rules of calculus.) By integrating over the time interval $[t_i,t_f]$, the heat and the work become functionals of the trajectories ${\bf X}$ and ${\bf Y}$, the latter being defined as in the preceding Section. The first law for the integrated quantities then reads 
\begin{align}
\label{Eq1law}
\Delta u=w[{\bf X},{\bf Y}]-q[{\bf X},{\bf Y}]\ ,
\end{align}
where
\begin{align}
\label{EqwXY}
w[{\bf X},{\bf Y}]:= \int_{t_i}^{t_f} dt F_{fb}(x_{t-\tau})\: v_t \ ,
\end{align}
\begin{align}
\label{EqqXY}
q[{\bf X},{\bf Y}]:= -\int_{t_i}^{t_f} dt\: \Big[m \dot v_t -F(x_t)-F_{fb}(x_{t-\tau})\Big] \: v_t  \ ,
\end{align}
and
\begin{align}
\Delta u=\frac{1}{2}m(v_f^2-v_i^2)+V(x_f)-V(x_i) \ .
\end{align}
As usual, the heat dissipated into the environment may be identified with an increase in entropy of the medium\cite{S2012}
\begin{align}
\Delta s_m[{\bf X},{\bf Y}]=\frac{q[{\bf X},{\bf Y}]}{T}\ .
\end{align}

\section{Second law-like inequalities}

We now address the main topic of this work, which concerns the second-law-like inequalities  associated with Eq. (\ref{EqL1}). Unsurprisingly, the non-Markovian character of the feedback control makes this issue nontrivial. 

For a Markovian process, the standard (and simplest) way of deriving the second law of thermodynamics is to consider the time evolution  of  the Gibbs-Shannon entropy of the system $S(t)$ (see \textit{e.g.} \cite{VDBE2010,TO2010}).  By using the  Fokker-Planck equation (or the master equation if the number of states is discrete), one finds that $dS(t)/dt$ splits into two contributions, the entropy flow from the system to the environment (\textit{i.e.}, the rate of heat dissipation in the medium) and the so-called irreversible entropy production (EP) rate, which is a non-negative quantity associated with time-reversal symmetry breaking of the dynamics. 

The statement of the second law is less obvious when only limited information on the system is available, either because the measurements are  performed at some finite set of times, or because one only observes a subset of the slow degrees of freedom participating in the  dynamics (which  is also what occurs when a Maxwell's demon is operating).  In such situations, the dissipation associated with the full process is inaccessible, and the observed EP depends on the level of coarse graining (see \textit{e.g.} \cite{RJ2007,GPVdB2008,AJM2009,HJ2009,CPV2012,E2012,MLBBS2012,BC2014,BKMP2014}).   An interesting issue is then to understand how energy and information are exchanged between the different interacting subsets. This is currently an active field of research\cite{IS2013,DE2014,HBS2014,HE2014,SS2014,HS2014}. 

We are facing a similar situation here since a complete description of the dynamics on the ensemble level would require to take into account all the $n$-time probability distribution functions $p(x,v,t)$, $p(x,v,t; y,t-\tau)$, etc. Choosing one of these probabilities to define the Shannon entropy of the system amounts to tracing out some dynamical variables, which in turn leads to a second-law-like inequality for the extracted work in the steady state. In addition, one can integrate out the position variable and observe the system in momentum space only.   The feedback manifests itself by the presence of additional contributions to the entropy production that will be referred to as ``entropy pumping'' rates, using the terminology introduced in \cite{KQ2004}.  From an information-theoretic perspective, it is also interesting  to study how the feedback exploits the information about the past position of the system to modify its present state.   Another viewpoint, at the level of individual stochastic trajectories, is to consider the behavior of path probabilities under time reversal. This leads to a fluctuation theorem and  to another second-law-like inequality. This issue deserves a separate and detailed study as the time-reversed dynamics is somewhat  unusual\cite{MR2014}.

\subsection{Coarse-graining, entropy pumping, and information flows}

\subsubsection{A first inequality}

We begin by considering the time evolution of the one-time Shannon entropy 
\begin{align}
\label{EqShannon1}
S^{xv}(t):=-\int dx dv\: p(x,v)\ln p(x,v) 
\end{align}
where $p(x,v):= p(x,v,t)$ is  solution of the FP equation (\ref{EqKramers1}) (we remind that  the time dependence of the various pdf's and   currents is dropped to simplify the notation). We stress again that $p(x,v)$ depends on the preparation of the system for $t<t_i$.  Taking the time derivative of $S^{xv}(t)$ and using Eq. (\ref{EqKramers1}), we get
\begin{align}
\label{EqdotSxv}
d_t S^{xv}(t)=\int dx dv\: \partial_v J^v(x,v)\ln p(x,v)
\end{align}
where  we have used the conservation of probability and partial integration (assuming that the boundary contributions vanish).  We then single out   from the current  $J^v(x,v)$  defined by Eq. (\ref{Eqflux1}) the piece 
\begin{align}
\label{EqJirr}
J_{irr}^v(x,v)= -\frac{\gamma}{m}vp(x,v)-\frac{\gamma T}{m^2}\partial_v p(x,v)
\end{align}
which is antisymmetric under time reversal and is traditionally called the  ``irreversible"  current \cite{R1989}. $J_{irr}^v(x,v)$ is responsible for the heat exchanged with the bath 
\begin{align}
\label{EqHeat}
\dot Q(t)=-m\int dx\:dv\: v J_{irr}^v(x,v)=\frac{\gamma}{m} (m\langle v^2_t\rangle -T)\ ,
\end{align}
as obtained by taking the ensemble average of Eq. (\ref{Eqheat0}). (From now on, we use the over-dot notation for the rates that may have a nonzero limit in a stationary regime, such as $\dot Q(t)$, whereas $d_t$ denotes the time derivative of average quantities whose only time-dependence comes from that of the pdf's: such time derivatives then vanish in a stationary regime.) 

The heat flow thus proceeds through the kinetic energy of the system (note that $\dot Q(t)$ has another expression in the absence of inertia\cite{Sbook2010}).  After repeating the standard mathematical manipulations that lead to the second law and treating  the effective force ${\overline F}_{fb}(x,v,t)$ as it were an external force, we obtain a first  entropy balance equation 
\begin{align}
\label{EqEP1}
d_tS^{xv}(t)=\dot S_i^{xv}(t)-\dot S_{pump}^{xv}(t)-\frac{1}{T}\dot Q(t)
\end{align}
where 
\begin{align}
\label{EqSixv}
\dot S_i^{xv}(t):= \frac{m^2}{\gamma T} \int dx\: dv\: \frac{[J_{irr}^v(x,v)]^2}{p(x,v)} \ge 0
\end{align}
and 
\begin{align}
\label{EqSpumpxv}
\dot S_{pump}^{xv}(t)&:= \frac{1}{m}\int dx \: dv \: {\overline F}_{fb}(x,v,t)\partial_v p(x,v) \nonumber\\
&=-\frac{1}{m}\langle \partial_v {\overline F}_{fb}(x,v,t)\rangle\ .
\end{align}
In line with the general concepts of linear irreversible  thermodynamics\cite{O1931,dGM1984}, the positive rate $\dot S_i^{xv}(t)$ is expressed as a product of the thermodynamic force $-\gamma [v +(T/m)\partial_v\ln p(x,v)]$ by the corresponding flux.
The unusual term $\dot S_{pump}^{xv}(t)$  describes the influence of the continuous feedback control. It is nonzero because  the effective force ${\overline F}_{fb}(x,v,t)$ is velocity-dependent and thus contains a piece that is antisymmetric under time reversal (see \textit{e.g.} Eq. (\ref{EqFfb1}) in Section V). Since this is very similar to the entropy reduction mechanism that takes place with a velocity-dependent (non-delayed) feedback control\cite{KQ2004}, we will adopt the same terminology  and call  $\dot S_{pump}^{xv}(t)$  an ``entropy pumping" rate (see also \cite{MR2013,HS2014}).  We stress, however, that the stochastic force $F_{fb}(t)$ in our model only depends on the position and that it is the delay $\tau$ which makes the entropy pumping effective. We will see in Sec. V that $\tau$ is chosen in an optimal way in the feedback cooling (or cold damping) technique.  The result of \cite{KQ2004} is actually recovered in the small-$\tau$ limit for a linear feedback $F_{fb}(x_{t-\tau})= k'x_{t-\tau}$. The effective force is  then proportional to the instantaneous velocity  at  first order in $\tau$, ${\overline F}_{fb}(x,v,t)= -k'\tau v +{\cal O}(\tau^2)$, which readily yields $\dot S_{pump}^{xv}=\gamma'/m $ with $\gamma'= k'\tau$. 
(Note that for the convenience of our presentation we here define the entropy pumping rate as in \cite{HS2014} with an opposite sign to that adopted in  \cite{KQ2004,MR2012,MR2013}.)

Eq. (\ref{EqEP1}) simplifies in the steady state, which is the normal operating regime of an autonomous system.  Then $d_tS^{xv}(t)=0$ and $\dot {\cal Q} =\dot {\cal W}$ from the first law, and the non-negativity of $\dot {\cal S}_i^{xv}$ leads to a first  second-law-like inequality 
\begin{align}
\label{IneqEP1}
\frac{\dot {\cal W}_{ext}}{T} \le \dot {\cal S}_{pump}^{xv} 
\end{align}
where $\dot {\cal W}_{ext}= - \dot {\cal W}$ is the extracted work rate. (From now on, calligraphic typefaces such as $\dot {\cal Q},\dot {\cal W},\dot {\cal W}_{ext}$, etc...  denote time-independent steady-state quantities.) This inequality is quite meaningful since it reveals that  work can be extracted from the environment (\textit{i.e.}, $\dot {\cal W}_{ext}>0$) provided that $\dot {\cal S}_{pump}^{xv}>0$. This is precisely what is done in the cold damping technique where this work is used to reduce the thermal fluctuations of the Brownian entity.  Of course, this inequality is predictive only if an expression of $\dot {\cal S}_{pump}^{xv}$ is available, which in general depends on the dynamical details of the system (see however Eq. (\ref{EqSvpump}) and the remark below it).

Note that an entropy balance equation can also be derived at the level of an individual trajectory, starting from the stochastic entropy $s^{xv}(t)=-\ln p(x_t,v_t)$ where  $p(x,v)$ is evaluated along the stochastic trajectory\cite{S2005,IP2006}. However, this equation is not very useful for the present purpose and is not reproduced here. As it must be, Eq. (\ref{EqEP1}) is recovered upon averaging.

\subsubsection{A second inequality}

We now perform a similar calculation by considering the evolution of the system in momentum space only. The starting point is the (one-time) coarse-grained Shannon entropy 
\begin{align}
\label{EqShannon0}
S^{v}(t):=-\int dv\: p(v)\ln p(v) 
\end{align}
where $p(v) := p(v,t)$ is solution of the Fokker-Planck equation (\ref{EqKramers0}).  Then,
\begin{align}
\label{EqdotSv}
d_tS^{v}(t)=\int dv\: \partial_v J(v) \ln p(v)\ ,
\end{align}
and similar manipulations lead to a second entropy balance equation 
\begin{align}
\label{EqEP0}
d_tS^{v}(t)=\dot S_i^v(t)-\dot S_{pump}^v(t)-\frac{1}{T}\dot Q(t)
\end{align}
where 
\begin{align}
\label{EqSirrv}
\dot S_i^v(t):=  \frac{m^2}{\gamma T}\int dv\: \frac{[J_{irr}(v)]^2}{p(v)}\ge 0
\end{align} 
and
\begin{align}
\label{EqSpumpv}
\dot S_{pump}^{v}(t)&:= \frac{1}{m}\int dv \:[{\overline F}(v,t)+ {\overline F}_{fb}(v,t)]\partial_v p(v)\nonumber\\
&=-\frac{1}{m}\langle \partial_v[{\overline F}(v,t)+ {\overline F}_{fb}(v,t)]\rangle\ .
\end{align}
Again, we have defined the irreversible current as if there were no feedback, \textit{i.e.}, $J_{irr}(v)= -(\gamma/m)vp(v)-(\gamma T/m^2)\partial_v p(v)$.  Note that there are now two contributions to the  ``entropy pumping'' rate $\dot S_{pump}^{v}(t)$, the effective force ${\overline F}(v,t)$  arising from the coarse graining over $x$ (see Eq. (\ref{EqbarFv})). One can readily check that this contribution vanishes at first order in $\tau$, whereas $\dot S_{pump}^{v}(t)=\gamma'/m$ like $\dot S_{pump}^{xv}(t)$. 

Although $\dot S_i^{xv}(t)$ and $\dot S_i^v(t)$ in Eqs. (\ref{EqEP1}) and (\ref{EqEP0}) are not defined in terms of relative entropies (see the discussion below in subsection B) and therefore one cannot use the powerful properties of the Kullback-Leibler distance\cite{KPVdB2007,GPVdB2008,RP2012}, one still has a coarse-graining inequality\cite{MR2013}
 \begin{align}
\label{Ineq0}
\int dv\: \frac{[J_{irr}(v)]^2}{p(v)}\le \int dx\: dv\:  \frac{[J_{irr}^v(x,v)]^2}{p(x,v)}\ ,
\end{align} 
which is the continuous version of the  inequality considered in \cite{VDBE2010} and is a direct consequence of the Cauchy-Schwarz inequality. Hence,
\begin{align}
\label{IeqSpump1}
 \dot S_{pump}^{v}(t)- \dot S_{pump}^{xv}(t)\le d_tS^{xv}(t)-d_tS^v(t)\ .
\end{align}
Interestingly, the right-hand side of this inequality is equal to the information flow $\dot I_{flow,v}^{xv}$ which is the information-theoretic measure  that quantifies the change in the mutual information $I^{xv}$ between the random variable $x_t$ and  $v_t$ due to the evolution of $v_t$\cite{AJM2009}. By definition, 
\begin{align}
I^{xv}(t):= \int dx\: dv\: p(x,v)\ln \frac{p(x,v)}{p(x)p(v)}\ ,
\end{align}
and taking the time derivative yields
\begin{align}
\label{EqdIxv}
d_t I^{xv}(t)=- [\dot I_{flow,x}^{xv}(t)+\dot I_{flow,v}^{xv}(t)]
\end{align}
where 
\begin{align}
\label{EqIflow1}
\dot I_{flow,x}^{xv}(t)&:=\int dx\: dv\:  \partial_x J^x(x,v)\ \ln \frac{p(x,v)}{p(x)p(v)}\nonumber\\
\dot I_{flow,v}^{xv}(t)&:=\int dx \:dv\: \partial_v J^v(x,v)\ln \frac{p(x,v)}{p(x)p(v)}\ ,
\end{align}
and we have used Eq. (\ref{EqKramers1}) and the conservation of probability. (In this work, we define information flows with the same sign convention as in \cite{HS2014}, which explains the minus sign in Eq. (\ref{EqdIxv}).) Whereas $I^{xv}(t)$ is a symmetric and non-negative quantity, the  flows $\dot I_{flow,x}^{xv}(t)$ and $\dot I_{flow,v}^{xv}(t)$ are directional and have no definite sign. From Eqs. (\ref{EqdotSxv}) and (\ref{EqdotSv}), we readily obtain
\begin{align}
\label{EqIflow30}
\dot I_{flow,v}^{xv}(t)=d_tS^{xv}(t)-d_tS^v(t)\ ,
\end{align}
and the inequality (\ref{IeqSpump1}) can thus be rewritten as
\begin{align}
\label{IeqSpump2}
\dot S_{pump}^{v}(t) \le \dot S_{pump}^{xv}(t)+\dot I_{flow,v}^{xv}(t)\ .
\end{align}
By subtracting Eq. (\ref{EqEP0}) from Eq. (\ref{EqEP1}) and using Eq. (\ref{EqIflow30}), we also obtain the  relation
\begin{align}
\label{EqSii3}
\dot S_i^{xv}(t)-\dot S_i^{v}(t)= \dot I_{flow,v}^{xv}(t)+[\dot S^{xv}_{pump}(t) -\dot S^{v}_{pump}(t)]  
\end{align}
which shows that the increase in the non-negative EP rate $\dot S_i(t)$ when going from the momentum space ($\dot S_i(t)= \dot S_i^v(t)$) to the full phase space ($\dot S_i(t)=\dot S_i^{xv}(t)$)  is due both to the information flow $\dot I_{flow,v}^{xv}(t)$ and to a modification of the entropy pumping rate.

In the steady state, all time derivatives $d_t$ are zero and Eq. (\ref{EqEP0}) leads to a second second-law-like inequality, 
\begin{align}
\label{IneqEP2}
\frac{\dot {\cal W}_{ext} }{T} \le  \dot {\cal S}_{pump}^{v}\ ,
\end{align}
while Eq. (\ref{IeqSpump2}) implies  
\begin{align}
 \dot {\cal S}_{pump}^{v}\le  \dot {\cal S}_{pump}^{xv}
\end{align}
since $\dot {\cal I}_{flow,v}^{xv}=0$ from Eq. (\ref{EqIflow30}). Therefore, $\dot {\cal S}_{pump}^{v}$ is in general  a tighter upper bound for the extracted work rate than $\dot {\cal S}_{pump}^{xv}$. Note that  $\dot {\cal S}_{pump}^{v}$ can be computed from the mere observation of the stationary velocity pdf, without any knowledge of the details of the feedback control such as the gain and the time delay. Indeed, since the current $J(v)$ vanishes in the steady state, one has from Eq. (\ref{Eqflux0})
\begin{align}
{\overline F}(v)+ {\overline F}_{fb}(v)=\gamma [v+\frac{T}{m}\frac{d}{dv}\ln p_{st}(v)]\ ,
\end{align}
 and thus from Eq. (\ref{EqSpumpv})
\begin{align}
\label{EqSvpump}
 \dot {\cal S}_{pump}^{v}&=-\frac{\gamma}{m}[1+\frac{T}{m}\int dv\: p_{st}(v)\frac{d^2}{dv^2}\ln p_{st}(v)]\nonumber\\
 &=\frac{\gamma T}{m^2}\big[\langle \big(\frac{d}{dv}\ln p_{st}(v)\big)^2\rangle_{st} -\frac{m}{T}\big] \ .
\end{align}
  
In fact, as will be seen in Sec. V, the two pumping rates $ \dot {\cal S}_{pump}^{xv}$ and  $\dot {\cal S}_{pump}^{v}$ turn out to be  identical  in the case of a linear SDDE as $p_{st}(x,v)$ is then a bivariate Gaussian distribution with $\langle xv\rangle_{st}=0$ (this also implies that ${\overline F}(v)=0$). Accordingly, one also has $\dot {\cal S}_i^{xv}=\dot {\cal S}_i^{v}$ and the two second-law-like inequalities (\ref{IneqEP1}) and (\ref{IneqEP2}) become identical.

\subsubsection{Other information-theoretic measures}

Instead of coarse-graining the description of the system, we may go in the opposite direction and take into account the correlations between the microstates  at time $t$, $t-\tau$, $t-2\tau$, etc. Such correlations already exist at equilibrium (see \textit{e.g.} \cite{MC2014}), but they are modified when the system is subjected to the continuous feedback control, and it is interesting to study these changes as they may reveal how the exchange of information between the system and the controller is affected by the time delay.  On the other hand, from the thermodynamic point of view,  it is expected that adding new dynamical variables to the description, such as $x_{t-\tau}$, $x_{t-2\tau}$, etc., leads to an increase of the observed dissipation  and to weaker bounds for the work that can be extracted in the stationary state.

From the information-theoretic point of view, it is useful to recast Eq.(\ref{EqL1}) as
\begin{align}
\label{EqL21}
m\dot v_t&=F(x_t)+F_{fb}(y_t)-\gamma v_t+\sqrt{2\gamma T}\xi_t\nonumber\\
y_t&=x_{t-\tau}
\end{align}
in order to emphasize that we are dealing with a feedback controller that performs measurements and actions on the system. The relevant information measures  are then  the mutual informations, for instance, 
 \begin{align}
 \label{EqMIxvy}
 I^{xv;y}(t):= \int dx dv dy\: p(x,v;y)\ln \frac{p(x,v;y)}{p(x,v)p(y)}
\end{align}
or, if the dependence on $x$ is traced out,
 \begin{align}
 \label{EqMIvy}
 I^{v;y}(t):= \int dv dy\: p(v;y)\ln \frac{p(v;y)}{p(v)p(y)}\ .
\end{align}
The corresponding information flows due to the evolution of $v_t$ are
 \begin{align}
 \label{EqIflow2}
\dot I_{flow,v}^{xv;y}(t):= \int dx dv dy\:  \partial_v J^v(x,v;y)\ln \frac{p(x,v;y)}{p(x,v)p(y)}\
\end{align}
and 
\begin{align}
\label{EqIflow3}
\dot {\cal I}_{flow,v}^{v;y}(t):= \int dv \: dy\: \partial_v J^v(v;y)\ln \frac{p(v;y)}{p(v)p(y)} 
\end{align}
where $p(v;y)=\int dx\: p(x,v;y)$ and $J^v(v;y)=\int dx\:J^v(x,v;y)$. Note that $\dot I_{flow,x}^{xv;y}(t)=0$ from the definition of $J^x(x,v;y)$ (see Eq. (\ref{Eqflux2})), so that $\dot {\cal I}_{flow,v}^{xv;y}(t)=-\dot{\cal  I}_{flow,y}^{xv;y}(t)$ in the stationary regime. 
 We do not consider information measures that involve  whole trajectories  such as the transfer entropy rate\cite{Sch2000,IS2013,HBS2014,IS2014,SDNM2014,HS2014} or the trajectory mutual information rate\cite{MK2006,TW2009,HS2014} because  these quantities  diverge when the feedback is deterministic. This can be seen for instance  in Fig. 1 of \cite{HS2014} (see also the discussion in \cite{SDNM2014}).

We can use $\dot I_{flow,v}^{xv;y}(t)$ to express the entropy balance equation  (\ref{EqEP1})  in a different  way.  To this aim, we   start again from the Shannon entropy $S^{xv}(t)$ but use the FP equation (\ref{EqKramers2}) instead of Eq. (\ref{EqKramers1}). Some simple manipulations then lead to an alternative expression of $d_tS^{xv}(t)$,
\begin{align}
\label{EqEP3}
d_tS^{xv}(t)=\dot S_i^{xv;y}(t)-\dot I_{flow,v}^{xv;y}(t)-\frac{1}{T}\dot Q(t) \,,
\end{align}
which involves the non-negative rate 
\begin{align}
\label{EqSirr2}
\dot S_i^{xv;y}(t):=  \frac{m^2}{\gamma T}\int dx\: dv\: dy \frac{[J_{irr}^v(x,v;y)]^2}{p(x,v;y)} 
\end{align}
where  $J_{irr}^v(x,v;y)= -(\gamma/m)vp(x,v;y)-(\gamma T/m^2)\partial_v p(x,v;y)$.
Comparing Eq. (\ref{EqEP3}) with Eq. (\ref{EqEP1}) then yields a  relation similar to Eq. (\ref{EqSii3}),
\begin{align}
\label{EqSii}
\dot S_i^{xv;y}(t)-\dot S_i^{xv}(t)=\dot I_{flow,v}^{xv;y}(t)-\dot S^{xv}_{pump}(t)\ \, ,
\end{align}
which in turn leads to the inequality
\begin{align}
\label{Eqin4}
\dot S^{xv}_{pump}(t)\le \dot I_{flow,v}^{xv;y} (t)
\end{align}
owing to a coarse-graining inequality similar to (\ref{Ineq0}). Therefore, as announced, $\dot {\cal I}_{flow,v}^{xv;y}$  is  a weaker bound for the extracted work rate than $\dot {\cal S}^{xv}_{pump}$ in the steady state. We  emphasize once again that  $\dot {\cal I}_{flow,v}^{xv;y}$ does not vanish when the feedback is turned off, since there still exists a nonzero information flow from the past to the present during the dynamical evolution.  

Likewise, by using $\dot {\cal I}_{flow,v}^{v;y}$, we obtain the entropy balance equation
\begin{align}
\label{EqEP4}
d_tS^{v}(t)=\dot S_i^{v;y}(t)-\dot I_{flow,v}^{v;y}(t) -\dot S^{v;y}_{pump}(t)-\frac{1}{T}\dot Q(t)
\end{align}
which  is similar to Eq. (\ref{EqEP3}) but contains an additional ``entropy pumping'' contribution induced by the  coarse-graining over $x$, 
\begin{align}
\label{Eqpump3}
\dot S^{v;y}_{pump}(t):=-\frac{1}{m}\langle \partial_v {\overline F}(v,y,t)\rangle \,,
\end{align}
where
\begin{align}
\label{EqeffForce}
{\overline F}(v,y,t):= \int dx\:  F(x)p(x\vert v;y)
\end{align}
and $p(x\vert v;y)=p(xv;y)/p(v;y)$. (Note that there is no entropy pumping rate directly induced by the feedback force $F_{fb}(x_{t-\tau})$ because $y_t=x_{t-\tau}$ is an explicit variable at this level of description.) Finally, as in Eq. (\ref{EqSii}), we  have the relation 
\begin{align}
\label{EqSii1}
\dot S_i^{v;y}(t)-\dot S_i^{v}(t)= \dot I_{flow,v}^{v;y}(t)+[\dot S^{v;y}_{pump}(t) -\dot S^{v}_{pump}(t)] \,,
\end{align}
from which follows the inequality
\begin{align}
\label{Eqin5}
\dot S^{v}_{pump}(t) \le \dot S^{v;y}_{pump}(t)+\dot I_{flow,v}^{v;y}(t) \ .
\end{align}
A similar inequality was derived in \cite{HS2014} without including the second entropy pumping rate $\dot S^{v;y}_{pump}$ as the force  $F(x)$ was absent from the Langevin equation.

\subsection{Inequality derived from time reversal}

\subsubsection{Generalized local detailed balance}

We have so far considered entropy production rates, heat or information flows at the ensemble level, using probability distribution functions and FP equations without any reference to time irreversibility.  Our approach was  quite similar to previous analyses of Markovian diffusion processes\cite{KQ2004,AJM2009,MR2013,HS2014} because the non-Markovian character of the dynamics was hidden in the effective forces ${\overline F}_{fb}(x,v,t), {\overline F}_{fb}(v,t)$ or in the information flows $\dot I_{flow,v}^{xv;y}(t),\dot I_{flow,v}^{v;y}(t)$. These quantities actually  depend on the two-time probability distribution functions $p(x,v,t;y,t-\tau)$ or $p(v,t;y,t-\tau)$, which themselves depend on three-time pdf's, etc.

On the other hand, the non-Markovian nature of the feedback  is immediately manifest when one examines the behavior of the path probabilities under time reversal. The basic relation that we need to generalize is the microscopic reversibility condition (or local detailed balance) that links dissipation to time-reversal symmetry breaking\cite{K1998,C1999,LS1999,M2003,G2004,S2012}.  For a typical Markovian stochastic dynamics, \textit{e.g.}, an underdamped Langevin equation  describing the time-evolution of a Brownian particle driven by an external time-dependent parameter $\lambda(t)$, this relation reads\cite{IP2006}
\begin{align}
\label{EqLBE}
 \frac{{\cal P}_{\lambda}[{\bf X}\vert {\bf x}_i]}{{\cal P}_{\lambda^\dag}[{\bf X}^\dag\vert {\bf x}_i^\dag]}=e^{\beta q_{\lambda}[{\bf X}]}\ ,
\end{align}
where $q_{\lambda}[{\bf X}]$ is the heat dissipated along the forward path ${\bf X}$ over the time interval $[t_i,t_f]$ (corresponding to a change $\Delta s_{m,\lambda}[{\bf X}]=\beta q_{\lambda}[{\bf X}]$ in the entropy of the medium),  ${\cal P}_{\lambda}[{\bf X}\vert {\bf x}_i]$ is the  probability to observe the path, given its initial point ${\bf x}_i=(x_{t_i},v_{t_i})$,  and ${\cal P}_{\lambda^\dag}[{\bf X}^\dag\vert {\bf x}_i^\dag]$ is the probability to observe its time-reversed image ${\bf X}^\dag$ (in which the sign of the momenta is reversed), given its initial point ${\bf x}_i^\dag=(x_{t_f},-v_{t_f})$. The subscript $\lambda^\dag$ refers to the reverse or ``backward'' process in which the protocol $\lambda(t)$ is time reversed.  In other words, Eq. (\ref{EqLBE}) tells us that  heat is recovered from the time-antisymmetric part of the Onsager-Machlup action\cite{IP2006}. It is of course crucial that the fluctuating heat itself be an odd quantity under time reversal, \textit{i.e.}, $q_{\lambda^\dag}[{\bf X}^\dag]=-q_{\lambda}[{\bf X}]$.

It is readily seen from the definition of the heat, Eq. (\ref{EqqXY}), that this latter property remains true in the presence of a time-delayed continuous feedback {\it provided that} the time-reversal operation is combined with the change  $\tau\rightarrow-\tau$, 
\begin{align}
q_{-\tau}[{\bf X}^\dag,{\bf Y}^\dag]=-q_{\tau}[{\bf X},{\bf Y}]
\end{align}
(the subscript $\lambda$ is now replaced by $\tau$ since the feedback is the only force that drives the system out of equilibrium). The inversion  $\tau\rightarrow-\tau$ transforms the original dynamics into a ``conjugate" dynamics, hereafter denoted by the  $\sim$ symbol, and this is not a benign mathematical operation, as we now discuss.

Indeed, the main characteristic of the conjugate dynamics is that it breaks causality since the force $\widetilde F_{fb}(t)=F_{fb}(x_{t+\tau})$ now depends on  the future position of the Brownian entity. Therefore, this dynamics does not correspond to any physical  process. This unpleasant feature was carefully avoided when defining the backward experiment corresponding to a step-by-step nonautonomous  feedback protocol\cite{SU2010,HV2010}. In this case, the reverse process can be implemented by carrying out each step of the forward process in the reverse order {\it without} performing the feedback  control. This protocol has been carried out  in actual\cite{TSUMS2010} and numerical experiments \cite{SU2012} to check the generalized Jarzynski equality involving the so-called efficacy parameter\cite{SU2010,SU2012}.  It is clear that such an operation  is not possible in autonomous systems since the feedback force is applied continuously and is therefore affected by the time-reversal operation.

 However, from a formal point of view, the acausal character of the conjugate dynamics does not forbid the definition of mathematical objects that can be interpreted as path probabilities. The only issue is to properly take care of the normalization. Indeed, as stressed in \cite{MR2014},  the Jacobian of the transformation $\xi(t) \rightarrow x(t)$ associated with the conjugate Langevin equation in the time interval $[t_i,t_f]$ is no longer trivial since acausality makes  the (discrete) Jacobian matrix no longer  lower triangular.  In consequence, the Jacobian cannot be treated as an irrelevant multiplicative factor and the  conditional  path probability (obtained as usual by inserting the Langevin equation into the Gaussian measure for the noise)  takes the form
\begin{align}
\label{Eqpath2}
\widetilde{\cal P}[{\bf  X^\dag}\vert {\bf x}_i^\dag; {\bf  Y^\dag}]\propto  \vert \widetilde {\cal J}[{\bf X}]\vert  \: e^{-\beta  \widetilde {\cal S}[{\bf  X^\dag},{\bf Y^\dag}]}\ ,
\end{align}
where 
\begin{align}
\label{Eqaction1}
\widetilde  {\cal S}[{\bf  X^\dag},{\bf Y^\dag}]=\frac{1}{4\gamma}\int_{t_i}^{t_f}dt \:\Big[m\ddot x_t-\gamma \dot x_t-F(x_t)-F_{fb}(x_{t-\tau})\Big]^2 
\end{align}
is a well-defined, though  unusual, generalized Onsager-Machlup action functional. (One recovers the original feedback force $F_{fb}(x_{t-\tau})$ in the action by performing the two operations ${\bf  X}\rightarrow {\bf  X^\dag}$ and $t \rightarrow -t$.) Again, this can be made more rigorous by discretizing the time.   An important feature is that the Jacobian $\widetilde {\cal J}[{\bf X}]$ is in general a (positive) functional of the path. We also note that $\widetilde {\cal J}[{\bf X}]={\cal J}$ when $t_f-t_i\le \tau$ because $x_{t-\tau}$ does not belong to the trajectory ${\bf X}$ (and thus $x_{t+\tau}$ does not belong to the trajectory ${\bf X}^\dag$). The corresponding Jacobian matrix is  then again lower triangular.

Dividing  Eq. (\ref{Eqpath2}) by Eq. (\ref{Eqpath0}), we  obtain the generalized  local detailed balance with continuous time-delayed feedback  control:
\begin{align}
\label{EqELB1}
\frac{{\cal P}[{\bf X}\vert {\bf x}_i; {\bf Y}]}{\widetilde {\cal P}[ {\bf X}^\dag\vert {\bf x}_i^\dag; {\bf Y}^\dag]}=\frac{{\cal J}}{\widetilde {\cal J}[{\bf X}]} e^{\beta  q[{\bf X},{\bf Y}] } \ ,
\end{align}
where $q[{\bf X},{\bf Y}]$ is defined in Eq. (\ref{EqqXY}) and where we have taken into account the (forward) Jacobian ${\cal J}$ in order to get rid of the infinite normalization factors. 

\subsubsection{Fluctuation relation, entropy production, and second-law inequality}

In order to derive a fluctuation relation and  a second-law inequality from the identity (\ref{EqELB1}), we need to introduce some initial probabilities ${\cal P}_{init}[ {\bf x}_i; {\bf Y}]$ and ${\cal P}^\dag_{init}[{\bf x}_i^\dag; {\bf Y}^\dag]$ for the forward and backward processes and to define normalized weights ${\cal P}[{\bf X},{\bf Y}]={\cal P}[{\bf X}\vert {\bf x}_i; {\bf Y}]{\cal P}_{init}[ {\bf x}_i; {\bf Y}]$ and ${\cal P}^\dag [{\bf X}^\dag,{\bf Y}^\dag]=\widetilde {\cal P}[ {\bf X}^\dag\vert {\bf x}_i^\dag; {\bf Y}^\dag]{\cal P}^\dag_{init}[{\bf x}_i^\dag; {\bf Y}^\dag]$.  Of course, ${\cal P}_{init}[ {\bf x}_i; {\bf Y}]$ and ${\cal P}^\dag_{init}[{\bf x}_i^\dag; {\bf Y}^\dag]$ must be probabilities  on the space of trajectories generated by the  forward dynamics since this is the only physical one. (In passing, note that the time-reversal operation causes the trajectory ${\bf Y}^\dag$ to lie in the future of the trajectory ${\bf X}^\dag$.) This  leads us to define the functional
\begin{align}
\label{EqR}
R[{\bf X},{\bf Y}]&:= \beta  q[{\bf X},{\bf Y}]+\ln \frac{{\cal J}}{\widetilde {\cal J}[{\bf X}]}+\ln \frac{{\cal P}_{init}[{\bf x}_i;{\bf Y}]}{{\cal P}^\dag_{init}[{\bf x}_i^\dag;{\bf Y}^\dag]}\ ,
\end{align}
which by construction satisfies the integral fluctuation theorem (IFT)
\begin{align}
\label{EqIFT}
\langle e^{-R[{\bf X},{\bf Y}]}\rangle=1 
\end{align}
where  $\langle...\rangle$ denotes an average over all paths ${\bf X}$ and ${\bf Y}$ weighted by the probability ${\cal P}[{\bf X},{\bf Y}]$. A problem, however, is  that $R[{\bf X},{\bf Y}]$ is not a well-defined mathematical quantity for $t_f-t_i\ge \tau$  since the two trajectories ${\bf X}$ and ${\bf Y}$ are then contiguous and thus  ${\cal P}_{init}[{\bf x}_i;{\bf Y}]={\cal P}_{init}[{\bf Y}]\delta ({\bf y}_f-{\bf x}_i)$ (see the remark inside the parenthesis after Eq. (\ref{Eqpath1})). This makes the last term in the right-hand side of Eq. (\ref{EqR})  highly singular. As suggested in \cite{MR2014}, this problem can be fixed by integrating out the dependence on the trajectory ${\bf Y}$ and considering the coarse-grained path functional $R_{cg}[{\bf X}]$ defined by 
\begin{align}
e^{-R_{cg}[{\bf X}]}:=\frac{1}{{\cal P}[{\bf X}]}\int {\cal D}{\bf Y}\: {\cal P}[{\bf X},{\bf Y}]e^{-R[{\bf X},{\bf Y}]} \ .
\end{align}
By using the definition of $R[{\bf X},{\bf Y}]$, this can be simply rewritten as
\begin{align}
\label{EqRcg}
R_{cg}[{\bf X}]= \ln \frac{{\cal P}[{\bf X}]}{\widetilde {\cal P}[{\bf X}^\dag]}\ ,
\end{align}
where 
\begin{align}
\label{Eqpath3}
\widetilde {\cal P}[{\bf X}^\dag]:= \int {\cal D}{\bf Y}^\dag\: \widetilde {\cal P}[{\bf X}^\dag\vert {\bf x}_i^\dag;{\bf Y}^\dag]{\cal P}_{init}^\dag [{\bf x}_i^\dag;{\bf Y}^\dag]\ .
\end{align}
By construction, $R_{cg}[{\bf X}]$ obeys the IFT
\begin{align}
\label{EqIFT1}
\langle e^{-R_{cg}[{\bf X}]}\rangle=1 \ ,
\end{align}
where $\langle...\rangle$ now denotes the average over the paths ${\bf X}$ weighted by ${\cal P}[{\bf X}]$, and its expectation is expressed as the relative entropy or Kullback-Leibler divergence $D({\cal P}\vert\vert \widetilde {\cal P})$ between the distributions ${\cal P}$ and $\widetilde {\cal P}$:
\begin{align}
\label{EqRcgav}
\langle R_{cg}[{\bf X}]\rangle=\int {\cal D}{\bf X}\: {\cal P}[{\bf X}] \ln \frac{{\cal P}[{\bf X}]}{\widetilde {\cal P}[ {\bf X}^\dag]}\ .
\end{align}
Therefore $\langle R_{cg}[{\bf X}]\rangle$ is a non-negative quantity, which makes it a good candidate for a definition of  the total entropy production in the feedback-controlled system\cite{MR2014}.  In this respect,  Eq. (\ref{EqRcgav}) may be viewed as a generalization of the relation $\langle \Delta s[{\bf X}]\rangle=\int {\cal D}{\bf X}\: {\cal P}[{\bf X}] \ln {\cal P}[{\bf X}]/{\cal P}[ {\bf X}^\dag]$  that is commonly adopted as a mathematical (if not physical) definition of entropy production\cite{C1999,M2003,G2004,S2005}, even for non-Markovian stochastic processes\cite{GPVdB2008,DE2014,RP2012}.

All these equations remains quite abstract so far, and in order to make contact with physical quantities the initial probabilities ${\cal P}_{init}[ {\bf x}_i; {\bf Y}]$ and ${\cal P}^\dag_{init}[{\bf x}_i^\dag; {\bf Y}^\dag]$ must be properly chosen. This is  a more delicate issue than for a Markovian Langevin dynamics\cite{S2005,IP2006}, especially in a transient regime, and to avoid such complications we will now assume that the system has settled into a stable stationary state. As already emphasized, this is actually the normal working mode of autonomous systems (at least of artificial devices). For convenience, we will also symmetrize the time interval $[t_i,t_f]$ by setting $t_i=-{\cal T}$ and $t_f={\cal T}$ as in \cite{MR2014} (${\cal T}$ is a time and should not be confused with the bath temperature $T$). 
The natural choice for the initial probabilities is then $p_{init}({\bf x}_i)=p_{st}(x_{-{\cal T}},v_{-{\cal T}})$, $p_{init}^\dag({\bf x}_i^\dag)=p_{st}(x_{{\cal T}},v_{{\cal T}})$, and ${\cal P}_{init}[{\bf Y}\vert  {\bf x}_i]={\cal P}_{st}[{\bf Y}\vert  {\bf x}_i]$, ${\cal P}^\dag_{init}[ {\bf Y}^\dag\vert {\bf x}_i^\dag]={\cal P}_{st}[ {\bf Y}^\dag\vert {\bf x}_i^\dag]$.

These are rather complicated quantities, though, and computing $\langle R_{cg}[{\bf X}]\rangle_{st}$ remains a difficult task even for a linear  Langevin equation. (Some perturbative calculations will be performed in paper II in the overdamped limit, in relation with fluctuation relations.) On the other hand, since  in this  work we are only interested in deriving second-law inequalities, we may consider the long-time limit ${\cal T}\gg \tau$, which allows us to neglect the influence of the past trajectory ${\bf Y}$. Indeed, since the  duration of ${\bf Y}$ is at most  $\tau$, the dependence  of $R_{cg}[{\bf X}]$  on ${\bf Y}$ is a temporal boundary term, as can be seen by splitting  the integration over $t$  in the  Onsager-Machlup actions $ {\cal S}[{\bf  X},{\bf Y}]$ and $\widetilde  {\cal S}[{\bf  X^\dag},{\bf Y^\dag}]$ into $\int_{-{\cal T}}^{-{\cal T}+\tau}$ and $\int_{-{\cal T}+\tau}^{{\cal T}}$, which yields
\begin{align}
\frac{{\cal P}[{\bf X}]}{\widetilde {\cal P}[{\bf X}^\dag]}&=\frac{{\cal J}}{\widetilde {\cal J}[{\bf X}]}e^{-\beta \int_{-{\cal T}+\tau}^{{\cal T}}dt \: [m\dot v_t-F(x_t)-F_{fb}(x_{t-\tau})]v_t}
\nonumber\\
&\times \frac{\int {\cal D}{\bf Y}\: e^{-\beta \int_{-{\cal T}}^{-{\cal T}+\tau}dt\: {\cal S}[{\bf  X},{\bf Y}]}{\cal P}_{st} [{\bf x}_i;{\bf Y}]}{\int {\cal D}{\bf Y}^\dag\: e^{-\beta \int_{-{\cal T}}^{-{\cal T}+\tau}dt \: \widetilde {\cal S}[{\bf  X^\dag},{\bf Y^\dag}]}{\cal P}_{st}^\dag [{\bf x}_i^\dag;{\bf Y}^\dag]}\ .
\end{align}
So long as we deal with expectation values, the last term can be neglected in the long-time limit and from Eq. (\ref{EqRcgav}) we finally obtain the equation 
\begin{align}
\label{Eqcentral}
\dot {\cal R}_{cg}=\frac{ \dot {\cal Q}}{T}+\dot {\cal S}_{\cal J}\ ,
\end{align}
where we have defined the two asymptotic  rates
\begin{align}
\label{EqdefRcg}
\dot {\cal R}_{cg}&:= \lim_{{\cal T}\rightarrow \infty}\frac{1}{2\cal T}\langle R_{cg}[{\bf X}]\rangle_{st}\nonumber\\
\dot {\cal S}_{\cal J}&:= \lim_{{\cal T}\rightarrow \infty}  \frac{1}{2{\cal T}}\langle\ln \frac{\cal J}{\widetilde {\cal J}[{\bf X}]}\rangle_{st}\ ,
\end{align}
and used the fact that the heat flow $ \dot {\cal Q}$ is constant in the stationary state.  (Here, for convenience we define $\dot {\cal S}_{\cal J}$ with the opposite sign to that adopted in  \cite{MR2014}.) Since $\langle R_{cg}[{\bf X}]\rangle_{st}$ is always non-negative, Eq. (\ref{Eqcentral}) leads to the sought second-law-like inequality associated with time reversal:
\begin{align}
\label{Eq2law3}
\frac{ {\cal {\dot W}}_{ext} }{T} \le \dot {\cal S}_{\cal J}\ .
\end{align}
 We stress that the new quantity $\dot {\cal S}_{\cal J}$ comes from a calculation that has nothing to do with a coarse-graining procedure performed at the level of a Fokker-Planck equation. Therefore, there is no obvious relationship between $\dot {\cal S}_{\cal J}$ and  the entropy pumping rates $\dot {\cal S}_{pump}^{xv}$ and $\dot {\cal S}_{pump}^v$ defined earlier. This will be confirmed by explicit calculations in Sec. V. However, these three quantities $\dot {\cal S}_{\cal J}, \dot {\cal S}_{pump}^{xv}, \dot {\cal S}_{pump}^v$ become equal in the case of a linear feedback $F_{fb}(t) = k' x_{t-\tau}$ and in the small-$\tau$ limit where one recovers the Markovian model of \cite{KQ2004} (see the remark after Eq. (\ref{EqSpumpxv})). Since the effective force is then proportional to the instantaneous velocity at first order in $\tau$,  changing $\tau$ into $-\tau$ is equivalent to changing $\gamma'=k'\tau$ into $-\gamma'$, which leads to ${\cal J}/\widetilde{\cal J}=e^{2{\cal T}\gamma'/m}$\cite{MR2012,IP2006}. Hence $\dot {\cal S}_{\cal J}=\dot {\cal S}^{xv}_{pump}=\dot{\cal S}^{v}_{pump}=\gamma'/m$ in this limit. Let us note in passing that the change $\gamma'\rightarrow -\gamma'$, although it does not involve any breaking of causality, is  not a benign mathematical operation either. As stressed in  \cite{MR2012}, thermal  fluctuations are then enhanced, and there exists no stationary state with the conjugate dynamics if $\gamma'>\gamma$.

\subsubsection{Expression of the Jacobian}

For the inequality (\ref{Eq2law3}) to be quantitative, we must have at our disposal an expression of the Jacobian $\widetilde {\cal J}[{\bf X}]$  suitable for explicit calculations of $\dot {\cal S}_{\cal J}$. This is obtained by using the operator representation of the conjugate Langevin equation in the continuum limit. Following \cite{OO2007,ABC2010}, we formally express the Jacobian as the functional determinant  
\begin{align}
\label{EqtildeJ}
\widetilde {\cal J}[{\bf X}]&=\mbox{det}_{tt'}\left[m\partial_t^2 \delta_{t-t'}+\gamma \partial_t \delta_{t-t'}-\widetilde F'_{tot}(t,t')\right]\nonumber\\
&=\mbox{det}_{tt'}\left[m\partial_t^2 \delta_{t-t'}+\gamma \partial_t \delta_{t-t'}\right]\mbox{det}_{tt'}\left [\delta_{t-t'}-\widetilde M_{tt'}\right]
\end{align}
where 
\begin{align}
\label{EqFptot}
\widetilde F_{tot}^{'}(t,t')&:= \frac{\delta}{\delta x_{t'}} [F(x_t)+F_{fb}(x_{t+\tau})]\nonumber\\
&=\delta(t-t')F'(x_{t})+\delta(t+\tau-t')F'_{fb}(x_{t+\tau})\ ,
\end{align}
and $\widetilde M(t,t')$  is the operator defined by
\begin{align}
\label{EqdefM}
\widetilde M_{tt'}&:= \{G\circ \widetilde F'_{tot}\}_{tt'}\nonumber\\
&= \int_{-{\cal T}}^{{\cal T}} dt''\: G(t-t'')\widetilde F'_{tot}(t'',t') \ .
\end{align}
$G(t)$ is the retarded Green function for the inertial and dissipative terms in the Langevin equation; it is solution to the equation  
\begin{align}
m\partial^2_t G(t-t')+\gamma \partial_tG(t-t')=\delta(t-t')\ ,
\end{align}
\textit{i.e.},
\begin{align}
G(t)=\gamma^{-1}[1-e^{-\gamma t/m}]\Theta(t)
\end{align} 
where $\Theta(t)$ is the Heaviside step function.  From Eqs. (\ref{EqFptot}) and (\ref{EqdefM}), it is natural to split $\widetilde M(t,t')$ into two pieces,
 \begin{align}
\widetilde M(t,t')=M^{(0)}(t,t')+\widetilde M^{(1)}(t,t')\ ,
\end{align}
where 
\begin{align}
\label{EqM0}
M^{(0)}(t,t')&=G(t-t')F'(x_{t'})
\end{align}
and 
\begin{align}
\label{EqM1}
\widetilde M^{(1)}(t,t')&=G(t+\tau-t')F'_{fb}(x_{t'})\Theta({\cal T}-\tau+t') \ .
\end{align}
In these equations, we have restricted $t$ and $t'$ to the time interval $[-{\cal T},{\cal T}]$. After dividing $\widetilde {\cal J}[\bf X]$ by the corresponding expression of ${\cal J}$ that  involves only $M^{(0)}(t,t')$ and using the matrix  identity $\ln \det =  \mbox{Tr} \ln$, we finally  arrive at the expression
\begin{align}
\label{EqdefJ}
\ln \frac{\widetilde {\cal J}[{\bf X}]}{{\cal J}}&=\mbox{Tr}_{t,t'}\ln [\delta_{t-t'}-\psi_{t,t'}]\nonumber\\
&\equiv-\sum_{n=1}^{\infty}\frac{1}{n}\int_{-{\cal T}}^{{\cal T}} dt\: \Big\{\underbrace{\psi\circ \psi\circ...\psi}_{n \: \mbox{times}}\Big\}_{tt}\ ,
\end{align}
where 
\begin{align}
\label{Eqpsi}
\psi(t,t')&:= (K\circ \widetilde M^{(1)})_{tt'}
 \end{align}
and $K$ is the operator inverse of $\delta -M^{(0)}$ with respect to the operation $\circ$ (which becomes the  standard convolution in the long-time limit ${\cal T}\rightarrow \infty$).
Eq. (\ref{EqdefJ}) will be the starting point of the calculations performed in the next section. (Note that the second line of Eq. (\ref{EqdefJ}) is {\it a priori} only valid when the series converges.)

\section {Application to linear systems}

In order to make the preceding general analysis more concrete, we now perform a detailed study of a stochastic harmonic oscillator driven by a linear feedback control. The dynamics is governed by the second-order linear Langevin equation
\begin{align}
\label{EqLlin}
m\dot v_t=-kx_t+ k' x_{t-\tau} -\gamma v_t+\sqrt{2\gamma T}\xi_t 
\end{align}
with $k>0$.
 As pointed out in the Introduction,  this model is both analytically tractable and relevant to practical applications. In particular, with a positive feedback  ($k'>0$), the equation  faithfully describes the motion of feedback-cooled nano-mechanical  resonators  in the vicinity of their fundamental mode resonance (\textit{e.g.}, the mirror of an interferometric detector\cite{PCBH2000}, a cavity optomechanical system\cite{LRHKB2010}, or the cantilever of an AFM\cite{Mont2012}). Similar equations have also been used to describe the electromechanical modes of a gravitational-wave detector\cite{V2008,B2009} or a macroscopic electrical resonator\cite{VBMF2010}. On the other hand,  Eq. (\ref{EqLlin}) with  $k'<0$ is an archetype of a negative feedback loop  operating in a biological regulatory network, whose role is to maintain homeostasis. For instance, the model may describe  some of the behavior observed in the neuromuscular regulation of movement and posture  (see \cite{BR2000,M2009,PFFBT2006} and references therein). 

Like most SDDEs\cite{M1984,M2008,L2010}, Eq (\ref{EqLlin}) has a rich dynamical behavior, but for the purpose of this work we restrict our attention  to the stationary regime in which all probability distribution functions and probability currents become time-independent. Even so, we first need to determine the domain(s) of existence of the stationary state(s) in the parameter space. This task is accomplished in subsection A where we also give the stationary expressions of the various rates and investigate some useful limits. In subsection B, we then compute the asymptotic rate $\dot S_{{\cal J}}$ that enters the second-law inequality (\ref{Eq2law3}).  Finally, in subsection C, we numerically study and discuss a few examples. 

To begin with,  let us briefly recall the role of time delay in the context of active feedback-cooling\cite{PZ2012}, since we will often refer to this experimental technique in the discussion. This also allows us to introduce the parameters that commonly describe the dynamical behavior of  a damped harmonic oscillator and to introduce a special limit in the parameter space that will play a significant role in the following. 

In the stationary state, the relevant experimental quantity for mechanical oscillators is the power spectral density (PSD) of the displacement induced by thermal excitation, $S_{xx}(\omega)=\vert \chi(\omega)\vert^2 S_{F_{th}}$, where $S_{F_{th}}=2\gamma T$ is the PSD of the thermal noise\cite{note3}  and
\begin{align}
\label{Eqchi}
\chi(\omega)=\frac{1}{(-m\omega^2+k-k'\cos \omega \tau)-i (\gamma \omega +k'\sin \omega \tau)}
\end{align}
is the frequency response function of the oscillator. By definition, $S_{xx}(\omega)=\int_{-\infty}^{+\infty} dt \: e^{i \omega t} \phi_{xx}(t) $ is the Fourier transform of the time-correlation function $\phi_{xx}(t)=\langle x(0)x(t)\rangle_{st}$ which is time-translation invariant in a stationary state. In the absence of feedback, $S_{xx}(\omega)$ has the familiar Lorentzian  shape
\begin{align}
\label{EqSequil}
S_{xx}^{g=0}(\omega)=\frac{(1/m^2)S_{F_{th}}}{(\omega_0^2-\omega^2)^2+\omega^2/\tau_0^2} \ ,
\end{align}
where  $\omega_0=\sqrt{k/m}$ is the angular resonance frequency of the oscillator and $\tau_0=m/\gamma$ is the viscous relaxation time. When the feedback is operating, the PSD becomes 
\begin{align}
\label{EqSxx}
S_{xx}(\omega)= \frac{(1/m^2) S_{F_{th}}}{[\omega_0^2(1-\frac{g}{Q_0} \cos \omega \tau)-\omega^2]^2+(\omega+g\omega_0\sin \omega \tau)^2/\tau_0^2} 
\end{align}
where $Q_0=\omega_0\tau_0$ is  the intrinsic quality factor of the resonator and $g=k'/(\gamma \omega_0)=(k'/k) Q_0$ is a dimensionless parameter that represents the gain of the feedback loop.

In experiments performed with high-quality resonators ($Q_0\gg 1$), aiming at studying the quantum regime of mechanical motion\cite{PZ2012}, the feedback control is  designed to reduce thermal fluctuations  as much as possible in a small frequency band around $\omega_0$. In practice, this is done by tuning the phase of the amplifier inside the feedback loop such that $\tau=\tau^{opt}=\pi/(2\omega_0)$\cite{PCBH2000,LRHKB2010,Mont2012,V2008,B2009,VBMF2010}.  The signal applied to the actuator (for instance, a piezoelectric element mechanically coupled to the resonator) is then in quadrature with the displacement, and in the vicinity of  $\omega_0$ one has
\begin{align}
\label{EqSxxopt}
S_{xx}^{opt}(\omega)\approx \frac{(1/m^2)S_{F_{th}}}{(\omega_0^2-\omega^2)^2+(1+g)^2\omega^2/\tau_0^2}\ .
\end{align}
 This PSD is identical to the one that would be obtained by directly applying to the resonator a viscous damping force $F_{fb}(t)=-g\gamma \dot x_t$ (compare with Eq. (\ref{EqSequil})). Integrating Eq. (\ref{EqSxxopt}) over frequency  then yields the effective temperature of the fundamental mode\cite{PCBH2000}
\begin{align}
\label{EqTeffx}
T_x^{opt}= k \langle x^2\rangle_{st}=\frac{k}{\pi} \int_{0}^{+\infty} d\omega \:S_{xx}^{opt}(\omega)\approx \frac{T}{1+g}  
\end{align}
which decreases monotonically as the gain $g$ increased. (In practice, the detector noise, which we neglect here,  imposes a minimum achievable mode temperature\cite{PDMR2007}.) Therefore, as regards the reduction of thermal fluctuations, the non-Markovian Langevin Eq. (\ref{EqLlin}) with $\tau=\pi/(2\omega_0)$ is equivalent in the limit $Q_0\gg 1$ to the Markovian equation
\begin{align}
\label{EqCD}
m\dot v_t=-kx_t-(1+g)\gamma v_t+\sqrt{2\gamma T}\xi_t
\end{align}
which is often used to describe the  cold-damping technique\cite{LMCSYS2000,KQ2004,MR2012}.

\subsection{Stationary distributions and stability analysis}

For analyzing the behavior of the system in the whole parameter space and performing  numerical studies, it is  convenient to recast Eq. (\ref{EqLlin})  into a dimensionless version.  This is done by using $\omega_0^{-1}$ and $x_c:= k^{-1}(2\gamma\omega_0T)^{1/2}$ as units of time and position\cite{NF2011}, which yields
\begin{align}
\label{EqLlinred}
\dot v_t= -x_t-\frac{1}{Q_0} v_t+\frac{g}{Q_0}x_{t-\tau}+\xi_t
\end{align}
where $\tau$ is  the rescaled time delay (\textit{i.e.},  $\omega_0\tau \rightarrow \tau$). This shows that there are only three independent dimensionless parameters in the problem, $Q_0,g$, and $\tau$, and we will always use these reduced units hereafter, except when stated explicitly.

\subsubsection{Stationary distributions}

Since Eq. (\ref{EqLlinred}) is linear and the noise is Gaussian, all stationary pdf's, solutions of the Fokker-Planck equations, are multivariate Gaussian distributions. In particular,  $p_{st}(x,v)$ is given by 
\begin{align}
\label{Eqpdf1}
p_{st}(x,v)&=\frac{1}{2\pi[\langle x^2\rangle_{st}\langle v^2\rangle_{st}]^{1/2}}e^{-\frac{x^2}{2\langle x^2\rangle_{st}}-\frac{v^2}{2\langle v^2\rangle_{st}}}\ ,
\end{align}
where $\langle x^2\rangle_{st}$ and $\langle v^2\rangle_{st}$ are the mean square position and velocity, from which we may define the  two effective temperatures $T_x:=(2T/Q_0)\langle x^2\rangle_{st}$ and $T_v:=(2T/Q_0) \langle v^2\rangle_{st}$ (\textit{i.e.}, $T_x=k\langle x^2\rangle_{st}$ and $T_v=m\langle v^2\rangle_{st}$ in real units). $T_x$ is the configurational temperature that is commonly measured in experiments\cite{PCBH2000,Mont2012,PDMR2007}(see Eq. (\ref{EqTeffx}) above), whereas $T_v$ is the kinetic temperature that determines the heat flow (and thus the extracted work) in the stationary state according to Eq. (\ref{EqHeat}), \textit{i.e.},
\begin{align}
\label{EqHeat1}
\frac{\dot {\cal Q}}{T}=-\frac{\dot {\cal W}_{ext}}{T}=\frac{1}{Q_0} (\frac{T_v}{T}-1)
\end{align}
in reduced units. As will be seen later, these two temperatures are in general different.

In order to compute the mutual informations ${\cal I}^{xv;y}$ and $ {\cal I}^{v;y}$, the corresponding information flows $\dot {\cal I}_{flow,v}^{xv;y}$ and $\dot {\cal I}_{flow,v}^{v;y}$, and the entropy contribution  $\dot {\cal S}_{pump}^{v;y}$ defined by Eq. (\ref{Eqpump3}), we also need the explicit expression of the two-time pdf $p_{st}(x,v;y)$. This one is obtained by substituting a  general 3-variate Gaussian distribution into Eqs. (\ref{EqKramers1})-(\ref{EqFfb0}) and solving the linear equations for the coefficients of $x^2, v^2,y^2,xv,xy,vy$.  This leads to
\begin{align}
\label{Eqpdf2}
p_{st}(x,v;y)&=\frac{T^{3/2}}{(\pi Q_0)^{3/2}T_v^{1/2}(T_x^2(1-\alpha)-\Delta T_{xv}^2)^{1/2}}\nonumber\\
&\times \exp\{-\frac{T}{Q_0(T_x^2(1-\alpha)-\Delta T_{xv}^2)}[T_x(1-\alpha)x^2\nonumber\\
&+\frac{1}{T_v}(T_x^2-\Delta T_{xv}^2)v^2+T_x y^2-2\Delta T_{xv} x\:y\nonumber\\
&-\frac{2}{g}\frac{T-T_v}{T_v}(\Delta T_{xv} x-T_xy)v]\}
\end{align}
with $\alpha=(1/g^2)(T-T_v)^2/(T_xT_v)$ and $\Delta T_{xv}=(Q_0/g)(T_x-T_v)$. This expression is rather unenlightening, but the important feature is that the coefficients in the quadratic form are functions of $T_x$ and $ T_v$. This illustrates the fact  that the FP equation  does not provide sufficient information to compute these two quantities. 
The resulting expression of the effective force defined by Eq. (\ref{EqFfb0}) is more physically transparent, 
\begin{align}
\label{EqFfb1}
{\overline F}_{fb,st}(x,v)=(1-\frac{T_v}{T_x})x+\frac{1}{Q_0}(1-\frac{T}{T_v})v\ ,
\end{align} 
as it shows that the effect of the continuous feedback (at least at the level of the first FP equation) is to modify both the spring constant and the viscous damping. Accordingly, the  probability current  $J^v(x,v)$ in the FP equation takes the simple form $J_{st}^v(x,v)=-(T_v/T_x)x\: p_{st}(x,v)$. Since ${\overline F}_{fb,st}(x,v)$ is linear in $x$ and $v$, the effective viscous damping does not change when integrating out the dependence on $x$, and the  effective force ${\overline F}_{fb,st}(v)$ defined by Eq. (\ref{EqbarFfbv}) is
\begin{align}
\label{EqFfb2}
{\overline F}_{fb,st}(v)=\frac{1}{Q_0}(1-\frac{T}{T_v})v \ .
\end{align}
From Eqs. (\ref{EqSpumpxv}) and (\ref{EqSpumpv}), we  immediately obtain the expression of the entropy pumping rates 
\begin{align}
\label{EqSpumpv1}
\dot {\cal S}^{xv}_{pump}=\dot {\cal S}^{v}_{pump}=\frac{1}{Q_0}(\frac{T}{T_v}-1) \,,
\end{align}
which is the same as in \cite{MR2013,HS2014} but with a different kinetic temperature.  This equation shows that the entropy pumping rate is maximum when $T_v$ is minimum, as is the case for the extracted work rate.
The equality of $\dot {\cal S}^{xv}_{pump}$ and $\dot {\cal S}^{v}_{pump}$ is again  due to the linearity of the model, which implies that ${\overline F}_{st}(v)=\int F(x)p_{st}(x\vert v)=0$ in Eq. (\ref{EqSpumpv}). It is clear that this property is no longer true when nonlinearities become significant, for instance when the displacement of the resonator from its equilibrium position has a large amplitude (see \textit{e.g.} \cite{DK1984}).   

The expressions of ${\cal I}^{xv;y}$, $ {\cal I}^{v;y}$, $\dot {\cal I}_{flow,v}^{xv;y}$, $\dot {\cal I}_{flow,v}^{v;y}$, and  $\dot {\cal S}_{pump}^{v;y}$ are somewhat more involved. After inserting Eq. (\ref{Eqpdf2}) into Eqs. (\ref{EqMIxvy})-(\ref{EqIflow3}) and (\ref{Eqpump3}), we obtain
 \begin{align}
 \label{EqMIxvy21}
{\cal I}^{xv;y}=-\frac{1}{2}\ln [1- \frac{Q_0^2(T_v-T_x)^2+\frac{T_x}{T_v}(T-T_v)^2}{g^2T_x^2}]
\end{align}
 \begin{align}
 \label{EqMIvy21}
 {\cal I}^{v;y}= -\frac{1}{2}\ln  [1-\frac{(T-T_v)^2}{g^2T_xT_v}]\ ,
\end{align}
 \begin{align}
 \label{EqIflow21}
\dot {\cal I}_{flow,v}^{xv;y}&= \frac{1}{Q_0}\frac{T-T_v}{T_v}\frac{g^2T_x^2 -Q_0^2(T_v-T_x)^2+T_x(T-T_v)}{g^2T_x^2-Q_0^2(T_v-T_x)^2
-\frac{T_x}{T_v}(T-T_v)^2}\ ,
\end{align}
and 
 \begin{align} 
\label{Eqin6}
\dot {\cal S}_{pump}^{v;y}+\dot {\cal I}_{flow,v}^{v;y}=\frac{1}{Q_0}(T-T_v)\frac{g^2T_x+T-T_v}{g^2T_xT_v-(T-T_v)^2}\ .
 \end{align}
For brevity, we do not give the separate expressions of  $\dot {\cal I}_{flow,v}^{v;y}$ and $\dot {\cal S}_{pump}^{v;y}$ since only their combination enters in inequality (\ref{Eqin5}). (Note that the effective force ${\overline F}(v,y,t)$ defined by Eq. (\ref{EqeffForce}), and resulting from the coarse-graining over $x$, does not vanish in the stationary state, in contrast with the other effective force $\overline F(v,t)$; this makes the entropy pumping contribution $\dot {\cal S}_{pump}^{v;y}$  nonzero.)
For future reference, note also that all the rates vanish when $T_v=T$ (for $g\neq 0$), as does the mutual information $ {\cal I}^{v;y}$.

The above expressions only make sense if the system operates in a steady-state regime. This amount to imposing that the two variances $\langle x^2\rangle_{st}$ and $\langle v^2\rangle_{st}$  (or, equivalently, the effective temperatures $T_x$ and $T_v$) remain finite and positive. 
The next task is thus to determine  the regions(s) of the parameter space $(Q_0,g,\tau)$ in which these conditions are satisfied.

\subsubsection{Effective  temperatures and stability of the stationary state}

In principle, the variances  $\langle x^2\rangle_{st}$ and $\langle v^2\rangle_{st}$  can be obtained by integrating the corresponding fluctuation spectra $S_{xx}(\omega)$ and $\omega^2S_{xx}(\omega)$ over frequency (see \textit{e.g.} Eq. (\ref{EqTeffx})). However, this does not yield straightforward analytical derivations. Instead, we  determine these quantities  by solving the linear differential equation obeyed by the stationary time-correlation function $\phi_{vv}(t)$ in the time interval $0\le t\le \tau$. This procedure  is similar to that used in \cite{PFFBT2006} and earlier works\cite{KM1992,FBF2003},  and  the calculation is detailed  in Appendix B. We eventually arrive at the following expressions
\begin{align}
\label{EqTx}
\frac{T_x}{T}=\frac{1}{Q_0}\: \frac{\omega_1 f(\omega_2 )-\omega_2 f(\omega_1)}{\omega_1\omega_2(\omega_2^2-\omega_1^2)} 
\end{align}
and
\begin{align}
\label{EqTv}
\frac{T_v}{T}=\frac{1}{Q_0}\: \frac{\omega_2f(\omega_2)- \omega_1f(\omega_1)}{\omega_2^2-\omega_1^2} \ ,
\end{align}
where $\omega_1, \omega_2$ are given by Eq. (\ref{Eqroots0}) and the function $f(\omega)$ is defined by Eq. (\ref{Eqfomega}). The quantities $\omega_1$ and $\omega_2$ may be complex but $T_x$ and $T_v$ are real and positive as long as a stationary solution exists. One can check that Eqs. (\ref{EqTx}) and  (\ref{EqTv}) are  in agreement with the numerical integration of $S_{xx}(\omega)$ and $\omega^2S_{xx}(\omega)$ over frequency. 

For given values of $Q_0$ and $g$, the stability of a stationary solution is determined by critical values of the delay at which the temperatures $T_x$ and $T_v$  diverge. These values correspond to Hopf bifurcations at which the fixed point $x^*=0$ of the  deterministic second-order differential equation associated with Eq. (\ref{EqLlinred}) loses its stability (see \textit{e.g.} \cite{PFFBT2006} for a detailed study  of another second-order linear SDDE). The local stability analysis of this  equation is reported in Appendix C, where we extend the  studies performed in \cite{CG1982,CBOM1995}  to the case of a positive feedback. The results are summarized in Fig. \ref{Figdiag}. 
\begin{figure}[hbt]
\begin{center}
\includegraphics[width=8cm]{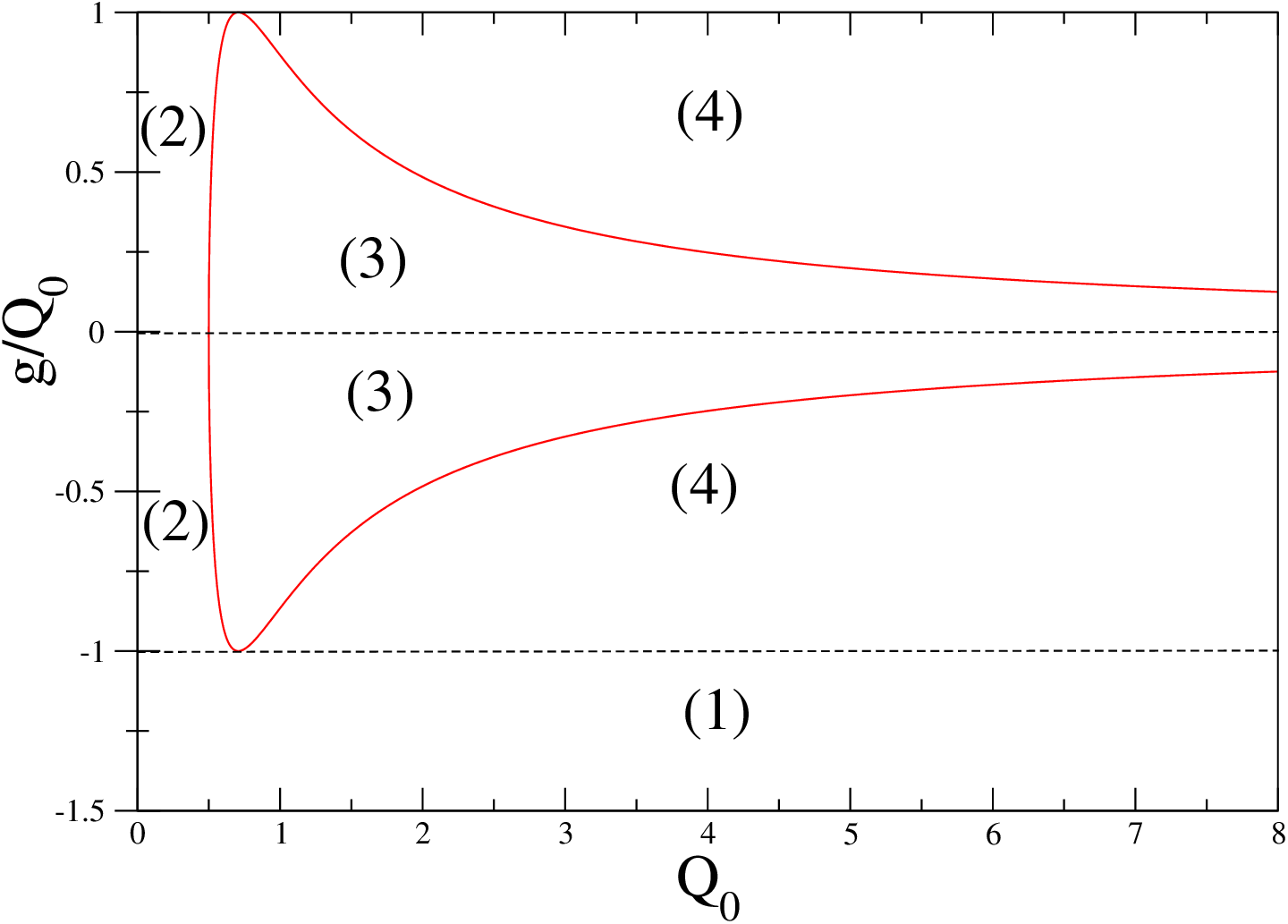}
\caption{\label{Figdiag} (Color on line) ``Phase" diagram of the stationary state in the plane $(Q_0,g/Q_0)$. The regions labeled 1-4 are each  characterized by a different  behavior of  the effective temperatures $T_x$ and $T_v$  as a function of $\tau$ (region 1 extends from $g/Q_0=-1$ to $-\infty$ but is arbitrarily truncated at $g/Q_0=-1.5$). }
\end{center}
\end{figure}

As explained in the Appendix, there are four regions in the plane $(Q_0,g/Q_0)$, each one corresponding to a different behavior of the effective temperatures $T_x$ and $T_v$ as a function of $\tau$.  In region 1, a stationary solution exists up to a certain critical value of $\tau$ at which the temperatures diverge.  In regions 2 and 3, a stationary solution exists for all $\tau\ge 0$ (but the behavior of the variances is different in the two regions). In region 4, there is an increasing  sequence of stability thresholds, up to a certain critical value beyond which there is no stationary solution. This  feature gives rises to a characteristic ``Christmas tree'' stability diagram\cite{CBOM1995} like the one displayed in Fig. \ref{FigChristmas_tree}. Such a multistability regime is observed in a wide variety  of biological systems and is also relevant to  feedback-cooled nano-mechanical resonators with a high quality factor, as discussed  in the next subsection. 
\begin{figure}[hbt]
\begin{center}
\includegraphics[width=9cm]{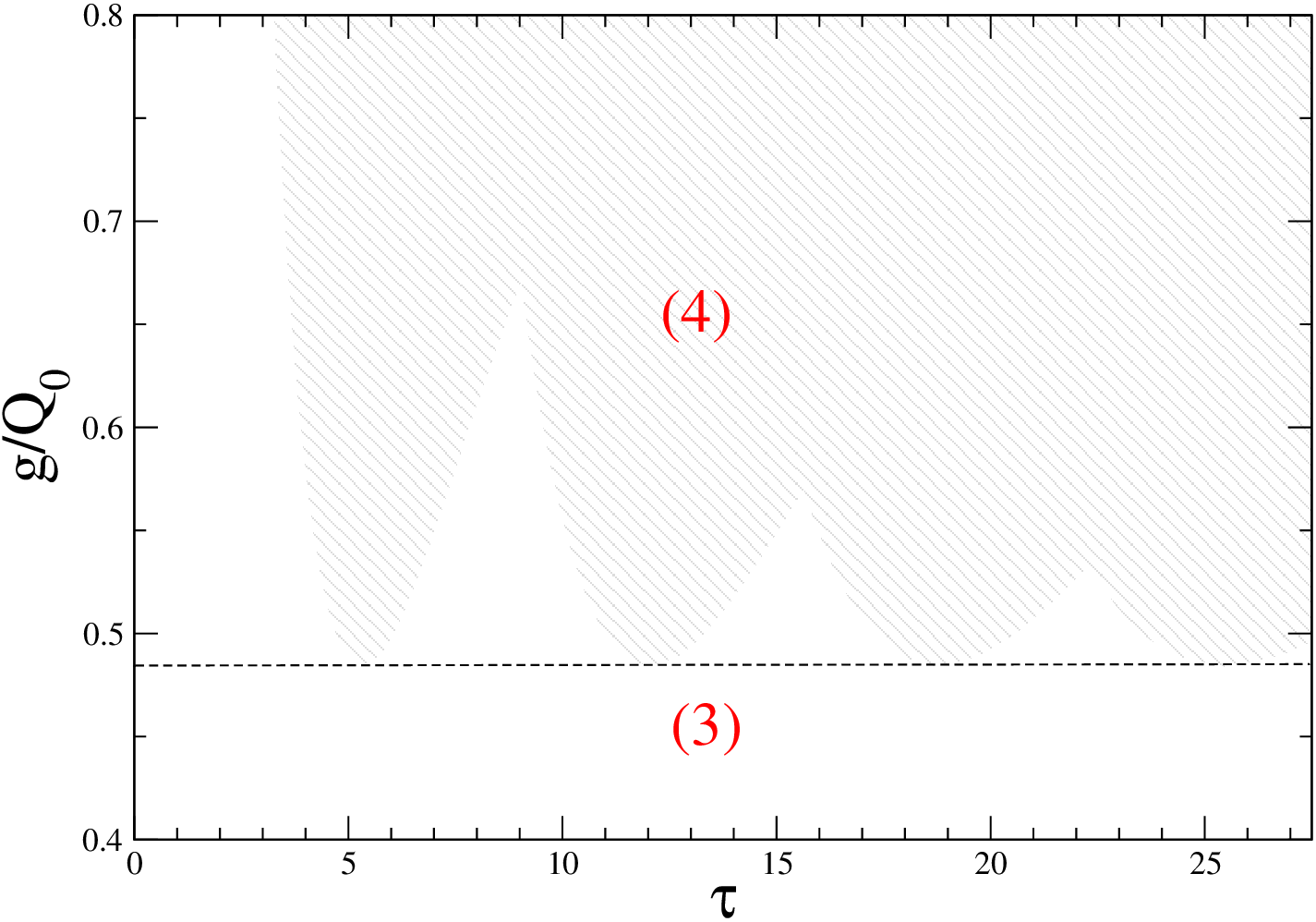}
 \caption{\label{FigChristmas_tree} (Color on line) ``Christmas-tree" stability diagram in the $(\tau,g/Q_0)$ plane  for $Q_0=2$. The horizontal dashed line $g/Q_0=(1/Q_0)\sqrt{1-1/(4Q_0^2)}\approx 0.484$ separates regions 3 and 4 in Fig. \ref{Figdiag} (see Appendix C). In the multistability region 4, the feedback-controlled oscillator reaches a stable stationary state in the zones located outside the shaded area.}.
\end{center}
\end{figure}

\subsubsection{Some useful limits}

Since the temperatures $T_x$ and $T_v$ given by Eqs. (\ref{EqTx})-(\ref{EqTv}) are complicated functions of $Q_0,g$ and $\tau$, the full behavior of the system can only be studied numerically, as will be illustrated in subsection C below. However, there are some limits in the parameter space for which an exact analysis  can be performed. These limits moreover may be relevant to actual situations. 

\vspace{0.5cm}

i) We first consider the limit $Q_0\gg 1$ which is relevant to high-quality-factor resonators\cite{PCBH2000,LRHKB2010,Mont2012,V2008,B2009,VBMF2010}.  In practice, feedback-cooling setups operate in region 4 of Fig. \ref{Figdiag} as soon as $g>\sqrt{1-4/Q_0^2} \approx 1$ (see Fig. \ref{FigChristmas_tree}). More precisely, they operate in the first stability lobe of the Christmas-tree stability diagram, before the first Hopf bifurcation that occurs for $\tau=\tau^*_{1,1}\approx \arcsin(1/g)+\pi$ when  $Q_0\gg 1$.

After expanding  Eqs. (\ref{EqTx}) and (\ref{EqTv}) in powers of $1/Q_0$,  we find 
\begin{align}
\label{EqTeffxQ}
\frac{T_x}{T}&= \frac{1}{1+g\sin\tau}+\frac{g}{2Q_0}\frac{\sin \tau (\tau +g\cos \tau)+2\cos \tau +g\tau}{(1+g\sin \tau)^2}\nonumber\\
&+{\cal O}(Q_0^{-2})\ ,
\end{align}
\begin{align}
\label{EqTeffvQ}
\frac{T_v}{T}&= \frac{1}{1+g\sin\tau}+\frac{g}{2Q_0}\frac{\sin \tau(\tau -g\cos \tau)+g\tau}{(1+g\sin \tau)^2}+{\cal O}(Q_0^{-2})\ .
\end{align}
Therefore, the two effective temperatures become equal in the limit $Q_0\rightarrow \infty$, and we recover the fact that  the optimal cooling for $g>0$  is obtained by choosing $\tau=\tau^{opt}=\pi/2$ (and more generally $\tau=\pi/2+2n\pi$). Then, $T_x^{opt}=T_v^{opt}=T/(1+g)$, in agreement with Eq. (\ref{EqTeffx}). This  can be traced back to the asymptotic behavior of the effective force ${\overline F}_{fb,st}(x,v)$,
\begin{align}
{\overline F}_{fb,st}(x,v)=\frac{g}{Q_0}(x\cos \tau -v\sin \tau ) +{\cal O}(Q_0^{-2}) \, ,
\end{align} 
which implies that the force becomes purely viscous for $\tau=\pi/2+2n\pi$. In this case, as already mentioned, the  Markovian equation (\ref{EqCD}) leads to the same reduction of  thermal fluctuations as the true Langevin equation. (Note that the expansions (\ref{EqTeffxQ})-(\ref{EqTeffvQ}) are not valid for $g\sin \tau \le -1$, as there is no stationary regime in this case.) 

\begin{figure}[hbt]
\begin{center}
\includegraphics[width=9.5cm]{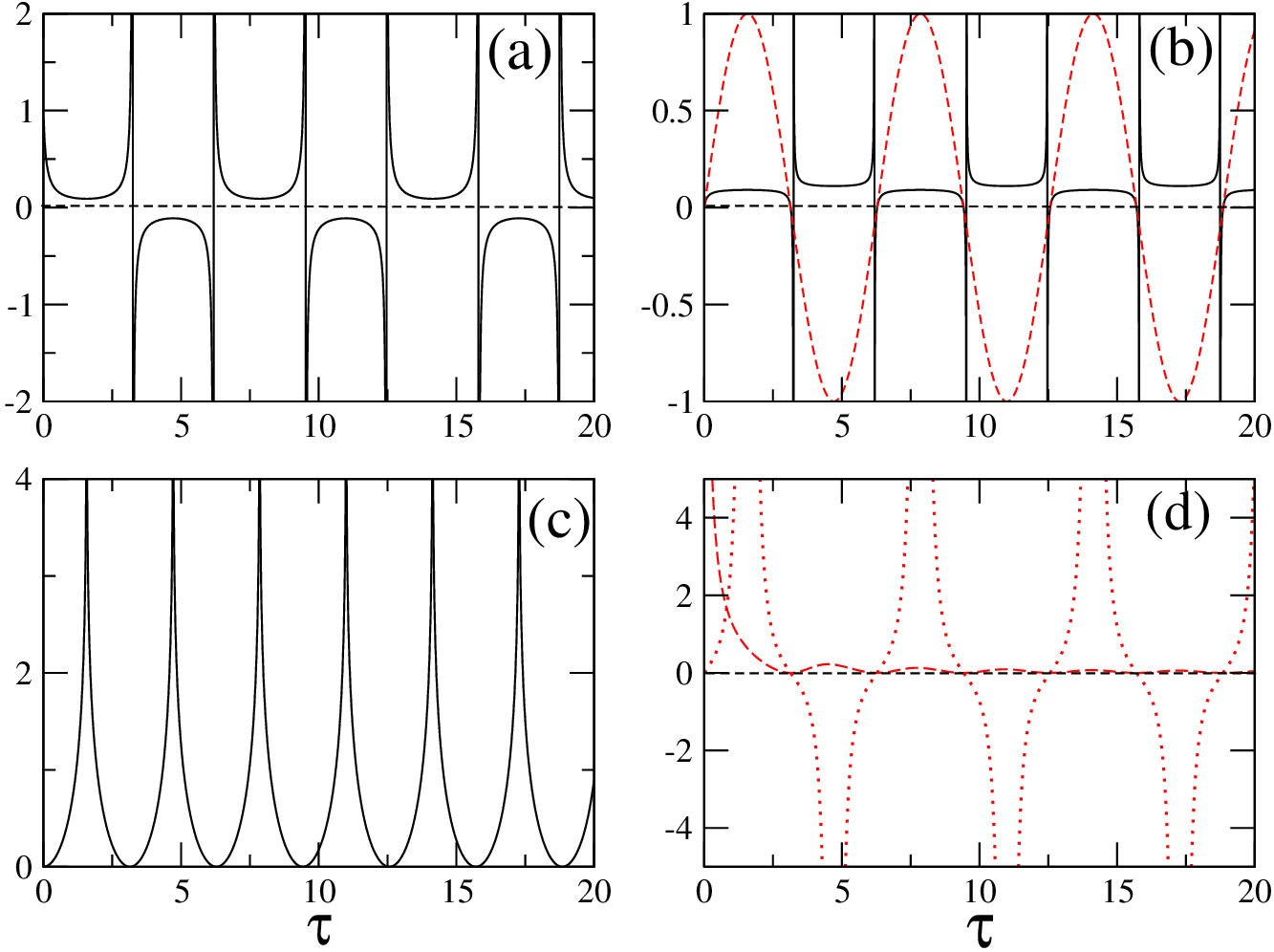}
 \caption{\label{FigAsymptoticQ} (Color on line) Asymptotic behavior in the large-$Q_0$ regime for $g=10$ as a function of $\tau$. A stationary state exists when the effective temperatures are positive. (a): $T_x$ or $T_v$ (they are equal when $Q_0\rightarrow \infty$). (b): $(Q_0/g)\dot{\cal W}_{ext}/T$ (solid black line) and  $(Q_0/g)\dot{\cal S}_{pump}^{v}$ (dashed red line). (c); ${\cal I}^{v;y}$. (d): $\dot{\cal I}_{flow,v}^{xv;y}$ (long-dashed red line) and $(Q_0/g)[\dot{\cal S}_{pump}^{v;y}+\dot {\cal I}_{flow,v}^{v;y}]$ (dotted red line).}
\end{center}
\end{figure}

This simple asymptotic behavior directly reflects on the thermodynamics and the second-law-like inequality (\ref{IneqEP1}), since the extracted work and the entropy pumping rate only depend on the kinetic temperature $T_v$. From Eqs. (\ref{EqHeat1}) and (\ref{EqSpumpv1}), we obtain
\begin{align}
\label{EqWQ}
\frac{\dot{\cal W}_{ext}}{T}&= \frac{g}{Q_0}\frac{\sin \tau}{1+g\sin \tau} - \frac{g}{2Q_0^2}\frac{\sin \tau(\tau-g\cos \tau)+g\tau}{(1+g\sin \tau)^2}\nonumber\\
&+{\cal O}(Q_0^{-3})\ ,
\end{align}  
\begin{align}
\label{EqSpumpQ}
\dot{\cal S}_{pump}^{v} (=\dot{\cal S}_{pump}^{xv})&= \frac{g}{Q_0}\sin \tau- \frac{g}{2Q_0^2}[\sin \tau(\tau-g\cos \tau)+g\tau]\nonumber\\
&+{\cal O}(Q_0^{-3})\ ,
\end{align} 
which shows that the two quantities, at the leading order in $Q_0^{-1}$, also reach their maximum value when $\tau=\pi/2+2n\pi$ (recall that we only consider the case of a positive feedback).  As it must be, the second-law-like inequality  (\ref{IneqEP1})  is satisfied at the leading order: $\dot{\cal W}_{ext}/(T\dot{\cal S}_{pump}^{v})=(1+g\sin \tau)^{-1}\le 1$. 

Similarly, from Eqs. (\ref{EqMIxvy21})-(\ref{Eqin6}), we  find that the mutual information ${\cal I}^{xv;y}$ diverges at the leading order in $1/Q_0$, whereas
\begin{align}
\label{EqMIvyQ}
{\cal I}^{v;y}=-\ln \vert\cos \tau\vert  +{\cal O}(Q_0^{-1})\ ,
\end{align}
\begin{align}
\label{EqIflow}
\dot{\cal I}_{flow,v}^{xv;y}=\frac{2\sin^2 \tau }{2\tau-\sin 2\tau}+{\cal O}(Q_0^{-1})\ ,
\end{align}
and 
\begin{align}
\label{Eqpump7}
\dot{\cal S}_{pump}^{v;y}+\dot {\cal I}_{flow,v}^{v;y}=\frac{g}{Q_0}\frac{\sin \tau (1+\frac{1}{g}\sin \tau)}{\cos^2 \tau}+{\cal O}(Q_0^{-2})\ .
\end{align}
The asymptotic behavior of all these quantities is shown in Fig. \ref{FigAsymptoticQ} where we have arbitrarily chosen a gain $g=10$. We also include the unstable regions in which the effective temperatures are negative, although, of course, the various quantities then loose their physical meaning.

We first observe that the mutual information ${\cal I}^{v;y}$ (which is zero for $\tau=0$ since $x_t$ and $v_t$ are uncorrelated) diverges for $\tau= \pi/2+2n\pi$ and vanishes for $\tau=2n\pi$. This means that the actual velocity of the resonator at time $t$ is perfectly predicted when measuring the position at time $t-(\pi/2 +2n\pi)$, which explains the behavior of the effective temperatures in the stability regions. However, ${\cal I}^{v;y}$ does not tell us if the controller has acquired an information that can be used to cool the system. This information is provided by $\dot{\cal S}_{pump}^{v}$ and the combined quantity  $\dot{\cal S}_{pump}^{v;y}+\dot {\cal I}_{flow,v}^{v;y}$,  which are indeed positive only in the cooling regime. (Moreover,  $\dot{\cal S}_{pump}^{v;y}+\dot {\cal I}_{flow,v}^{v;y}$ also diverges for  $\tau=\pi/2+2n\pi$ and is larger than $\dot{\cal S}_{pump}^{v}$, as predicted by inequality (\ref{Eqin5}).) On the other hand, the information flow $\dot {\cal I}_{flow,v}^{xv;y}$  is always positive when $Q_0\rightarrow \infty$, in agreement with inequality (\ref{Eqin4}) (since $\dot{\cal S}_{pump}^{v}$ is of order $1/Q_0$).  In fact, it appears that  $\dot {\cal I}_{flow,v}^{xv;y}$, which also takes into account the correlations between $x_t$ and $x_{t-\tau}$, does not carry a clear information about the functioning of the feedback.

Of course, this simple physical picture is only valid asymptotically when $Q_0\rightarrow \infty$. In particular, the two effective temperatures $T_x$ and $T_v$ differ at the order $Q_0^{-1}$, as do the optimal values of $\tau$ obtained by expanding  the equations $\partial T_x/\partial \tau=0$ and $\partial T_v/\partial \tau=0$:
\begin{align}
\label{Eqtauoptx}
\tau_x^{opt}=\frac{\pi}{2}+\frac{1}{2Q_0}+{\cal O}(Q_0^{-2}) \ .
\end{align} 
\begin{align}
\label{Eqtauoptv}
\tau_v^{opt}=\frac{\pi}{2}-\frac{1+2g}{2Q_0}+{\cal O}(Q_0^{-2}) \ .
\end{align}
(Alternatively, one could define the optimal cooling condition  by minimizing the sum of the two temperatures\cite{GVTGA2008}.)

\vspace{0.5cm}

ii) We next consider the small-delay limit $\tau\ll 1$  (\textit{i.e.}, $\tau\ll \omega_0^{-1}$ in real units), which is often used to approximate a SDDE by a nondelayed equation\cite{GLL1999}.  Expanding Eqs. (\ref{EqTx}) and (\ref{EqTv}) in powers of $\tau$ leads to
\begin{align}
\label{EqTxtau}
\frac{T_x}{T}&=\frac{1}{1-\frac{g}{Q_0}}[1-g\tau+g^2\tau^2+{\cal O}(\tau^3)]
\end{align}
and
\begin{align}
\label{EqTvtau}
\frac{T_v}{T}&=1-g\tau+g(g+\frac{1}{2Q_0})\tau^2+{\cal O}(\tau^3) \ . 
\end{align}
(The factor $1-g/Q_0=1-k'/k$ in Eq. (\ref{EqTxtau}) is due to the modification of the spring constant when $\tau=0$.) These equations show that $T_v=T_x(1-g/Q_0)\sim T/(1+g\tau)$ at the first order in $\tau$,  and the molecular refrigerator model of \cite{KQ2004,MR2012} is thus recovered in this limit by identifying $g\tau$ with $\gamma'/\gamma$, where $\gamma'$ is the additional friction coefficient, as already pointed out. This can again be related to the expansion of the effective force,
\begin{align}
{\overline F}_{fb,st}(x,v)&=\frac{g}{Q_0}[1-(1-\frac{g}{Q_0})\frac{\tau^2}{2}+{\cal O}(\tau^3)]x\nonumber\\
&-\frac{g\tau}{Q_0}[1-\frac{\tau}{2Q_0}+{\cal O}(\tau^2)]v\ .
\end{align}
Note that  the next-order contribution to the viscous term is positive, so that the feedback force  less effectively opposes the motion of the Brownian particle as $\tau$ increases. 

Likewise, the small-$\tau$ expansions of the various rates and information-theoretic quantities in the stationary state are given by
\begin{align}
\frac{\dot {\cal W}_{ext}}{T}=\frac{g\tau}{Q_0}-\frac{g}{Q_0}(g+\frac{1}{2Q_0})\tau^2+{\cal O}(\tau^3)\ ,
\end{align}
\begin{align}
\label{EqexpSptau}
\dot {\cal S}_{pump}^{xv}=\frac{g\tau}{Q_0}-\frac{g\tau^2}{2Q_0^2}+{\cal O}(\tau^3)\ ,
\end{align}
\begin{align}
\label{EqMIyv}
{\cal I}^{v;y}= \frac{1}{2}(1-\frac{g}{Q_0})\tau^2+{\cal O}(\tau^3)
\end{align}
\begin{align}
\label{EqIf1Q2}
\dot {\cal I}_{flow,v}^{xv;y}= \frac{1}{1+gQ_0}\tau^{-1}+{\cal O}(1)
\end{align}
\begin{align}
\label{EqIf2Q2}
\dot{\cal S}_{pump}^{v;y}+\dot {\cal I}_{flow,v}^{v;y}=\frac{g\tau}{Q_0}-\frac{3g-2Q_0}{2Q_0^2}\tau^2+{\cal O}(\tau^3)\ .
\end{align}
Note that  ${\cal I}^{v;y}$ initially increases with $\tau$, in line with the behavior of the effective temperatures, whereas $\dot {\cal I}_{flow,v}^{xv;y}$ diverges as $\tau \rightarrow 0$, since $x_{t-\tau}$ then coincides with $x_t$. One has $\dot{\cal W}_{ext}/(T\dot{\cal S}_{pump}^{v})=1-g\tau+{\cal O}(\tau^2)\le 1$ in accord with the second-law-like  inequality  (\ref{IneqEP1}). As required, the other inequalities  (\ref{Eqin4}) and  (\ref{Eqin5}) are  also satisfied.

\vspace{0.5cm}

iii) Finally, we consider the overdamped limit $m\rightarrow 0$ which was studied in \cite{MIK2009,JXH2011}. When the inertial term is small, it is convenient to work with  the original  Langevin equation (\ref{EqLlin}) and  keep the parameter $k'$  to quantify the feedback strength. The expansion in powers of $m$ is not analytic since one of the roots  of Eq. (\ref{Eqomega0}) diverges as $m \rightarrow 0$ (specifically,  $\omega_2\sim i\gamma /m$). However,  the non-analytic factors of the type $e^{-\gamma \tau /(2m)}$ can be neglected if $\tau$ is such that $m/(\gamma\tau) \rightarrow 0$, and we obtain
\begin{align}
\label{EqTeffxm}
\frac{T_x}{T}&= \frac{k-(kk'/\bar k)\sinh \frac{\bar k\tau}{\gamma}}{k-k'\cosh \frac{\bar k\tau}{\gamma}}+{\cal O}(m)\ ,
\end{align}
\begin{align}
\label{EqTeffvm}
\frac{T_v}{T}&=1-\frac{k'}{\gamma^2}\frac{k\cosh \frac{\bar k\tau}{\gamma}-k'-\bar k\sinh \frac{\bar k\tau}{\gamma}}{k-k'\cosh \frac{\bar k\tau}{\gamma} }m+{\cal O}(m^2).
\end{align}
where $\bar k=\sqrt{k^2-k'^2}$.
Therefore, in the limit $m\rightarrow 0$, $T_x$ goes to the nontrivial value  originally obtained in \cite{KM1992} (see also \cite{FBF2003} and Eq. (\ref{Eqphi0})), whereas $T_v\rightarrow T$. However, since $T_v-T$ is proportional to $m$, $\dot {\cal W}_{ext}/T=(\gamma/m) (1-T_v/T)$ and  $\dot {\cal S}_{pump}^{xv}=(\gamma/m) (T/T_v-1)$ are finite, and equal,  in this  limit:
\begin{align}
\label{Eqmzero}
\lim_{m\rightarrow 0}\frac{\dot {\cal W}_{ext}}{T}=\lim_{m\rightarrow 0} \dot {\cal S}_{pump}^{xv}=\frac{k'}{\gamma}\frac{k\cosh \frac{\bar k\tau}{\gamma}-k'-\bar k\sinh \frac{\bar k\tau}{\gamma}}{k-k'\cosh \frac{\bar k\tau}{\gamma} }\ .
\end{align}
On the other hand, the rate $\dot {\cal S}_i^{xv}=\dot {\cal Q}/T+\dot {\cal S}_{pump}^{xv}=(\gamma/m) (T-T_v)^2/(TT_v)$ defined by Eq. (\ref{EqSixv})  goes to zero. As  we shall see later, this is not the case for the rate $\dot {\cal R}_{cg}=\dot {\cal Q}/T+\dot {\cal S}_{\cal J}$  defined by Eq. (\ref{EqdefRcg}) and obtained from time reversal. Finally, for the sake of completeness, let us indicate that ${\cal I}^{v;y}= {\cal O}(m)$ whereas $\dot {\cal I}_{flow,v}^{xv;y}$ and $\dot{\cal S}_{pump}^{v;y}+\dot {\cal I}_{flow,v}^{v;y}$ go to finite (and nontrivial) limits. We also stress that the small-$m$ expansions are only valid for $\tau$ strictly positive, since the limits $m\rightarrow 0$ and $\tau \rightarrow 0$ do not commute. More generally, inertial effects cannot be neglected at a time scale of the order of, or smaller than, the viscous relaxation time $\tau_0=m/\gamma$. 

While discussing the overdamped limit, we  take the opportunity to point out that the statements of the second law in \cite{MIK2009} and  \cite{JXH2011} are incorrect. In \cite{MIK2009}, the entropy production rate, in the absence of an external periodic forcing, is identified with the heat flow rate $\dot {\cal Q}$ (Eq. (\ref{Eqmzero}) with $\gamma=1$ indeed coincides with Eq. (38) of this reference). However, the heat flow is negative in the cooling regime so that it cannot represent an entropy production obeying the second law. (This problem was overlooked in \cite{MIK2009} as only the case of a negative feedback where $\dot {\cal Q}$ is always positive was considered.) 
In \cite{JXH2011}, the entropy production rate is defined from a log-ratio of forward and backward path probabilities (see Eq. (21) in this reference). It then vanishes in the absence of an external periodic forcing. However, as we have already mentioned (see also Appendix A), this calculation is based on an erroneous expression of the path probability in which the actual feedback force in the Onsager-Machlup functional is replaced by the effective force defined at the level of the first FP equation.

\subsection{Calculation of $\dot S_{{\cal J}}$}

We now tackle the calculation of the Jacobian associated with the conjugate, ``acausal" Langevin equation
\begin{align}
\label{EqLlcon}
\dot v_t= -x_t-\frac{1}{Q_0} v_t+\frac{g}{Q_0}x_{t+\tau}+\xi_t\ .
\end{align}
Our main objective is to compute the asymptotic rate $\dot S_{{\cal J}}$ that enters the second-law-like inequality (\ref{Eq2law3}) obtained from time reversal.  Although this subsection is rather technical, we believe that it is useful to present the detailed arguments since it is quite unusual to consider quantities associated with an acausal dynamics.  However, the reader  only interested in the discussion from the physical viewpoint may only take notice of Eq. (\ref{Eqfinal}), which is the main result of this subsection, and go directly to subsection C.

\subsubsection{Perturbative expansion for a finite observation time}

The first task is to compute the operator $\psi(t,t')$ defined by Eq. (\ref{Eqpsi}). The crucial simplification due to the linearity of the Langevin equation is that the functional derivative $\widetilde F_{tot}^{'}(t,t')$ and hence the Jacobian $\widetilde {\cal J}$ itself become path-independent. Indeed, we  have 
\begin{align}
\widetilde F_{tot}^{'}(t)=-\delta(t)+\frac{g}{Q_0}\delta(t+\tau)
\end{align} 
in reduced units, so that
 \begin{align}
M^{(0)}(t,t')=-G(t-t')
\end{align}
and 
\begin{align}
\widetilde M^{(1)}(t,t')=\frac{g}{Q_0}G(t-t'+\tau)\Theta({\cal T}-\tau+t') \ ,
\end{align}
with
\begin{align}
G(t)=Q_0[1-e^{-t/Q_0}]\Theta(t)\ .
\end{align} 
 The operator $K=[\delta -M^{(0)}]^{-1}$ is then solution of the linear integral equation 
\begin{align}
\label{EqK}
K(t,t')+\int_{-{\cal T}}^{t}dt''\: G(t-t'')K(t'',t')=\delta(t-t')\ ,
\end{align}
and it is easy to see that $K(t,t')=0$ for $t'>t$. The integral in the above equation is therefore from $t'$ to $t$, provided $t'<t$, and this in turn implies  that $K(t,t')$ is just a function of $t-t'$. Eq. (\ref{EqK}) is then solved by first going to Laplace space, which  finally yields 
\begin{align}
K(t)=\delta(t)-\frac{e^{s_0^+t}-e^{s_0^-t}}{s_0^+-s_0^-} \ , \ t\ge 0
\end{align}
where $s_0^{\pm}$ are the roots of the quadratic equation $s^2+s/Q_0+1=0$, \textit{i.e.},
 \begin{align}
\label{spm}
s_0^{\pm}=\frac{1}{2Q_0}[-1\pm i\sqrt{4Q_0^2-1}]\ .
\end{align}
This allows us to compute  $\psi(t,t')$ from Eq. (\ref{Eqpsi}), leading to
\begin{align}
\label{Eqpsi1}
\psi(t,t')&=\frac{g}{Q_0}\frac{e^{s_0^+(t-t'+\tau)}-e^{s_0^-(t-t'+\tau)}}{s_0^+-s_0^-}\Theta(t-t'+\tau)\nonumber\\
&\times \Theta({\cal T}-\tau+t')\ .
 \end{align}
For a finite observation time, it is not possible to sum the infinite series  (\ref{EqdefJ}), but since $\psi(t,t')$ is proportional to $g$, one can perform a perturbative calculation in powers of $g$. It is worth taking a look at the first two terms for gaining some insight into the general behavior of the Jacobian $\widetilde {\cal J}$ as a function of  ${\cal T}$. 

Since $\psi(t,t')=0$ for $t'<-{\cal T}+\tau$, we immediately note that  $\ln \widetilde {\cal J}/{\cal J}=0$ (\textit{i.e.}, $ \widetilde{\cal J}={\cal J}$) for $2{\cal T}\le \tau$. As pointed out earlier, the Jacobian matrix is indeed lower triangular for $2{\cal T}\le \tau$ as  $x_{t+\tau}$ does not belong to the time-reversed trajectory ${\bf X}^\dag$. For $2{\cal T}\ge \tau$, the $g$-expansion  reads 
\begin{align}
\label{Eqexp1}
\ln \frac{\widetilde {\cal J}}{{\cal J}}&=-\int_{-{\cal T}+\tau}^{{\cal T}}dt \: \psi(t,t)\nonumber\\
&- \frac{1}{2}\int_{-{\cal T}+\tau}^{{\cal T}}dt \:\int_{-{\cal T}+\tau}^{{\cal T}}dt'\:\psi(t,t')\psi(t',t)+{\cal O}(g^3)\ , 
\end{align}
and after some tedious but straightforward calculations we obtain
\begin{align}
\label{Eqexp10}
&\ln \frac{\widetilde {\cal J}}{{\cal J}}=-\frac{g}{Q_0}(2{\cal T}-\tau)\frac{e^{s_0^+\tau}-e^{s_0^-\tau}}{s_0^+-s_0^-}\nonumber\\
&-\frac{g^2}{2Q_0^2}\frac{e^{-\tau/Q_0}}{(s_0^+-s_0^-)^2}[F_1({\cal T})\Theta (\tau-{\cal T})+F_2({\cal T})\Theta ({\cal T}-\tau)]\nonumber\\
&+{\cal O}(g^3)\ ,
\end{align}
 where 
\begin{align}
F_1({\cal T})&=(2{\cal T}-\tau)^2[e^{(s_0^+-s_0^-)\:\tau}+e^{-(s_0^+-s_0^-)\:\tau}]\nonumber\\
&-2\frac{[e^{(s_0^+-s_0^-)({\cal T}-\tau/2)}-e^{-(s_0^+-s_0^-)({\cal T}-\tau/2)}]^2}{(s_0^+-s_0^-)^2}
\end{align}
and 
\begin{align}
F_2({\cal T})&=4\tau({\cal T}-\frac{3\tau}{4})[e^{(s_0^+-s_0^-)\:\tau}+e^{-(s_0^+-s_0^-)\:\tau}]\nonumber\\
&-4({\cal T}-\tau)\frac{e^{(s_0^+-s_0^-)\: \tau}-e^{-(s_0^+-s_0^-)\:\tau}}{s_0^+-s_0^-}\nonumber\\
&-2\frac{[e^{(s_0^+-s_0^-) \:\tau/2}-e^{-(s_0^+-s_0^-) \:\tau/2}]^2}{(s_0^+-s_0^-)^2}\ .
\end{align}
We observe that $\partial^3F_1({\cal T})/\partial {\cal T}^3\vert_{{\cal T}=\tau}\ne \partial^3F_2({\cal T})/\partial {\cal T}^3\vert_{{\cal T}=\tau}$ so that the second-order term is continuous but non-analytic at ${\cal T}=\tau$.  More generally, the $g$-expansion shows that $\widetilde {\cal J}/{\cal J}$ as a function of ${\cal T}$ is non-analytic at ${\cal T}=\tau,3\tau/2,2\tau,$ etc. 

Finally, by taking the limit ${\cal T}\rightarrow \infty$ in Eq. (\ref{Eqexp10}), we obtain the $g$-expansion of the asymptotic rate $\dot S_{\cal J}=\lim_{{\cal T}\rightarrow \infty}(1/2\cal T)\ln J/\widetilde J$,
\begin{align}
\label{Eqexp11}
\dot S_{\cal J}&=\frac{g}{Q_0}\frac{e^{s_0^+\tau}-e^{s_0^-\tau}}{(s_0^+-s_0^-)}+\frac{g^2}{Q_0^2}\frac{e^{-\tau/Q_0}}{(s_0^+-s_0^-)^2}\nonumber\\
&\times\big[\tau(e^{(s_0^+-s_0^-)\:\tau}+e^{-(s_0^+-s_0^-)\:\tau})\nonumber\\
&-\frac{e^{(s_0^+-s_0^-)\: \tau}-e^{-(s_0^+-s_0^-)\:\tau}}{s_0^+-s_0^-}\big]+{\cal O}(g^3)\ .
\end{align}

\subsubsection{Calculation of the asymptotic rate $\dot S_{{\cal J}}$}

We now derive an explicit expression of  $\dot S_{{\cal J}}$ that is valid beyond the perturbative regime.  This requires a careful analysis.

We first note from Eq. (\ref{Eqpsi1}) that $\psi$ becomes a function of a single variable in the  limit ${\cal T}\rightarrow \infty$,
\begin{align}
\label{Eqpsit}
\psi(t)=\frac{g}{Q_0}\frac{e^{s_0^+(t+\tau)}-e^{s_0^-(t+\tau)}}{s_0^+-s_0^-}\Theta(t+\tau)\ ,
 \end{align}
as $\Theta({\cal T}-\tau+t')\rightarrow 1$. This implies from Eq. (\ref{EqdefJ}) that $\ln \widetilde {\cal J}/{\cal J}$ becomes proportional to $2 {\cal T}$ asymptotically, as is also clear from Eq. (\ref{Eqexp1}).  The formal power series in $g$,
\begin{align}
\label{Eqexp2}
\dot S_{\cal J}&=\sum_{n=1}^{\infty}\frac{1}{n}\Big\{\underbrace{\psi\circ \psi\circ...\psi}_{n \: \mbox{times}}\Big\}_{t=0}\ ,
\end{align}
can then be  expressed  as 
\begin{align}
\label{Eqexp3}
\dot S_{\cal J}&=\sum_{n=1}^{\infty}\frac{1}{2 \pi i}\int_{c-i\infty}^{c+i\infty}ds\:[\frac{1}{n}\psi(s)^n]\ ,
\end{align}
where $s= \sigma+i\omega$ and 
\begin{align}
\label{Eqpsis}
\psi(s)=\int_{-\infty}^{\infty} dt\: \psi(t)e^{-st}=\frac{g}{Q_0}\frac{e^{s\tau}}{(s-s_0^+)(s-s_0^-)}
\end{align}
is the bilateral Laplace transform of $\psi(t)$. Accordingly, the integration line $Re(s)=c$  must belong to the region of convergence (ROC) of $\psi(s)$, that is the region of the complex $s$-plane where the transform exists\cite{P1999}. Since $\psi(t)=0$ for $t\le -\tau$, the ROC is  defined by $\sigma>\sigma_0^+$,  where $\sigma_0^+$ is the real part of $s_0^+$, \textit{i.e.}, $\sigma_0^+=-1/(2Q_0)$ for $Q_0\ge 1/2$ and $\sigma_0^+=-1/(2Q_0)[1-(1-4Q_0^2)^{1/2}]$ for $Q_0\le 1/2$.

One can readily check that the expansion (\ref{Eqexp11}) is recovered by computing each term of the series (\ref{Eqexp3}) by contour integration, invoking Jordan's lemma and Cauchy's residue theorem: one closes the so-called Bromwich contour (the vertical line cutting the real axis in $c$, with $c$ in the ROC) with a large semicircle to the left-hand side of the complex plane and sums the two residues of $\psi(s)^n$ at $s_0^+$ and $s_0^-$.

In general, however, the power series (\ref{Eqexp3}) does not converge for arbitrary values of the gain $g$. Moreover, this is not a convenient route for computing  $\dot S_{\cal J}$ numerically. What is needed is a closed-form expression for the sum of the series that can be used in the whole parameter space $(Q_0,\tau,g)$ or at least in the regions where a stationary state exits. An obvious candidate for such a formula is the integral representation
\begin{align}
\label{EqdefJs0}
\dot S_{\cal J}=-\frac{1}{2 \pi i}\int_{c-i\infty}^{c+i\infty}ds\:\ln [1-\psi(s)]
\end{align}
obtained by interchanging the  sum and the integral and summing the series $\sum_{n=1} (1/n)\psi(s)^n$. Of course, this presupposes that  the integration line $Re(s)=c$ (with $c>\sigma_0^+$) is such that $\vert\psi(s)\vert <1$ all along the line. The series then converges uniformly to the principal value of $\ln[1-\psi(s)]$. 

By analytic continuation, however, one can still use Eq. (\ref{EqdefJs0}) when $\vert\psi(s)\vert >1$ to compute  $\dot S_{\cal J}$ provided one stays on the same branch of the function $\ln[1-\psi(s)]$. In other words, the  Bromwich contour $Re(s)=c$ must not cross any of the cuts that define the branch in the complex $s$-plane. Although this requires a careful (and rather tedious) study of the branch points of $\ln[1-\psi(s)]$ as a function of the system's parameters, the reward of these efforts will be the derivation of a simple formula for $\dot S_{\cal J}$: see Eq. (\ref{Eqfinal}). We stress that the naive choice in terms of Fourier transforms that corresponds to taking $c=0$ is not the correct solution in general\cite{note2}.

For this study it is convenient to first rewrite Eq. (\ref{EqdefJs0}) as 
\begin{align}
\label{EqdefJs}
\dot S_{\cal J}=\frac{1}{2 \pi i}\int_{c-i\infty}^{c+i\infty}ds\:\ln\frac{\widetilde \chi(s)}{\chi_{g=0}(s)} 
\end{align}
where 
\begin{align}
\label{Eqchis0}
\chi_{g=0}(s):=[(s-s_0^+)(s-s_0^-)]^{-1}=[s^2+s/Q_0+1]^{-1}
\end{align}
is the standard response function (in Laplace representation) of the  harmonic oscillator (cf. Eq. (\ref{Eqchi}) with $\omega=is$ and $k'=0$), and 
\begin{align}
\label{Eqchis}
\widetilde \chi(s):=[s^2+\frac{s}{Q_0}+1-\frac{g}{Q_0}e^{s\tau}]^{-1} \ .
\end{align}
An interpretation of $\widetilde \chi(s)$ will be given later on. As a result, the branch points of the integrand in Eq. (\ref{EqdefJs})  are  $s_0^+,s_0^-$ on the one hand, and  $s=+\infty$ and the poles of $\widetilde \chi(s)$ on the other hand\cite{note4}. The location of these poles in the complex plane evolves in an intricate manner with the system's parameters. From a practical standpoint it is convenient to fix the values of $Q_0$ and $\tau$ and take $g$ as the control variable, as in feedback-cooling experimental setups (see \textit{e.g.} \cite{PDMR2007}). The results of this analysis are summarized in Appendix D (to simplify the discussion, we only consider the case of a positive feedback). The study is numerical for the most part, since analytical calculations can be performed only in the initial perturbative regime $g/Q_0\ll 1$ or for $Q_0\gg1$. 
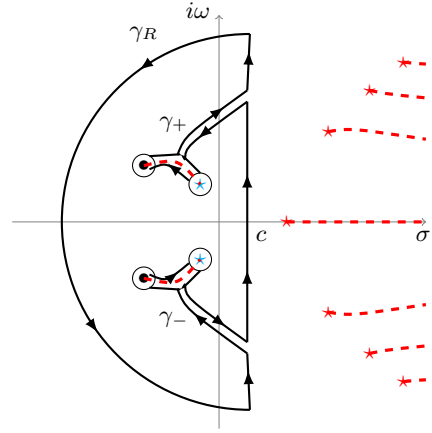
\begin{figure}
\begin{tikzpicture}[scale=0.5]

\coordinate (A1) at (-2,1.5);
\coordinate (A2) at (-2,-1.5);
\coordinate (B1) at (-0.5,1);
\coordinate (B2) at (-0.5,-1);
\coordinate (C1) at (2.9,2.4);
\coordinate (C2) at (2.9,-2.4);
\coordinate (D1) at (4,3.5);
\coordinate (D2) at (4,-3.5);
\coordinate (E1) at (4.9,4.25);
\coordinate (E2) at (4.9,-4.25);
\coordinate (R) at (1.8,0);

\draw [help lines,->] (0,-5.5) -- (0,5.5);  
\draw [help lines,->] (-5.5,0) -- (5.5,0);   
\node at (0,5.2)[above left]{$i\omega$};
\node at (5,0)[below right] {$\sigma$};
\node at (0.75,0)[below right] {$c$};


\node at (A1) {\textbullet}; 
\node at (A2) {\textbullet};
\node[cyan] at (B1) {$\star$}; 
\node[cyan] at (B2) {$\star$};
\node[red] at (C1) {$\star$}; 
\node[red] at (C2) {$\star$};
\node[red] at (D1) {$\star$}; 
\node[red] at (D2) {$\star$};
\node[red] at (E1) {$\star$}; 
\node[red] at (E2) {$\star$};
\node[red] at (R) {$\star$};

\draw (A1) circle(0.3);
\draw (B1) circle(0.3);
\draw (A2) circle(0.3);
\draw (B2) circle(0.3);

\draw [dashed,red, very thick]  (A1).. controls (-1.1,1.7) ..(B1); 
\draw [dashed,red, very thick]  (A2).. controls (-1.1,-1.7) ..(B2);

\draw[dashed,red, very thick] (C1)..controls (3.5,2.5)..(5.5,2.2); 
\draw[dashed,red,very thick] (C2)..controls (3.5,-2.5)..(5.5,-2.2);
\draw[dashed,red,very thick] (D1)..controls (4.5,3.4)..(5.5,3.3);
\draw[dashed,red,very thick] (D2)..controls (4.5,-3.4)..(5.5,-3.3);
\draw[dashed,red,very thick] (E1)..controls (5.2,4.22)..(5.5,4.2);
\draw[dashed,red,very thick] (E2)..controls (5.2,-4.22)..(5.5,-4.2);
\draw[dashed,red,very thick] (R)..controls (4,0)..(5.5,0);

\coordinate (N1) at (0.75,3.5);
\coordinate (N2) at (0.75,-3.5);
\coordinate (Q1) at (0.75,3.2);
\coordinate (Q2) at (0.75,-3.2);

\draw[thick,black,xshift=2pt,
decoration={ markings,  
      mark=at position 0.7 with {\arrow{latex}}}, 
      postaction={decorate}]
  (0.75,-5) -- (N2);
\draw[thick,black,xshift=2pt,
decoration={ markings,  
      mark=at position 0.3 with {\arrow{latex}}, 
      mark=at position 0.7 with {\arrow{latex}}}, 
      postaction={decorate}]
  (Q2) -- (Q1);  
\draw[thick,black,xshift=2pt,
decoration={ markings,  
      mark=at position 0.7 with {\arrow{latex}}}, 
      postaction={decorate}]
  (N1) -- (0.75,5);  
    
 \draw[thick,black,xshift=2pt,
decoration={ markings,
      mark=at position 0.2 with {\arrow{latex}}, 
      mark=at position 0.7 with {\arrow{latex}}}, 
      postaction={decorate}]
 (0.75,5) arc (90:270:5) -- (0.75,-5);
     
  
  \node at (-2,5) {$\gamma_R$};
   \node at (-1.2,2.6) {$\gamma_+$}; 
   \node at (-1.2,-2.6) {$\gamma_-$};
\coordinate (M1) at (-1.1,1.8);
\draw [thick, 
decoration={ markings,  
      mark=at position 0.7 with {\arrow{latex}}}, 
      postaction={decorate}]
(M1).. controls (-0.9,2.3) ..(N1); 

\coordinate (P1) at (-0.9,1.65);
\draw [thick,
decoration={ markings,  
      mark=at position 0.7 with {\arrow{latex}}}, 
      postaction={decorate}]
  (Q1).. controls (-0.9,2) ..(P1);

\coordinate (M2) at (-1.1,-1.8);
\draw [thick, 
decoration={ markings,  
      mark=at position 0.7 with {\arrow{latex}}}, 
      postaction={decorate}]
(N2).. controls (-0.9,-2.3) ..(M2); 

\coordinate (P2) at (-0.9,-1.65);
\draw [thick,
decoration={ markings,  
      mark=at position 0.7 with {\arrow{latex}}}, 
      postaction={decorate}]
 (P2).. controls (-0.9,-2) ..(Q2);

\coordinate (A10) at (-1.85,1.75);
\coordinate (B10) at (-0.5,1.3); 
 \draw [thick]  (A10).. controls (-1,1.8) ..(M1);
 \draw [thick]  (P1).. controls (-1,1.8) ..(B10);

 \coordinate (A11) at (-1.85,1.4);
\coordinate (B11) at (-0.8,1.1); 
 \draw [thick,
 decoration={ markings,  
      mark=at position 0.7 with {\arrow{latex reversed}}}, 
      postaction={decorate}]
        (A11).. controls (-1.4,1.6) ..(B11);

\coordinate (A20) at (-1.85,-1.75);
\coordinate (B20) at (-0.5,-1.3); 
 \draw [thick]  (A20).. controls (-1,-1.8) ..(M2);
  \draw [thick]  (P2).. controls (-1,-1.8) ..(B20);

 \coordinate (A21) at (-1.85,-1.4);
\coordinate (B21) at (-0.8,-1.1); 
 \draw [thick,
 decoration={ markings,  
      mark=at position 0.7 with {\arrow{latex}}}, 
      postaction={decorate}]
       (A21).. controls (-1.4,-1.6) ..(B21);

\end{tikzpicture}
\caption{\label{FigCuts} (Color on line) Schematic distribution of the branch points of $\ln [\widetilde \chi(s)/\chi_{g=0}(s)]$ in the complex s-plane. The black dots are the two poles $s_0^{\pm}$ of $\chi_{g=0}(s)$ and the stars are the poles  $\tilde s^{\pm}$ of type 0 (blue) and of type 1 (red)  of $\widetilde \chi(s)$. Only the first  poles of type 1 are shown in the figure. 
The  dashed (red) lines represent the branch cuts (on the right-hand side of the $s$-plane, they go to $+ \infty$). The solid black lines, including the vertical line that represents the Bromwich contour $Re(s)=c$, form the contour used for calculating $\dot S_{\cal J}$ from Eq. (\ref{EqdefJs}).}
\end{figure}

A schematic distribution of the branch points and of the corresponding branch cuts is shown in Fig. \ref{FigCuts}. It is clear in this example that there is only one possibility for placing the Bromwich contour, \textit{i.e.}, in the interval between the  two  poles $\tilde s^{\pm}$ of type 0 and the leftmost pole of type 1 (this terminology for the poles of $\widetilde \chi(s)$ is justified in Appendix D). Although there can be more complicated situations than this one (see the example in Fig. \ref{Fig17} of Appendix D), one can show that the choice for the location of the integration line is always unambiguous and can be stated as follows:  {\it there must be two and only two poles of $\widetilde \chi (s)$ on the left side of the line}. This allows  a straightforward  calculation of the integral in Eq. (\ref{EqdefJs}) by contour integration, which is  similar to the calculation that leads to the classical Bode's integral formula\cite{B1945}. In the example shown in Fig. \ref{FigCuts} (in which  the  poles  $s_0^{\pm}$ and  $\tilde s^{\pm}$ are complex),  the closed contour consists of the original line $Re(s)=c$, a large semi-circle  $\gamma_R$ with a radius going to infinity on the left, and the two contours $\gamma_+$ and $\gamma_-$ that start from the line $Re(s)=c$ and enclose the branch cuts on the left. Since there are no other singularities within the contour, the integrand is analytic and from Jordan's lemma and Cauchy's theorem we obtain
\begin{align}
&\frac{1}{2i\pi}\int_{c-i\infty}^{c+i\infty}ds\:\ln \frac{\widetilde \chi(s)}{\chi_{g=0}(s)}+\frac{1}{2i\pi}\oint_{\gamma_+}ds\:\ln \frac{\widetilde \chi(s)}{\chi_{g=0}(s)} \nonumber\\
&+\frac{1}{2i\pi}\oint_{\gamma_-}ds\:\ln \frac{\widetilde \chi(s)}{\chi_{g=0}(s)} =0\ .
\end{align}
The contributions along the upper and lower halves of the contours $\gamma_{\pm}$ cancel each other and the only  contribution comes from the change in the argument of the logarithm when circumventing the poles. In consequence,
\begin{align}
\oint_{\gamma_{\pm}}ds\:\ln \frac{\widetilde \chi(s)}{\chi_{g=0}(s)} =-2i\pi (\tilde s^{\pm}-s_0^{\pm}) \ ,
\end{align}
which eventually leads to the formula 
\begin{align}
\label{Eqfinal}
\dot S_{\cal J}=(\tilde s^{+}-s_0^{+}) +(\tilde s^{-}-s_0^{-})
\end{align}
which is also valid if the poles are real. This simple expression of $\dot S_{\cal J}$ was not given in \cite{MR2014}.   By using the $g$-expansion of the  poles of $\tilde \chi(s)$ (see Eq. (\ref{Eqpoleexp})), one can readily check that the $g$-expansion of $\dot S_{\cal J}$ given by Eq. (\ref{Eqexp11}) is  recovered. 
The $1/Q_0$ expansion, which is relevant to actual feedback-cooling setups, is discussed below.

Finally, it is interesting to note that fixing the Bromwich contour is a way to (uniquely) define an ``acausal response function" in the time domain:
\begin{align}
\label{EqLaplace_response}
\widetilde \chi(t)= \frac{1}{2 \pi i}\int_{c-i\infty}^{c+i\infty}ds\: \frac{e^{st}}{s^2+\frac{s}{Q_0}+1-\frac{g}{Q_0}e^{s\tau}}\ ,
\end{align}
where one considers $\widetilde \chi(s)$, given by Eq. (\ref{Eqchis}), as a {\it bilateral} Laplace transform. We emphasize that $\widetilde \chi(t)$ is neither causal nor anti-causal. (Note that the term ``acausal response function" may be viewed as an abuse of language since the acausal dynamics cannot be implemented physically.) The acausal character induces a quite different behavior from that observed with standard response functions which are solution of causal integral or integro-differential equations of Volterra type and have been thoroughly investigated in the mathematical and engineering literature (\textit{e.g.},  in control theory)\cite{C2002}.  This is further discussed and illustrated in Appendix \ref{App_acausal_response}.

\subsubsection{Useful limits}

As we did previously with the various rates and information-theoretic quantities, we now consider the behavior of $\dot S_{\cal J}$ for large $Q_0$, for small $\tau$, and in the overdamped limit.  According to Eq. (\ref{Eqfinal}), this amounts to studying the behavior of the poles $\tilde s^{\pm}$ in these limits. For brevity, we only give the corresponding expansions for $\dot S_{\cal J}$.
\vspace{0.5cm}

i) For $Q_0\gg 1$, the location of the poles of type $0$ of $\tilde \chi(s)$ in the complex $s$-plane can be  obtained by expanding Eqs. (\ref{eq:subeq1})-(\ref{eq:subeq2}) in powers in $1/Q_0$. From Eq. (\ref{Eqfinal}), this leads to 
\begin{align}
\label{EqSdotJQ0}
\dot {\cal S}_{{\cal J}}&=\frac{g}{Q_0}\sin \tau- \frac{g}{2Q_0^2}[\sin \tau(\tau-g\cos \tau)+g\tau\cos 2\tau]\nonumber\\
&+{\cal O}(Q_0^{-3})\ ,
\end{align}
which can be compared to the expansion (\ref{EqSpumpQ}) of the entropy pumping rate $\dot {\cal S}_{pump}^{v}$. This shows that the two rates (which are both upper bounds to the extracted work rate) coincide at the leading order in  $1/Q_0$. Accordingly, for $g>0$, $\dot {\cal S}_{{\cal J}}$ also reaches a maximum for $\tau=\pi/2+2n\pi$.  Moreover, one has
\begin{align}
\label{Eqcompar}
\dot {\cal S}_{{\cal J}}-\dot {\cal S}_{pump}^{v}=\frac{g^2}{2Q_0^2}\tau (1-\cos 2\tau) +{\cal O}(Q_0^{-3})\ ,
\end{align}
so that $\dot {\cal S}_{{\cal J}}\ge \dot {\cal S}_{pump}^{v}$ at the next order. Although a general analytical argument is still lacking, numerical studies (see subsection C) seem to show that this inequality is always true. On the other hand, despite the fact that $\dot {\cal S}_{{\cal J}}\le \dot {\cal I}_{flow,v}^{xv;y}$  and  $\dot {\cal S}_{{\cal J}}\le \dot{\cal S}_{pump}^{v;y}+\dot {\cal I}_{flow,v}^{v;y}$ at the corresponding leading orders in the stability regions, these two inequalities are not always valid (see \textit{e.g.} Fig. \ref{FigTemp_vs_gain} below).

\vspace{0.5cm}

ii) The small-$\tau$ expansion of $\dot S_{{\cal J}}$ is obtained by reordering the small-$g$ expansion (\ref{Eqexp3}) (the term of order $n$ in $g$ is correct up to order $2n$  in $\tau$). This yields
\begin{align}
\label{EqSjtau}
\dot S_{{\cal J}}&=\frac{g}{Q_0}\tau-\frac{g}{2Q_0^2}\tau^2+{\cal O}(\tau^3) \ ,
\end{align}
so that  $\dot S_{{\cal J}}=\dot {\cal S}_{pump}^{v}$ at the first two orders in $\tau$ (see expansion (\ref{EqexpSptau})). As noted earlier at the end of section IV.B.2, the result of \cite{KQ2004} is recovered at the lowest order in $\tau$ by identifying $k' \tau$ (in real units) with a  friction coefficient $\gamma'$ (since changing $\tau$ into $-\tau$ amounts to changing $\gamma'$ into $-\gamma'$). Going to the next order, we find
\begin{align}
\label{Eqcompar1}
\dot {\cal S}_{{\cal J}}-\dot {\cal S}_{pump}^{v}=\frac{g^2}{Q_0}\tau^3+{\cal O}(\tau^{4})\ ,
\end{align}
which reinforces the conjecture that  the inequality $\dot {\cal S}_{{\cal J}}\ge \dot {\cal S}_{pump}^{v}$  is always true. (We stress again that  $\dot {\cal S}_{{\cal J}}$ is in general distinct from the entropy pumping rate; this point was  perhaps unclear in \cite{MR2014}.)

\vspace{0.5cm}

iii) Finally, we consider the overdamped  limit $m=0$, keeping the original parameters of Eq. (\ref{EqLlin}). In this case, one simply has $\tilde \chi(s)=[\gamma s+k-k'e^{s\tau}]^{-1}$ and it turns out that for $k-k'>0$ (which is the condition for a stable NESS to exist\cite{KM1992}), this function has a single real pole of type $0$ given by $\tilde s=-k/\gamma-(1/\tau)W_0\left(-(k'\tau/\gamma) e^{-k \tau/\gamma }\right)$ where $W_0$ is the Lambert function  of order $0$\cite{CGHJK1996} (see also Appendix D). This readily yields the explicit formula
\begin{align}
\label{EqSjm}
\dot S_{{\cal J}}=- \frac{1}{\tau}W_0(-\frac{k'\tau}{\gamma} e^{-k\tau/\gamma}) \ .
\end{align}
Note that this expression goes to the finite value $-k'/\gamma$ as  $\tau \rightarrow 0$ whereas Eq. (\ref{EqSjtau}) gives $\dot S_{{\cal J}}(\tau=0)=0$. This is due to a discrepancy between  It\^o and Stratonovich calculus in this limit. As mentioned earlier, the limits $m\rightarrow 0$ and $\tau \rightarrow 0$ do not commute and the overdamped model ceases to be valid for $\tau\lesssim \tau_0$ where   $\tau_0=m/\gamma$ is the viscous relaxation time.
  
By comparing with Eq. (\ref{Eqmzero}), one can check that $\dot S_{{\cal J}}\ge\dot {\cal S}_{pump}^{v}$ also in this case (and equality is only realized when $\tau=0$). This  is actually a mere consequence of the second-law-like inequality (\ref{Eq2law3}), since $\dot {\cal S}_{pump}^{xv}$ and $\dot{\cal W}_{ext}/T$ are equal in the overdamped limit.

\subsection{Numerical studies}

To illustrate the essential features of the above general analysis, we now provide a numerical study of a few cases that describe  the thermodynamic behavior of feedback-controlled damped oscillators with low, intermediate, and large quality factors. We take either the feedback gain $g$ (with $g>0$) or the delay $\tau$ as the control variable. 

\subsubsection{$Q_0=2$}

We first choose $\tau=2.5$ as an example of a continuous feedback with a time lag shorter than $2\pi$, the period of oscillation (recall that we use the angular resonance frequency $\omega_0$ to define the unit of time). From the stability diagram of Fig. \ref{FigChristmas_tree}, we see that the system can settle in a stationary state  for all values of the gain $g$. (Although the system operates in the multistability region for $g/Q_0\gtrsim 0.484$, there is no Hopf bifurcation for $\tau=2.5$.)
Accordingly, the variances $\langle x^2\rangle_{st}$ and $\langle v^2\rangle_{st}$,  associated with the configurational and kinetic temperatures $T_x$ and $T_v$, stay finite for all values of $g/Q_0<1$, as shown in Fig. \ref{FigTemp_vs_gain}.
\begin{figure}[hbt]
\begin{center}
\includegraphics[width=9cm]{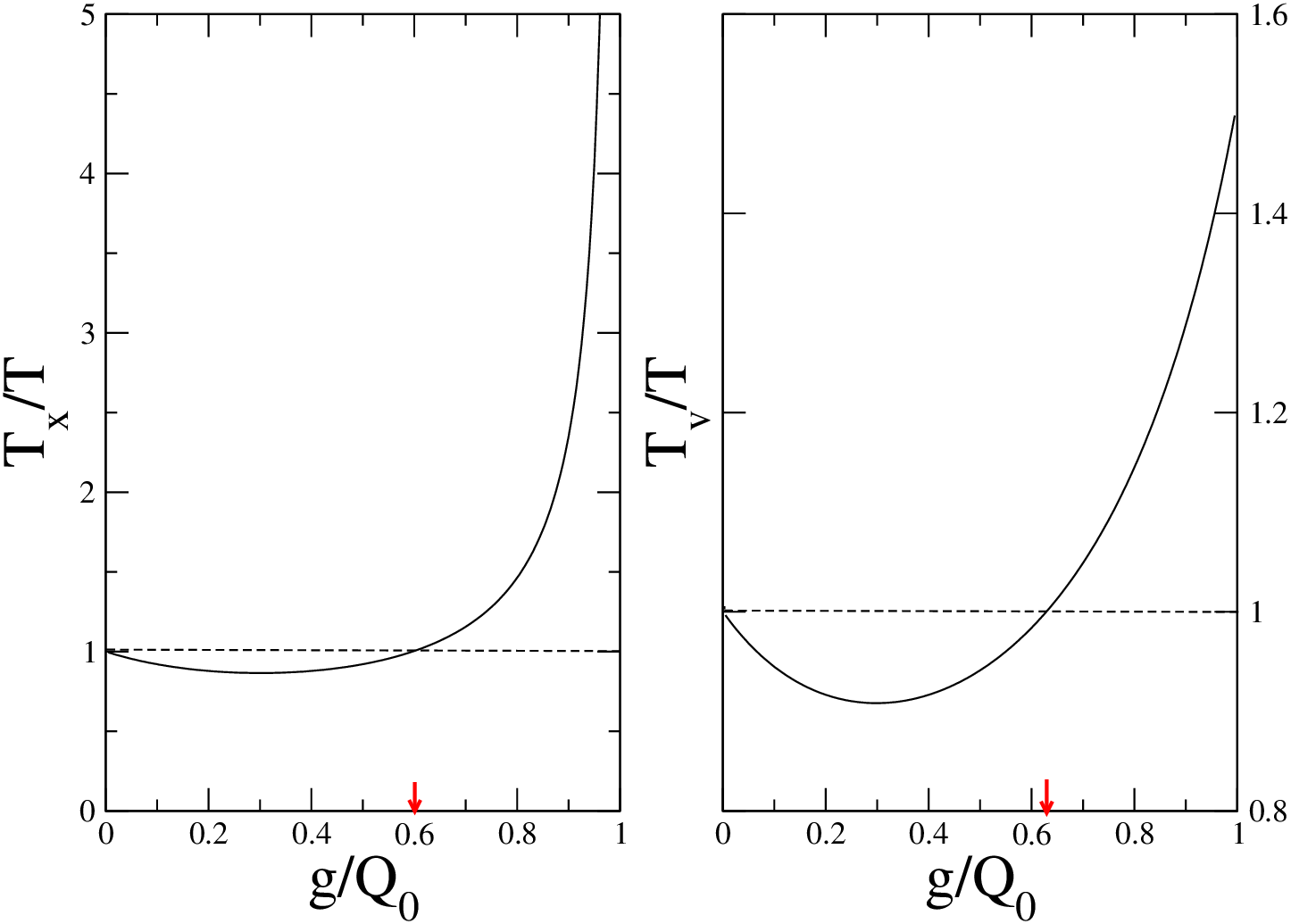}
 \caption{\label{FigTemp_vs_gain} (Color on line) Feedback-controlled harmonic oscillator: configurational ($T_x$) and kinetic ($T_v$) temperatures as a function of the feedback gain for $Q_0=2$ and $\tau=2.5$. One has $T_x/T=1$ for $g/Q_0\approx 0.60$ and  $T_v/T=1$ for $g/Q_0\approx 0.63$ (red arrows).}
\end{center}
\end{figure}

The main 	lesson  from Fig. \ref{FigTemp_vs_gain} is that the two temperatures have a nonmonotonic behavior as a function of the feedback gain. As $g$ increases, the system is first cooled then heated.  In this respect, the delay plays a role similar to that of measurement errors (see \textit{e.g.} Fig. 1 in \cite{MR2013}). We stress that the values $g^*_x$ and $g^*_v$ (indicated by red arrows in the figure) for which $T_x=T$ and $T_v=T$  are not identical. More generally, the configurational and kinetic temperatures are distinct as long as the feedback operates. Hence, the equipartition theorem is never satisfied and the system is  always out of equilibrium (in other words, there is no uniquely defined temperature, as stressed in\cite{GVTGA2008}).  We also observe in Fig. \ref{FigTemp_vs_gain} that the fluctuations of  $x$ grow indefinitely as $g\rightarrow Q_0$ (\textit{i.e.}, $k'\rightarrow k$ in the original Langevin equation (\ref{EqLlin})) whereas the fluctuations of $v$ remain finite\cite{note9}. In consequence, the heat flow is also finite.
\begin{figure}[hbt]
\begin{center}
\includegraphics[width=9cm]{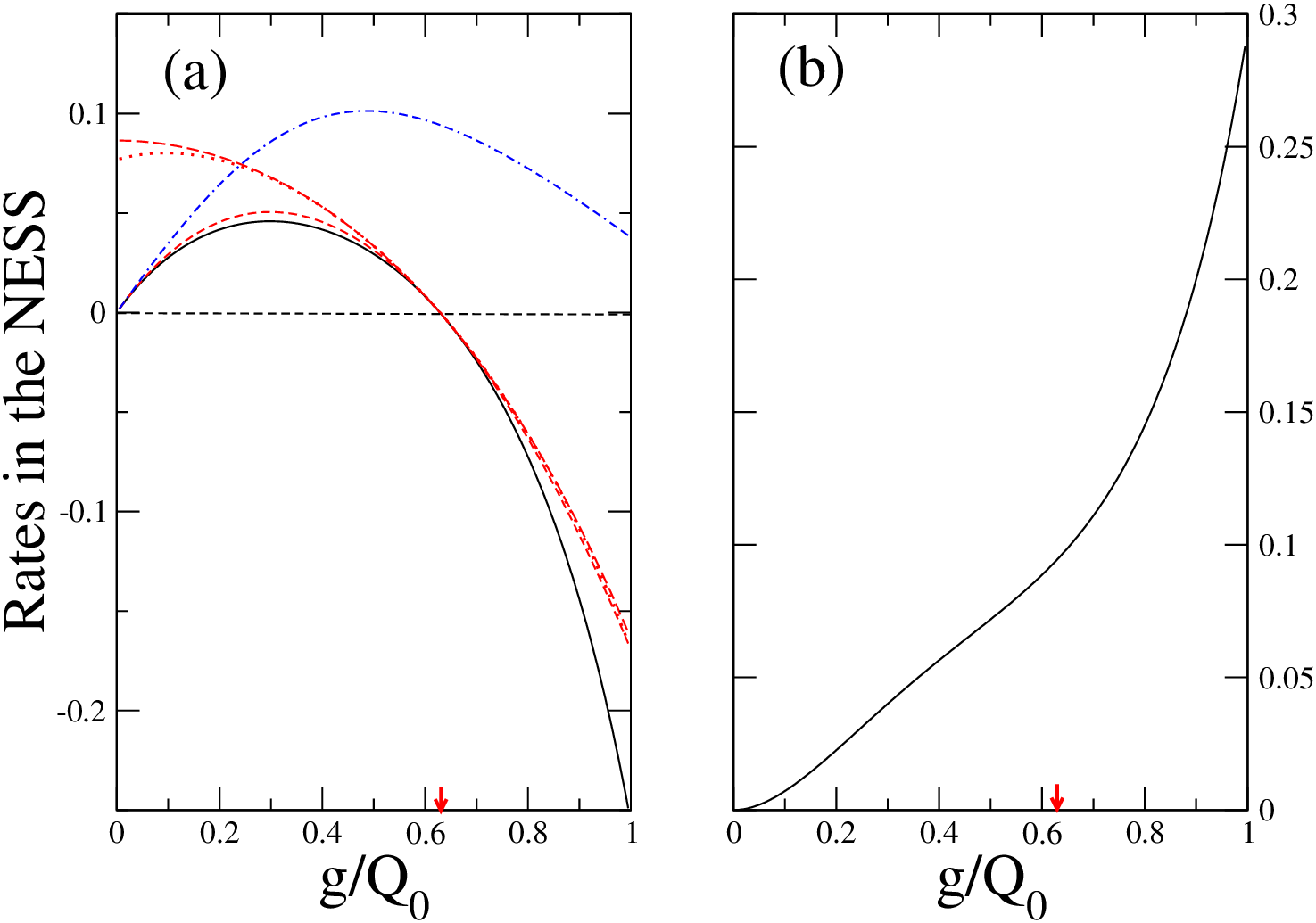}
 \caption{\label{Fig_rates_vs_gain} (Color on line) Feedback-controlled harmonic oscillator: second-law-like inequalities as a function of the feedback gain for $Q_0=2$ and $\tau=2.5$. (a): Comparison of the rates $\dot {\cal W}_{ext}/T$ (solid black line),  $\dot {\cal S}_{pump}^{v}$ (short-dashed red line),  $\dot {\cal I}_{flow,v}^{xv;y}$ (long-dashed red line),  $\dot {\cal I}_{flow,v}^{v;y}+\dot {\cal S}_{pump}^{v;y}$ (dotted red line), and $\dot {\cal S}_{\cal J}$ (dashed-dotted blue line); (b): Plot of  the ``entropy production" rate $\dot {\cal R}_{cg}=-\dot {\cal W}_{ext}/T+\dot {\cal S}_{\cal J}$.}
\end{center}
\end{figure}

For $g< g^*_v$, the system operates in the feedback-cooling regime: thanks to the information  continuously acquired by the measurement of the position, heat is transferred from the bath to the system and, according to the first law, this energy is released to the controller as work\cite{KQ2004}. However, as can be seen in Fig. \ref{Fig_rates_vs_gain}(a), $\dot {\cal W}_{ext}/T$ cannot be larger than $\dot {\cal S}_{pump}^{v}$ and $\dot {\cal S}_{\cal J}$ in agreement with the second-law-like inequalities (\ref{IneqEP1}) and (\ref{Eq2law3}) (we remind the reader that the two entropy pumping rates $\dot {\cal S}_{pump}^{v}$ and $\dot {\cal S}_{pump}^{xv}$ are equal when the Langevin equation is linear).  $\dot {\cal S}_{pump}^{v}$ turns out to be a rather tight bound for the extracted work and it is significantly smaller than $\dot {\cal S}_{\cal J}$. It is also  a better bound than $\dot {\cal I}_{flow,v}^{xv;y}$ and $\dot {\cal I}_{flow,v}^{v;y}+\dot {\cal S}_{pump}^{v;y}$, in agreement with  inequalities (\ref{Eqin4}) and (\ref{Eqin5}). In addition note that $\dot {\cal I}_{flow,v}^{v;y}+\dot {\cal S}_{pump}^{v;y}$ is a slightly better bound than $\dot {\cal I}_{flow,v}^{xv;y}$.  

The value $g^*_v$ of the gain is remarkable because there is no heat exchanged with the bath in this case. The viscous part of the effective force vanishes, as can be seen from Eq. (\ref{EqFfb1}), and the entropy pumping rates and the information flows vanish, as we noticed earlier. On the other hand,  the probability currents $J^{x}_{st}(x,v)$ and $J^{v}_{st}(x,v)$ are not zero and the system is {\it not} at equilibrium ($T_x\ne T$). This important feature is correctly captured by the bound  $\dot {\cal S}_{\cal J}$ obtained from time reversal. 
 This implies that the rate $\dot {\cal R}_{cg}=-\dot {\cal W}_{ext}/T+\dot {\cal S}_{\cal J}$, which is by construction nonnegative, is moreover always {\it strictly} positive, as illustrated in Fig. \ref{Fig_rates_vs_gain}(b). We take this as an indication that  $\dot {\cal R}_{cg}$ properly accounts for the irreversible character of the feedback process and  can be thus  regarded as a measure of the entropy production rate in the NESS.

\begin{figure}[hbt]
\begin{center}
\includegraphics[width=9cm]{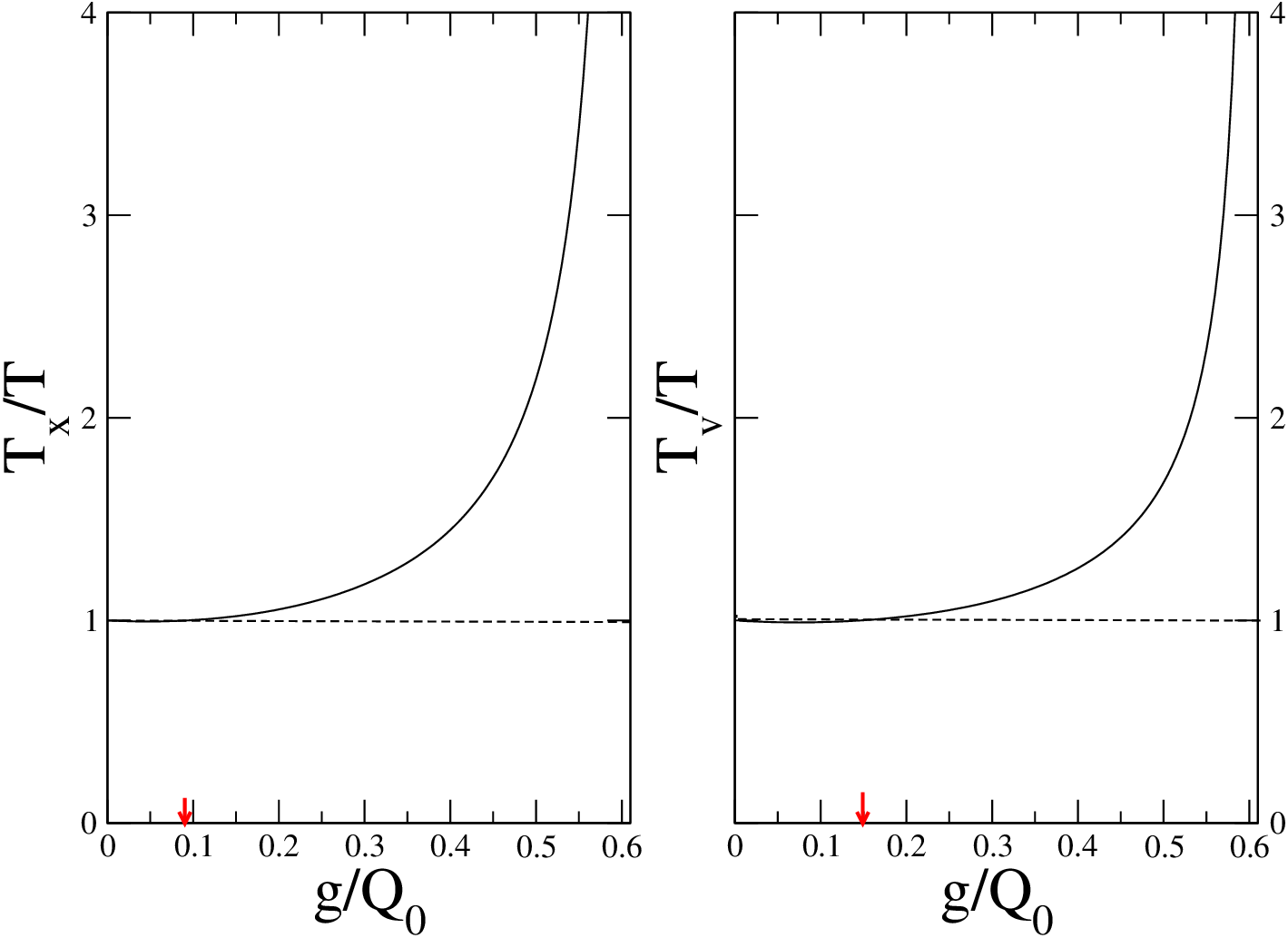}
 \caption{\label{FigTemp_vs_gain_tau=8} (Color on line) Same as Fig. \ref{FigTemp_vs_gain} for $\tau=8$. The temperatures $T_x$  and $T_v$ diverge at the stability limit of the stationary state ($g/Q_0\rightarrow g_c/Q_0\approx 0.61$). One has $T_x/T=1$ for $g/Q_0\approx 0.09$ and  $T_v/T=1$ for $g/Q_0\approx 0.15$ (red arrows).}
\end{center}
\end{figure}

We next consider a larger delay  $\tau=8$. The main difference with the preceding case is that the stationary state looses its stability when the gain reaches the critical value $g_c\approx 1.22$, as can be read off the diagram of Fig. \ref{FigChristmas_tree} (for $g/Q_0\gtrsim 0.484$, the system operates in the second stability lobe).  The fluctuations of both $x$ and $v$ then blow up  as $g\rightarrow g_c$ and the temperatures $T_x$ and $T_v$ plotted in Fig. \ref{FigTemp_vs_gain_tau=8} diverge, as does the heat flow (see the behavior of $\dot {\cal W}_{ext}$ in Fig. \ref{Fig_rates_vs_gain_tau=8}). There is still a cooling regime for very small values of the gain, although this is hardly seen on the scale of Fig. \ref{FigTemp_vs_gain_tau=8}.  Again, we observe on Fig. \ref{Fig_rates_vs_gain_tau=8} that the entropy pumping rate $\dot {\cal S}_{pump}^{v}$ is a tighter bound for the extracted work than $\dot {\cal S}_{\cal J}$. Note that these two quantities remain finite at the stability limit ($\dot {\cal S}_{pump}^{v} \rightarrow -1/Q_0$), as do the bounds $\dot {\cal I}_{flow,v}^{xv;y}$ and $\dot {\cal I}_{flow,v}^{v;y}+\dot {\cal S}_{pump}^{v;y}$.
As in the case $\tau=2.5$,  $\dot {\cal S}_{\cal J}$  is positive in the whole stability domain and does not vanish when $T=T_v$. 

\begin{figure}[hbt]
\begin{center}
\includegraphics[width=9cm]{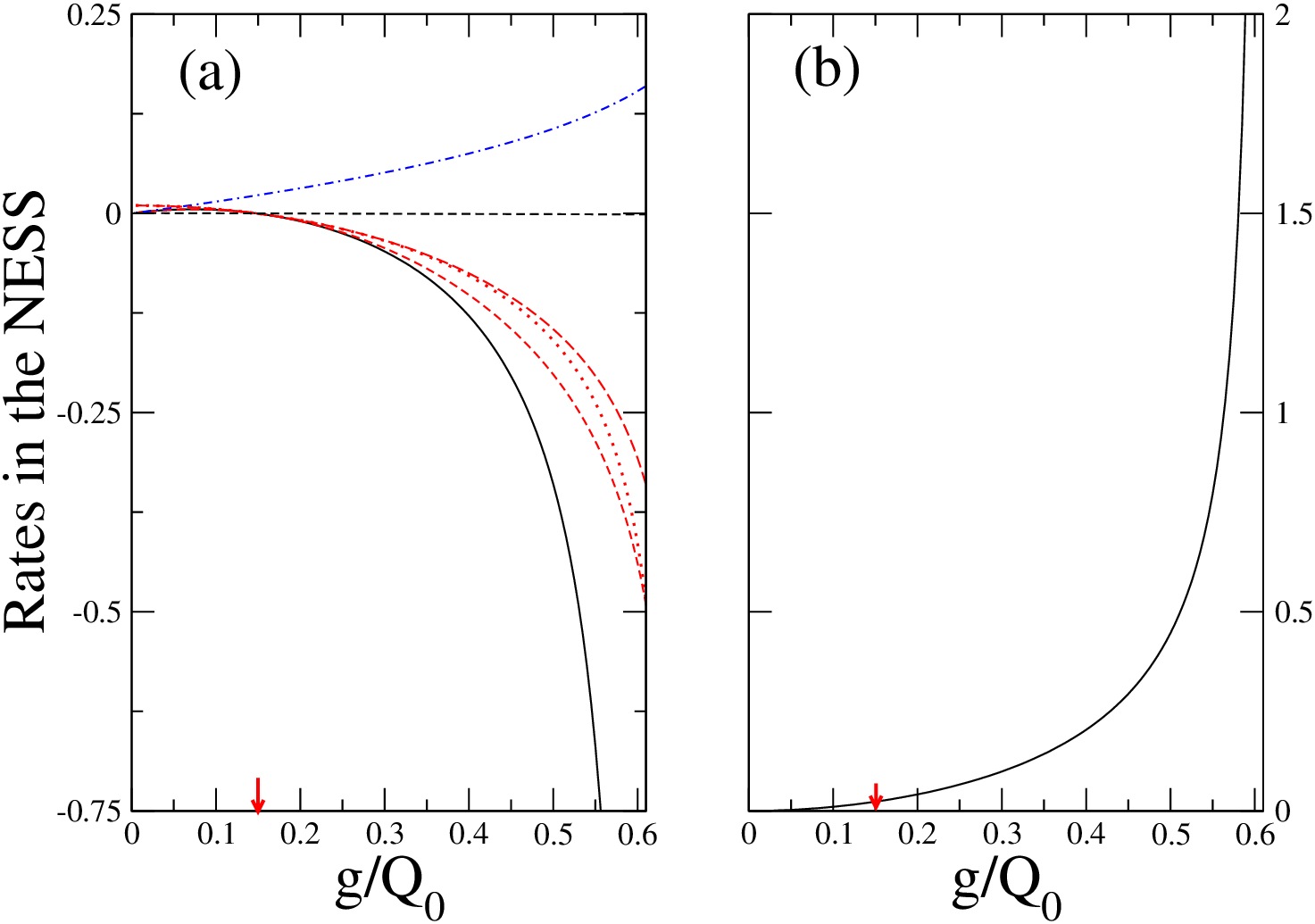}
 \caption{\label{Fig_rates_vs_gain_tau=8} (Color on line) Same as Fig. \ref{Fig_rates_vs_gain} for $\tau=8$. $\dot {\cal W}_{ext}$ (solid black line) diverges as $g/Q_0\rightarrow g_c/Q_0\approx 0.61$ whereas $\dot {\cal S}_{pump}^{xv}$ (short-dashed red line) and $\dot {\cal S}_{\cal J}$ (dashed-dotted blue line) remain finite. The other bounds built from information flows (long-dashed and dotted red lines) also stay finite.}
\end{center}
\end{figure}
\begin{figure}[hbt]
\begin{center}
\includegraphics[width=9cm]{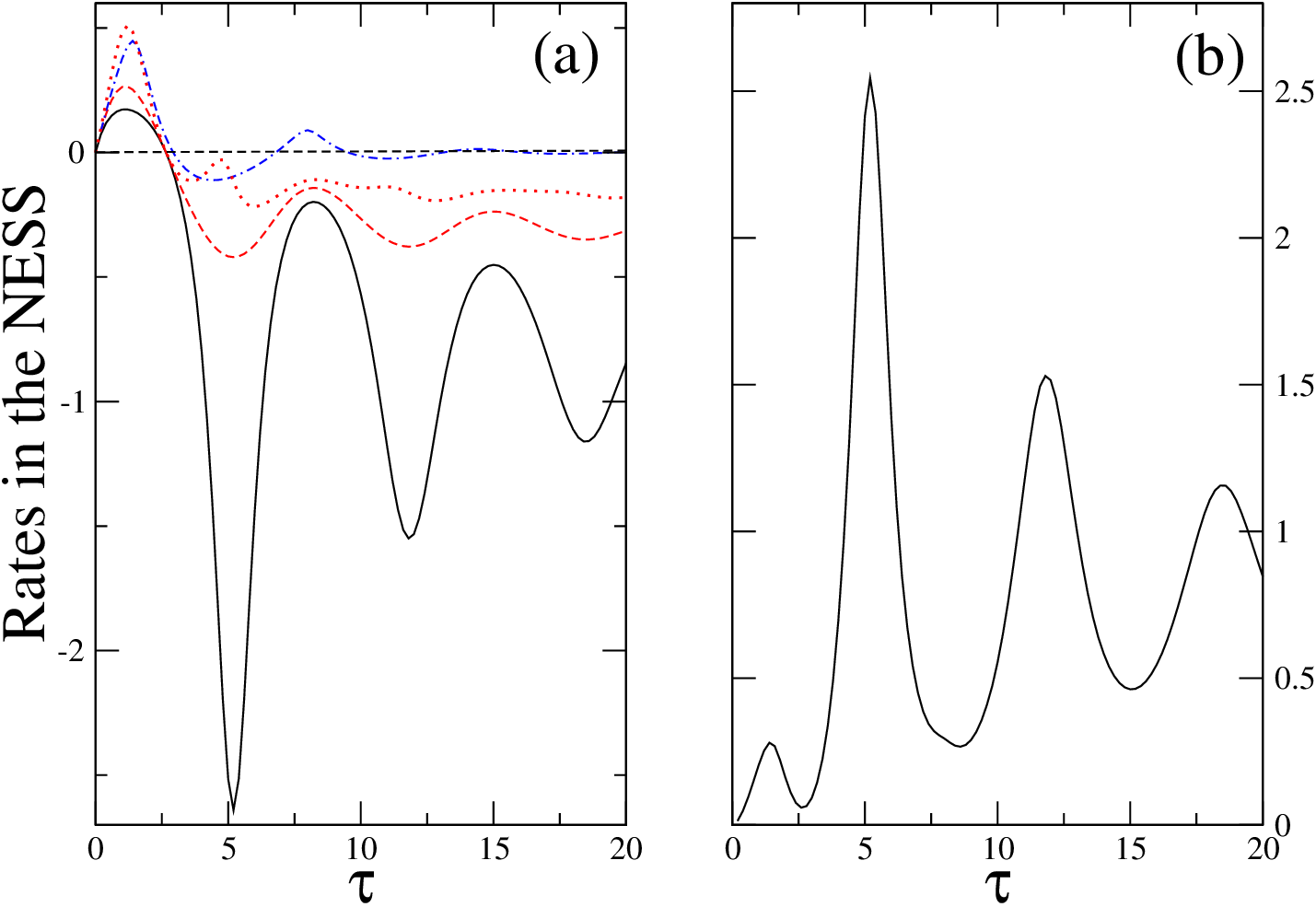}
 \caption{\label{Fig_rates_vs_tau_2} (Color on line) (a): Second-law-like inequalities (see caption of Fig. \ref{Fig_rates_vs_gain}) as a function of the delay for $Q_0=2$ and $g/Q_0=0.45$. The information flow $\dot {\cal I}_{flow,v}^{xv;y}$ is not represented as it diverges for $\tau=0$. Note that $\dot {\cal S}_{\cal J}$ (dashed-dotted blue line) changes sign with $\tau$, in contrast with its behavior with $g/Q_0$ for $\tau=2.5$ and $\tau=8$. (b): ``Entropy production" rate $\dot {\cal R}_{cg}$.}
\end{center}
\end{figure}

To gain more insight into the influence of the delay on the functioning of the feedback and on the thermodynamics of the process, we now fix the value of $g$ and vary $\tau$.  Indeed, the behavior  as a function of $g$ has the inconvenient feature that the random variables $y_t=x_{t-\tau}$ and $x_t$ or $v_t$ are already correlated in the absence of feedback (accordingly, and with our choice of sign, the corresponding information flows are strictly positive). Specifically, we consider the two values $g/Q_0=0.45$ and $g/Q_0=0.55$. As can be seen from the diagram in Fig. \ref{FigChristmas_tree}, the stationary state is stable for arbitrary $\tau\ge 0$ in the first case whereas there is a series of stability switches due to Hopf bifurcations in the second case. The corresponding variations of the temperatures $T_x$ and $T_v$ are shown in Figs. \ref{Fig15} and \ref{Fig16} of Appendix C. 
 
As shown in Fig.  \ref{Fig_rates_vs_tau_2} for  $g/Q_0=0.45$,  all quantities  display oscillations as a function of $\tau$, and the cooling regime only exists for small delays  ($\tau\lesssim 2.65$). The maximum in the extracted work  (corresponding to the minimal value of $T_v$) occurs for $\tau\approx 1.19$. Remarkably, this value is not so far from the optimal value $\pi/2$ valid in the limit $Q_0\rightarrow \infty$. As it must be, all second-law-like  inequalities are satisfied for all values of $\tau$.  Again, we observe that $\dot {\cal S}_{pump}^{v}$ is the tightest bound to $\dot {\cal W}_{ext}/T$. Note  that the sign of $\dot {\cal S}_{\cal J}$ changes with $\tau$ and that its first zero is close but distinct from $\tau^*(g)$, the value for which $T_v=T$ and  $\dot {\cal W}_{ext}=0$. In all cases, provided $\tau>0$, the rate $\dot{\cal R}_{cg}$ plotted in Fig. \ref{Fig_rates_vs_tau_2}(b) is always strictly positive.

An interesting feature is that  $\dot {\cal S}_{\cal J}\rightarrow 0$ as $\tau \rightarrow \infty$ whereas  $T_x$ and $T_v$ (see Eqs. (\ref{Eqlim3})), and therefore $\dot {\cal W}_{ext}$ and the other rates, go to finite values. When $\tau$ is very large, the feedback is indeed completely inefficient  since measuring the position at time $t-\tau$ does not bring any information about the state of the system at time $t$. The feedback force then acts as a purely random force and the system is heated ($\dot {\cal W}_{ext}<0$). It is remarkable that $\dot {\cal S}_{\cal J}$ captures this effect by going to zero. Accordingly, one has $\dot{\cal R}_{cg}\sim -\dot {\cal W}_{ext}/T$ asymptotically. As will be discussed in a forthcoming paper\cite{RTM2015}, $\dot {\cal S}_{\cal J}$ has some relation, at least in the case of linear systems, with the logarithm of the so-called ``efficacy parameter" defined in \cite{SU2010,SU2012}  for nonautonomous Maxwell's demons. The fact that the efficacy parameter converges to $1$ as the delay between the measurement  and the control action becomes very large has been observed experimentally\cite{TSUMS2010}.

The behavior for $g/Q_0=0.55$ is shown in  Fig. \ref{Fig_rates_vs_tau_2bis}. The main difference with the preceding case is the occurrence of Hopf bifurcations that result in the presence of three stability lobes (see Fig. \ref{Fig16} in Appendix C; for clarity, only two of them are shown in Fig. \ref{Fig_rates_vs_tau_34.2}). The only new piece of information is that $\dot {\cal S}_{\cal J}$, as a well-defined mathematical quantity,  remains finite in the instability zones (blue dotted line in the figure). This of course has no special physical meaning.

\begin{figure}[hbt]
\begin{center}
\includegraphics[width=9cm]{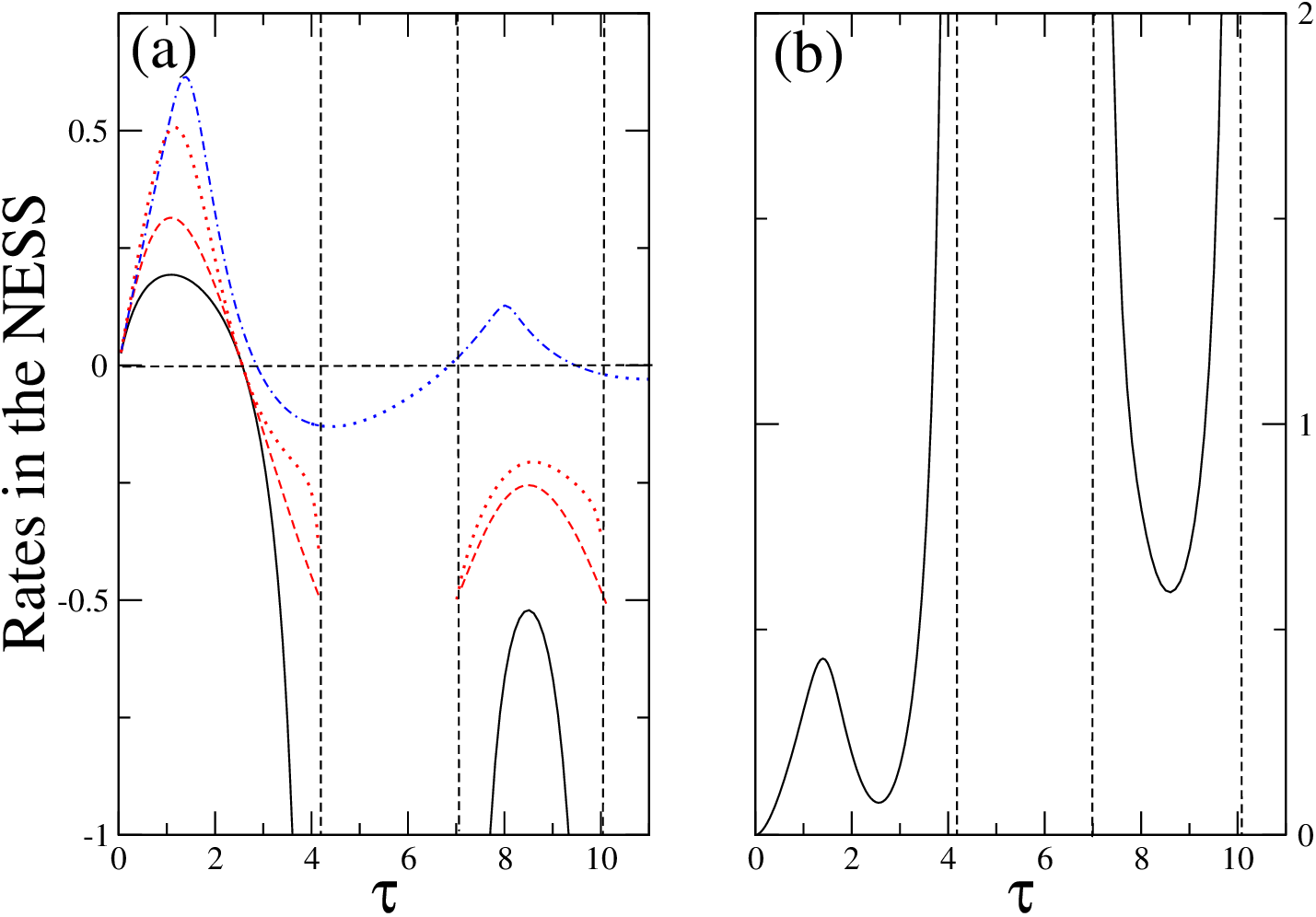}
 \caption{\label{Fig_rates_vs_tau_2bis} (Color on line) Same as Fig. \ref{Fig_rates_vs_tau_2} for  $Q_0=2$ and $g/Q_0=0.55$. The stability lobes are delimited by the dashed vertical lines; the third stability lobe (for $15.02<\tau< 15.97$) is not represented. The dotted blue line represents the continuation of $\dot S_{\cal J}$ in the instability regions.}
\end{center}
\end{figure}

\subsubsection{$Q_0=34.2$}

The above cases describe the behavior of oscillators with a low quality factor, such as a torsion pendulum in a viscous fluid (see \textit{e.g.} \cite{JGC2007}). 
We now  consider a resonator with a larger $Q_0$ and take as an example the AFM micro-cantilever used in the experiments of \cite{GBPC2010}  for which  $Q_0\approx 34.2$ ($2\pi/\omega_0=116 \: \mu s$ and $\tau_0=632\: \mu s$). The corresponding rates are plotted in Fig. \ref{Fig_rates_vs_tau_34.2} as a function of $\tau$, for an arbitrarily value of the gain  $g=0.25Q_0$. 
\begin{figure}[hbt]
\begin{center}
\includegraphics[width=9cm]{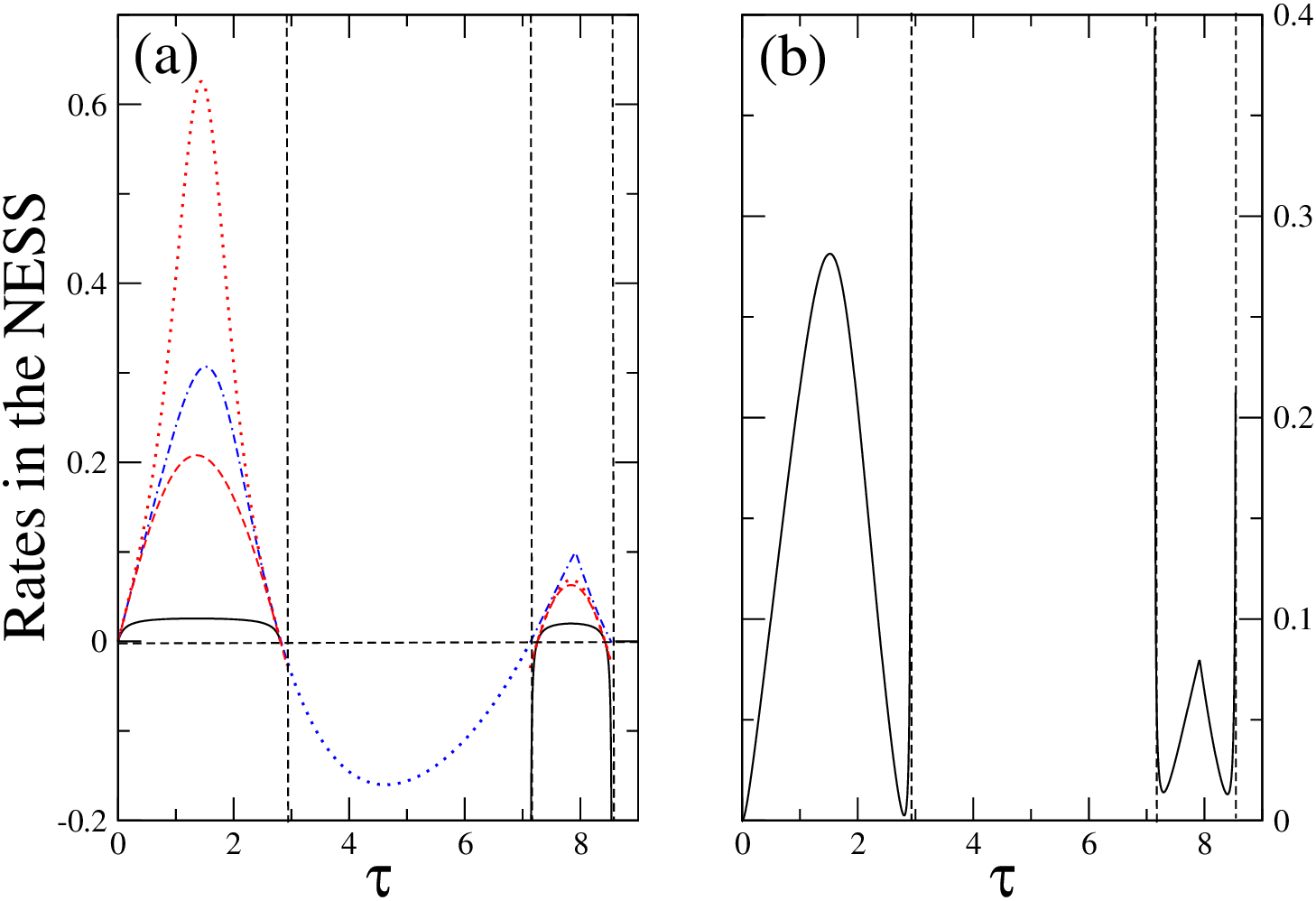}
 \caption{\label{Fig_rates_vs_tau_34.2} (Color on line) Same as Fig. \ref{Fig_rates_vs_tau_2} for $Q_0=34.2$ and $g/Q_0=0.25$ (in this case, there are only two stability lobes). (a): $\dot {\cal W}_{ext}/T$ (solid black line) and various upper bounds. (b): ``Entropy production" rate $\dot {\cal R}_{cg}$.}
\end{center}
\end{figure}

Although there are only two stability lobes in this case, the behavior starts to resemble the one observed for $Q_0\rightarrow \infty$ (see Figs. 3 (b) and 3(d)). In particular, the bounds $\dot {\cal S}_{pump}^{v}$ and $\dot {\cal I}_{flow,v}^{v;y}+\dot {\cal S}_{pump}^{v;y}$ significantly overestimate the extracted  work rate. If one were to define the corresponding  ``feedback efficiencies" as $\epsilon_1=\dot {\cal W}_{ext}/(T\dot {\cal S}_{pump}^{v})$ and  $\epsilon_2=\dot {\cal W}_{ext}/[T(\dot {\cal I}_{flow,v}^{v;y}+\dot {\cal S}_{pump}^{v;y})]$, they would be very small. The same is true for the bound $\dot S_{\cal J}$, and therefore the main contribution to the positive ``entropy production" rate $\dot {\cal R}_{cg}=-\dot {\cal W}_{ext}/T+\dot {\cal S}_{\cal J}$ in Fig. \ref{Fig_rates_vs_tau_34.2}(b) comes from $\dot S_{\cal J}$. Note that the ranking $\dot {\cal W}_{ext}/T\le \dot {\cal S}_{pump}^{v}\le \dot {\cal S}_{\cal J}\le \dot {\cal I}_{flow,v}^{v;y}+\dot {\cal S}_{pump}^{v;y} $ in the first stability lobe is the one predicted in the limit $Q_0\rightarrow \infty$. On the other hand,  $\dot {\cal S}_{\cal J}\ge\dot {\cal I}_{flow,v}^{v;y}+\dot {\cal S}_{pump}^{v;y} $ in the second lobe.

As discussed in Appendix D, a remarkable mathematical property  of  the rate  $\dot {\cal S}_{\cal J}$  is that  it is not an analytic function of $g$ (for fixed $\tau$) or of $\tau$ (for fixed $g$) because of a crossing phenomenon in the poles of the acausal response function $\tilde \chi(s)$. This is the origin of the cusp observed in $\dot {\cal S}_{\cal J}$ (and thus also in $\dot {\cal R}_{cg}$) in the second stability lobe for $\tau \approx 7.9$ (see Fig. 17).

\subsubsection{$Q_0=250$}

Finally, we consider an even larger value of the quality factor, $Q_0=250$. (Note, however, that this is still very far from the quality factors of the resonators used in feedback-cooling setups which may be of the order of $10^4$, see \textit{e.g.} \cite{PCBH2000,PDMR2007,Mont2012}.) 
The rates as a function of $\tau$ for $g/Q_0=0.04$ are plotted in Fig. \ref{Fig_rates_vs_tau_250} where we only show the first two stability lobes, for clarity. In these lobes, the behavior is now fully in agreement with the asymptotic behavior described earlier, with $\dot {\cal S}_{\cal J}\le \dot {\cal I}_{flow,v}^{v;y}+\dot {\cal S}_{pump}^{v;y} $. Note also that the difference between $\dot {\cal S}_{\cal J}$ and $\dot {\cal S}_{pump}^{v} $ increases with $\tau$, as predicted by Eq. (\ref{Eqcompar}).  
\begin{figure}[hbt]
\begin{center}
\includegraphics[width=9cm]{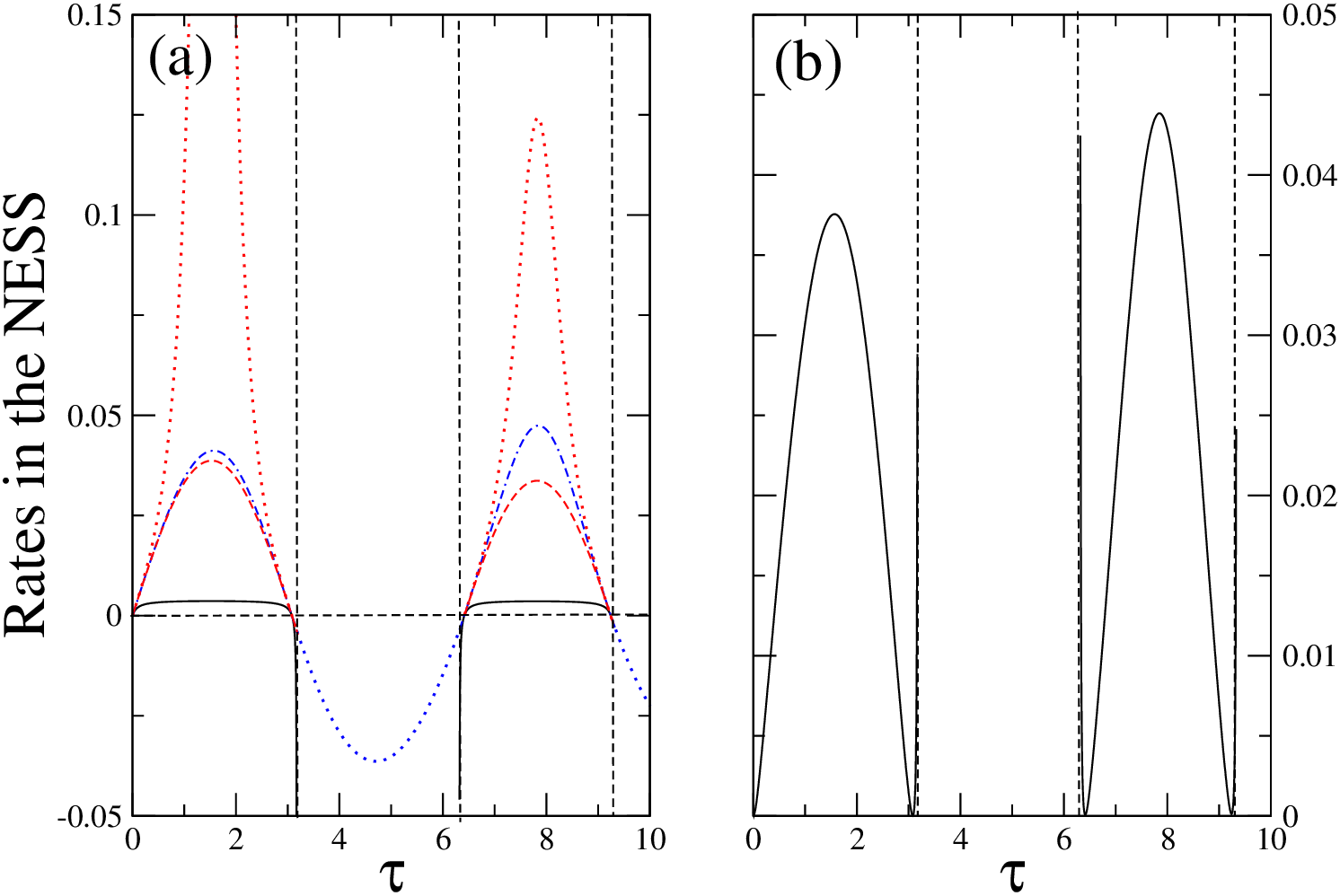}
 \caption{\label{Fig_rates_vs_tau_250} (Color on line) Same as Fig. \ref{Fig_rates_vs_tau_2} for $Q_0=250$ and $g/Q_0=0.04$. There are many other stability lobes that are not represented.} 
\end{center}
\end{figure}

\section{Closing remarks and future directions}

In this paper, we have developed an ensemble of theoretical tools enabling a comprehensive analysis of the nonequilibrium thermodynamics associated to  Langevin systems subjected to continuous time-delayed feedback control. The non-Markovian character of the dynamics raises new conceptual and technical  issues that are beyond the current framework of stochastic thermodynamics. We have shown that there are two important consequences. 

First, since the probabilistic description of the system on the ensemble level requires the knowledge of the whole Kolmogorov hierarchy, one can work  at different levels of the Fokker-Planck description  and use different definitions of the Shannon entropy. One then obtains a set of  nonequilibrium inequalities that generalize the standard second law of thermodynamics and that involve additional contributions characterizing the reduction of entropy production due to the continuous measurement process. When the system settles in a nonequilibrium steady state, these inequalities provide bounds to the work that can be extracted from the surrounding heat bath. The best bound appears  to be the so-called entropy pumping contribution, originally introduced  in the case of a Markovian velocity-dependent feedback.

Second, and perhaps more importantly from a fundamental standpoint, the microscopic reversibility condition (or local detailed balance) obtained by comparing the probabilities of the actual and time-reversed stochastic trajectories, is modified. The acausal character of the (fictitious) reverse process  introduces  a new functional whose general expression has been derived. This in turn leads to another second-law-like inequality, a new bound to the extracted work, and a sensible mathematical (if not physical) definition of the entropy production associated to the time-delayed Langevin equation. 

Of course, the present description is still a reduced one that adopts the only point of view of the feedback-controlled system and does not take into account the implementation cost of the controller. Moreover, measurement noise is not yet included in the formalism.

These issues will be investigated in a forthcoming paper\cite{RTM2015}. There, we will also study in more detail the asymptotic integral fluctuation theorem conjectured in \cite{MR2014} and establish its connection with the existing results for non-autonomous feedback. Finally, we will consider the statistics of the extracted work, focusing on the large deviation behavior.
  
\begin{acknowledgments}
We are grateful to E. Kierlik for useful discussions and to J. Horowitz and H. Sandberg for sending us their preprint\cite{HS2014} before publication, which inspired some parts of this work.
\end{acknowledgments}

\appendix

\section{Path probability for a linear overdamped SDDE in the stationary state}

In this Appendix, we illustrate the general formalism of Section II.C by deriving the explicit expression of the  path probability  associated with the linear SDDE
\begin{align}
\label{EqLov}
\gamma \dot x_t =-kx_t +k'x_{t-\tau}+\sqrt{2\gamma T}\xi(t) 
\end{align}
in the stationary state and for $t_f-t_i\le \tau$. This equation, which corresponds to the overdamped limit of Eq. (\ref{EqLlin}), is well documented in the literature\cite{KM1992,FB2001,FBF2003,BC2004}. In particular, the stability range  of the stationary state and the corresponding time-correlation function have been exactly determined.

Instead of trying to solve Eq. (\ref{Eqpathst}) directly, we use the fact that the process is Gaussian, which implies that the path probability of a trajectory ${\bf X}$ observed during the time interval $[t_i,t_f]$ is given by
\begin{align}
\label{EqPI1}
{\cal P}_{st}[{\bf X}]\propto e^{ -\frac{1}{2}\int_{t_i}^{t_f} dt \int_{t_i}^{t_f}dt'\: x_t\phi^{-1}(t,t') x_{t'}}
\end{align}
where  $\phi^{-1}(t,t')$ is the inverse of the stationary time-correlation function $\phi(t):= \langle x(0)x(t)\rangle_{st} $ defined according to 
\begin{align}
\label{EqinvC}
\int_{t_i}^{t_f} dt'' \phi( t-t'')\phi^{-1}(t'',t')=\delta(t-t') \ .
\end{align} 
The problem thus boils down to finding the solution of this integral equation.
First of all, let us examine the simple Markovian case $k'=0$ that corresponds to the Ornstein-Uhlenbeck process. In this case,
\begin{align}
\label{Eqco0}
\phi(t)=\frac{T}{k} e^{-\frac{k}{\gamma}\vert t\vert}\ ,
\end{align}
and the solution of Eq. (\ref{EqinvC}) is known. Indeed,  the same equation has to be solved to find the path-integral representation of a process driven by an Ornstein-Uhlenbeck noise (with the correlation function of the noise playing the role of $\phi(t)$). The result is\cite{WPMPR1989}
\begin{align}
\label{Eqco1}
\phi^{-1}(t,t')&=\frac{k^2}{2\gamma T}\big\{\delta(t-t')-\frac{\gamma^2}{k^2}\delta''(t-t')\nonumber\\
 & +\frac{4\gamma}{k}[\delta (t-t_i)\delta (t'-t_i)+\delta (t-t_f)\delta (t'-t_f)]\nonumber\\
 &+\frac{4\gamma^2}{k^2}[\delta'(t-t_i)\delta(t'-t_i)-\delta'(t-t_f)\delta(t'-t_f)]\big\}\ ,
\end{align}
where $\delta' $ and $\delta ''$ denote the first and second derivatives of $\delta$.  By substituting into Eq. (\ref{EqPI1}), a few manipulations lead to\begin{align}
\label{EqOM}
{\cal P}_{st}[{\bf X}]\propto p_{st}(x_i)e^{-\frac{1}{4\gamma T}\int_{t_i}^{t_f} dt\: (\gamma \dot x_t + k x_t)^2} \ .
\end{align}
with $p(x_i)\propto e^{-kx_i^2/(2T)}$. One thus  recovers the classical Onsager-Machlup action functional\cite{OM1953} in the absence of external force.
Of course, this result is more directly obtained by inserting the Langevin equation into the probability density functional of the noise realizations.

We now consider the case $k'\ne 0$. The time-correlation function $\phi(t)$ for $\vert t\vert \le \tau$ is  now given by\cite{KM1992,FBF2003} 
\begin{align}
\label{Eqco2}
\phi(t)=A_+ e^{-\frac{\bar k}{\gamma}\vert t\vert}+A_- e^{\frac{\bar k}{\gamma} \vert t \vert}
\end{align}
where $ \bar k=\sqrt{k^2-k'^2}$, $ A_{\pm}=(1/2)[\phi(0) \pm T/\bar k]$, and
\begin{align}
\label{Eqphi0}
 \phi(0)=\langle x^2\rangle= \frac{T}{k}\frac{1-(k'/\bar k)\sinh (\frac{\bar k}{\gamma} \tau)}{1-(k'/k)\cosh(\frac{\bar k}{\gamma} \tau)}\ .
\end{align}
(Hereafter, for brevity, we only consider the case $k>\vert k'\vert$ so that $\bar k$ is real.)
After inserting the expression of $\phi(t)$ into Eq. (\ref{EqinvC}), one immediately finds  that the ``bulk" term of $\phi^{-1}(t,t')$ (involving $\delta(t-t')$ and $\delta''(t-t')$) has the same form as in Eq. (\ref{Eqco1}) with $k$ replaced by $\bar k$. On the other hand, the boundary terms that explicitly depend on $t_i$ and $t_f$ (called the ``surface'' terms  in \cite{WPMPR1989}) are much more elusive.  In order  to obtain the full expression of $\phi^{-1}(t,t')$,  it is actually  more convenient to discretize the problem and take the continuum limit at the end. We thus  divide the time interval $[t_i,t_f]$ into $N$  infinitesimal slices of width $\epsilon=(t_f-t_i)/N$ with $t_k=t_i+k \epsilon$ ($k=0...N$), and $x_k=x(t_k)$ with $x_0:= x_i$, $x_N:= x_f$.  Eq. (\ref{EqPI1})  then reads
\begin{align}
\label{EqP}
P(x_0,t_0; x_1,t_1...x_N,t_N)\propto e^{-\frac{1}{2}\sum_{k,l=0}^N x_k\phi^{-1}_{kl}x_l} \ ,
\end{align}
and Eq. (\ref{EqinvC}) is replaced by the matrix equation
\begin{align}
\label{EqinvC1}
\sum_{l'=0}^N\phi_{kl'}\phi^{-1}_{l'l}=\delta_{k,l}\ .
\end{align} 
Since $\phi(t)$ is even, the matrix $\phi_{kl}$ is symmetric with $\phi_{kl}=A_+e^{-(\bar k/\gamma)(t_l-t_k)}+A_-e^{(\bar k/\gamma)(t_l-t_k)}$ for $t_l\ge t_k$. The expression of the quadratic form in Eq. (\ref{EqP}) is found by generalizing the solution obtained for the Ornstein-Uhlenbeck noise in Ref.\cite{D1942} (see also \cite{PRS1983}), using the fact that only the boundary terms  for $k,l=0,N$ are different. As can be checked explicitly for small values of $N$, the result is
\begin{align}
\label{Eqquad}
\sum_{k,l=0}^N x_k\phi^{-1}_{kl}x_l& =\frac{1}{A_+-A_-}\Big\{\sum_{k=1}^N \frac{(x_k-e^{-(\bar k/\gamma) \epsilon}x_{k-1})^2}{1-e^{-2(\bar k/\gamma)\epsilon}}\nonumber\\
&+\frac{[A_+e^{-N(\bar k/\gamma)\epsilon}x_0-A_-x_N]^2}{A_+^2e^{-2N(\bar k/\gamma) \epsilon}-A_-^2}\Big\} \ .
\end{align}
Expanding in $\epsilon$ leads to
\begin{align}
\frac{(x_k-e^{-(\bar k/\gamma)  \epsilon}x_{k-1})^2}{1-e^{-2(\bar k/\gamma) \epsilon}}&=\frac{\gamma}{\bar k} \frac{(x_k-x_{k-1})^2}{2\epsilon}+\frac{1}{2}(x_k^2-x_{k-1}^2)\nonumber\\
&+\frac{\bar k}{\gamma} \epsilon\frac{x_k^2+x_{k-1}^2+x_kx_{k-1}}{6}+{\cal O}(\epsilon^2)
\end{align}
which in the continuum limit  $\epsilon  \rightarrow 0$, $N \rightarrow \infty$, with $N\epsilon=t_f-t_i$, gives
\begin{align}
&\lim_{\epsilon  \rightarrow 0,N \rightarrow \infty}\sum_{k=1}^N\frac{(x_k-e^{-(\bar k/\gamma) \epsilon}x_{k-1})^2}{1-e^{-2(\bar k/\gamma) \epsilon}}\nonumber\\
&=\frac{1}{2}\int_{t_i}^{t_f} dt\:(\frac{\gamma}{\bar k}\dot x_t^2+\frac{d}{dt}x_t^2+\frac{\bar k}{\gamma}x_t^2)\nonumber\\
&=\frac{1}{2\gamma \bar k}\int_{t_i}^{t_f} dt\:(\gamma \dot x+\bar k x_t)^2 \ .
\end{align}
After using $A_+-A_-=T/\bar k$, we finally obtain the sought expression of the path probability:
\begin{align}
\label{Eqpathexact}
{\cal P}_{st}^{(1)}[{\bf X}]&\propto e^{-\frac{1}{4 \gamma T}\int_{t_i}^{t_f} dt\: [\gamma \dot x_t +\bar k x_t]^2}\nonumber\\
&\times e^{ -\frac{\bar k}{2T}\frac{[A_+e^{-(\bar k/\gamma)(t_f-t_i)}x_i-A_-x_f]^2}{A_+^2e^{-2(\bar k/\gamma) (t_f-t_i)}-A_-^2}} \ .
\end{align}
Remarkably, the `action'  still looks like an Onsager-Machlup functional, but with a renormalized spring constant $\bar k$. Moreover, the second exponential factor introduces an explicit dependence on the initial and final microstates $x_i$ and $x_f$ that reflects the non-Markovian nature of the process. This factor identifies with $p_{st}(x_i)\propto \exp[-x_i^2/(2\langle x^2\rangle)]$ only when $\tau=0$, as, then,  $A_+=T/k$ and $A_-=0$. 

This exact expression of the path probability may be compared with the one given in \cite{JXH2011} (see Eq. (19) in this reference) in which the `renormalized' force $-\bar k x$ is replaced by the  effective force ${\overline F}_{tot,st}(x)=F(x)+{\overline F}_{fb,st}(x)=-k(T/T_x^{(0)}) x$ where $T_x^{(0)}$ is the $m=0$ limit of $T_x$ given by Eq. (\ref{EqTeffxm}). As we already noticed, this amounts to replacing the original time-delayed Langevin equation by an effective Markovian equation which yields the same stationary probability density $p_{st}(x)$ but does not generate the true trajectories. This invalidates the analysis of the stochastic thermodynamics performed in  \cite{JXH2011}.

One can check that Eq. (\ref{Eqpathexact}) is indeed the solution of the integral equation (\ref{Eqpathst}). This requires the computation of the Gaussian path integral over ${\bf Y}$ for ${\bf X}$ fixed, which is a routine but tedious exercise that is not reproduced here. (In particular, one needs to solve  the Euler-Lagrange equation that gives the ``optimal'' path ${\bf Y}^{^*}[{\bf X}]$.)
On the other hand, it is easy to verify that the stationary probability distribution $p_{st}(x_i)$ is recovered from the definition
\begin{align}
p_{st}(x_i)=\int dx_f\int_{x_i}^{x_f} {\cal D}[{\bf X}]{\cal P}_{st}^{(1)}[{\bf X}]
\end{align}
by  using the well-known expression of the transition probability for the Smoluchowski process\cite{R1989}
\begin{align}
\int_{x_i}^{x_f} {\cal D}[{\bf X}]e^{-\frac{1}{4 \gamma T}\int_{t_i}^{t_f} dt\: [\gamma \dot x_t +\bar k x_t]^2}\propto e^{-\frac{\bar k}{2T} \frac{[x_f-x_ie^{-(\bar k/\gamma) (t_f-t_i)}]^2}{1-e^{-2(\bar k/\gamma) (t_f-t_i)}}}\ .
\end{align} 
Performing the integration over $x_f$ readily yields
\begin{align}
p_{st}(x_i)&\propto e^{-\frac{A_+-A_-}{A_++A_-}\frac{\bar k x_i^2}{2T}}\propto e^{-\frac{x_i^2}{2\langle x^2\rangle}}\ .
\end{align}

Finally, it is important to recall that Eq. (\ref{Eqpathexact}) is only valid for $t_f-t_i\le \tau$. The key feature that allows us to solve Eq. (\ref{EqinvC1})  is that the two exponentials in the time-correlation function $\phi(t)$ involve a single characteristic time $\gamma/\bar k$.  This is no longer  true for $t>\tau$. For instance, in the time interval $\tau\le t\le 2\tau$, one can show that 
\begin{align}
\phi(t)&=k'\big[\frac{A_+}{k- \bar k}e^{-(\bar k/\gamma)(t-\tau)}+\frac{A_-}{k+ \bar k}e^{(\bar k/\gamma)(t-\tau)}\big]\nonumber\\
&-\frac{2T}{k'}e^{-\frac{k}{\gamma}(t-\tau)}\ ,
\end{align}
and computing $\phi_{kl}^{-1}$ for $t_f-t_i\le 2\tau$ is a daunting task. For the same reason, we have not succeeded in obtaining the expression of the path probability for $m\ne 0$ and $t_f-t_i\le \tau$ because the time-correlation functions involve two distinct relaxation times, $\omega_1^{-1}$ and $\omega_2^{-1}$, as shown  in Appendix B.

 \section{Determination of the effective temperatures $T_x$ and $T_v$}

Here, we detail the procedure that leads to Eqs. (\ref{EqTx}) and (\ref{EqTv}) for the effective temperatures $T_x$ and $T_v$.  To this aim, we calculate the stationary time correlation functions $\phi_{xx}(t)$ and $\phi_{vv}(t)=-\ddot \phi_{xx}(t)$  in the time interval $0\le t\le \tau$. The two  temperatures are then obtained from the respective mean square amplitudes as $T_x/T= (2/Q_0)\phi_{xx}(0)$ and $T_v/T=(2/Q_0)\phi_{vv}(0)$  (see \cite{PFFBT2006} of a similar calculation). 

Multiplying  Eq. ({\ref{EqLlinred}) by $v(0)$ and averaging over disorder leads to the deterministic differential equation
\begin{align}
\label{Eqdifvv01}
{\dot \phi}_{vv}(t)+\frac{1}{Q_0} \phi_{vv}(t)+\phi_{vx}(t)-\frac{g}{Q_0}\phi_{vx}(t-\tau)=0\ ,
\end{align}
where we have used the fact that $\langle v(0)\xi(t)\rangle=0$ for $t>0$ by causality. After differentiating  with respect to $t$ and using the symmetry relation $\phi_{vv}(t)=\phi_{vv}(-t)$, we derive a closed equation for $\phi_{vv}(t)$,
\begin{align}
\label{Eqdifvv}
{\ddot \phi}_{vv}(t)+\frac{1}{Q_0} {\dot \phi}_{vv}(t)+\phi_{vv}(t)-\frac{g}{Q_0} \phi_{vv}(\tau-t)=0\ .
\end{align}
This implies to restrict the calculation to the time interval $0\le t\le \tau$ so that $\tau-t$ remains positive.   We now seek a  solution of Eq. (\ref{Eqdifvv}) in the form
\begin{align}
\phi_{vv}(t)=A_v\sin \omega (t-\frac{\tau}{2})+B_v\cos\omega (t-\frac{\tau}{2}) 
\end{align}
where $\omega$ is a quantity that may be real or complex depending on the values of  $g$ and $Q_0$. Inserting this form into Eq. (\ref{Eqdifvv}) leads to the two equations
\begin{align}
\label{EqAB}
(1-\omega^2-\frac{g}{Q_0})B_v&=-\frac{\omega}{Q_0} A_v\nonumber\\
(1-\omega^2+\frac{g}{Q_0})A_v&=\frac{\omega}{Q_0} B_v \ ,
\end{align}
from which we obtain
\begin{align}
\label{Eqomega0}
\omega^4 -(2-\frac{1}{Q_0^2})\omega^2+1-\frac{g^2}{Q_0^2}=0\ .
\end{align}
The roots of this  polynomial equation are 
\begin{align}
\label{Eqroots0}
\omega_{1,2}=\left[1-\frac{1}{2Q_0^2}\pm \frac{1}{Q_0} \sqrt{\Delta}\right]^{1/2}
\end{align}
with
\begin{align}
\label{EqDis}
\Delta=g^2-1+\frac{1}{4Q_0^2} \ .
\end{align}
(Since the final expressions of the temperatures turn out to be even functions of  $\omega_1$ and $\omega_2$, we only take the positive sign in front of the square bracket in Eq. (\ref{Eqroots0}).) The general solution of Eq. (\ref{Eqdifvv}) is then a linear combination of the two particular solutions with frequencies $\omega_1$ and $\omega_2$, 
\begin{align}
\phi_{vv}(t)&=A_v^{(1)}\sin \omega_1 (t-\frac{\tau}{2})+B_v^{(1)}\cos\omega_1 (t-\frac{\tau}{2})\nonumber\\
&+A_v^{(2)}\sin \omega_2 (t-\frac{\tau}{2})+B_v^{(2)}\cos\omega_2 (t-\frac{\tau}{2}) \ .
\end{align}

In order to fully determine $\phi_{vv}(t)$, we need two additional  relations. The first one is obtained by taking the limit $t\rightarrow 0^+$ in Eq. (\ref{Eqdifvv01}) and using  the symmetry property $\phi_{vx}(-t)=-\phi_{vx}(t)$. This yields
\begin{align}
\label{Eqrel4}
{\dot \phi}_{vv}(0^+)+\frac{1}{Q_0} \phi_{vv}(0)+\frac{g}{Q_0}\phi_{vx}(\tau)=0 \ . 
\end{align}
The second relation is obtained by multiplying the FP equation (\ref{EqKramers1}) in the stationary case by $v^2$ and then integrating over $x$ and $v$. This gives 
\begin{align}
\label{Eqphipp0}
 \phi_{vv}(0)+g\phi_{vx}(\tau)=\frac{Q_0}{2} \ . 
\end{align}
(These two relations imply that $\dot \phi_{vv}(0^+)=-1/2$). Since $\phi_{vx}(t)=-\phi_{xv}(t)=-\int_0^t \phi_{vv}(t') dt'$,  we then obtain after some algebraic manipulations
\begin{align}
\label{Eqphixx}
2\phi_{vv}(t)&=\frac{\omega_1[\sin \omega_1 t-f(\omega_1)\cos\omega_1t]}{\omega_2^2-\omega_1^2}\nonumber\\
&- \frac{\omega_2[\sin \omega_2 t-f(\omega_2)\cos\omega_2t]}{\omega_2^2-\omega_1^2}
\end{align}
where the function $f(\omega)$ is defined by 
\begin{align}
\label{Eqfomega}
f(\omega)=\frac{\omega+\big[Q_0(1-\omega^2)-g\big]\tan(\omega\tau/2)}{Q_0(1-\omega^2)-g-\omega\tan(\omega\tau/2)} \ .
\end{align} 
This readily leads to Eq. (\ref{EqTv}).

Finally, in order to compute $\langle x^2\rangle_{st}=\phi_{xx}(0)$, we use a third relation obtained by multiplying the  FP equation by $xv$ and  integrating over $x$ and $v$, which yields
\begin{align}
\label{Eqphixx0}
\phi_{vv}(0)-\phi_{xx}(0)+\frac{g}{Q_0} \phi_{xx}(\tau)=0\ .
\end{align}
Substituting  $\phi_{xx}(\tau)=\phi_{xx}(0)+\int_0^{\tau} \phi_{xv}(t) dt$ then gives the expression of $\phi_{xx}(0)$. This leads to Eq. (\ref{EqTx}).
 
We recall that Eq. (\ref{Eqphixx})  is only valid for $t\in[0,\tau]$, but the expressions of the time-correlation functions in the successive intervals  $[\tau,2\tau],[2\tau,3\tau],...$ are easily obtained by solving Eq. (\ref{Eqdifvv01}).

\section{Stability analysis}

In this Appendix, we report the results of the local stability analysis of the deterministic equation 
\begin{align}
\label{EqLdeter}
\ddot x(t)+\frac{1}{Q_0} \dot x(t)+ x(t)- \frac{g}{Q_0}x(t-\tau)=0
\end{align}
which is needed  to determine the regions of the parameter space $(Q_0,g,\tau)$ in which a stationary solution to the Langevin equation  (\ref{EqLlinred}) exists. As  indicated in the main text,  the dynamical behavior of Eq. (\ref{EqLdeter}) has been previously investigated in the case of a negative feedback\cite{CG1982,CBOM1995}, with a full characterization of limit cycles and Hopf bifurcations.  We refer the interested reader to \cite{CBOM1995} for more details. Here, we  extend the determination of the stability domains to the case $g>0$ which is relevant to the feedback cooling regime (this implies that $g/Q_0<1$ so that the system is stable for $\tau=0$).

The stability analysis amounts to studying the roots of the  characteristic equation obtained by  substituting $x=e^{\lambda t}$ into Eq. (\ref{EqLdeter})  
\begin{align}
\label{Eqcharac}
\lambda^2+\frac{\lambda}{Q_0}+1-\frac{g}{Q_0}e^{-\lambda \tau}=0 \ .
\end{align}
The  mechanism for the loss of stability is that a root of  this equation acquires a positive real part. The stability curves are thus obtained by substituting $\lambda=i\omega$ into Eq. (\ref{Eqcharac}) and solving the coupled equations
\begin{align}
\label{Eqcoupled}
\cos \omega \tau&= \frac{Q_0}{g}(1-\omega^2)\nonumber\\
\sin \omega \tau&=-\frac{\omega}{g}\ .
\end{align}
It follows that $\omega$ is solution of Eq. (\ref{Eqomega0}) and that  $\tau$ takes one of the values 
\begin{align}
\label{Eqstau}
\tau_{n,1}^*&=\frac{2}{\omega_1}\tan^{-1}\Big(\frac{1}{\omega_1}\big[ Q_0(1-\omega_1^2)-g\big]\Big)+n\frac{2\pi}{\omega_1} \nonumber\\
\tau_{n,2}^*&=\frac{2}{\omega_2}\tan^{-1}\Big(\frac{1}{\omega_2}\big[ Q_0(1-\omega_2^2)-g\big]\Big)+n\frac{2\pi}{\omega_2}
\end{align}
with $n=0,1,2,...$.  It is readily seen from Eq. (\ref{Eqfomega}) that  either $f(\omega_1)$  or $f(\omega_2)$  then goes to infinity, which from Eqs. (\ref{EqTx}) and (\ref{EqTv}) makes the effective temperatures $T_x$ and $T_v$ diverge.
This leads to dividing  the plane $(Q_0,g/Q_0)$ into the four regions shown in Fig. \ref{Figdiag} of the main text. These regions correspond to the following four cases, illustrated by Figs. \ref{Fig13}-\ref{Fig16} below:

\vspace{0.25cm}

Region 1: $g/Q_0<-1$. The discriminant $\Delta$ defined by Eq. (\ref{EqDis}) is  positive and  $\omega_1$ is the only real root of Eq. (\ref{Eqomega0}). The deterministic system is stable for $\tau <\tau^*_{0,1}$ and unstable beyond $\tau^*_{0,1}$. Accordingly, a stationary solution to Eq. (\ref{EqLlinred}) only exists in the interval $0\le \tau<\tau^*_{0,1}$, as illustrated in Fig. \ref{Fig13} below.  We recall that $T_x/T \rightarrow (1-g/Q_0)^{-1}$ as $\tau \rightarrow 0$ because the spring constant is modified (see Eq. (\ref{EqTxtau})).
\begin{figure}[hbt]
\begin{center}
\includegraphics[width=8cm]{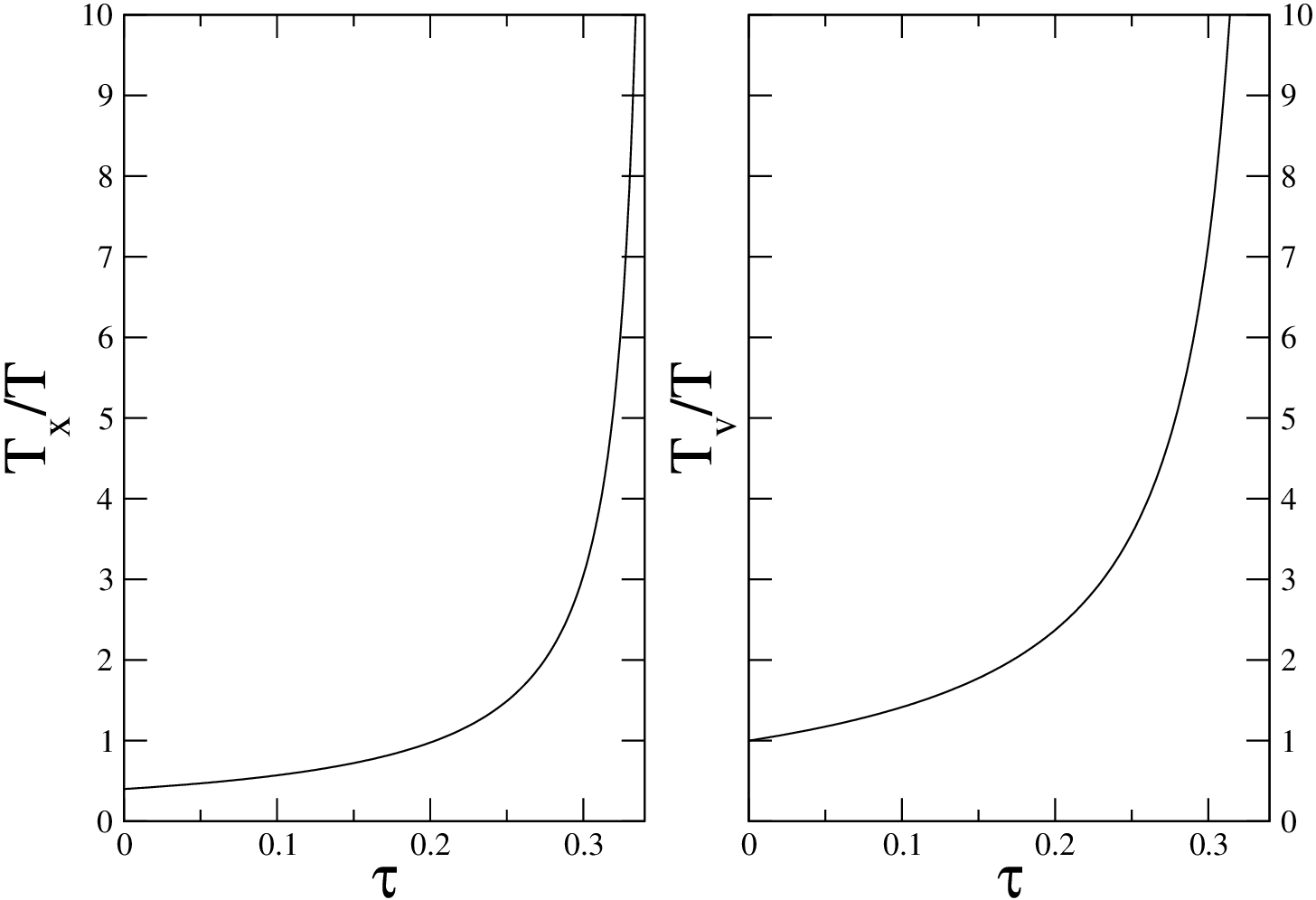}
\caption{\label{Fig13} Reduced temperatures $T_x/T$ and $T_v/T$ as a function of $\tau$ for $Q_0=2$ and $g/Q_0=-1.5$ (region 1).The two temperatures (or variances $\langle  x^2\rangle$ and $\langle v^2\rangle$) diverge at the critical delay $\tau_{0,1}^*\approx 0.349$ beyond which a stationary solution no longer exists.}
\end{center}
\end{figure}

\vspace{0.25cm}

Region 2:  $Q_0<\frac{1}{2}$ and $\vert \frac{g}{Q_0}\vert <1$, or $\frac{1}{2}<Q_0<\frac{1}{\sqrt{2}}$ and $\frac{1}{Q_0}\sqrt{1-\frac{1}{4Q_0^2}}<\vert g/Q_0\vert<1$. Then $\Delta >0$ and $\omega_1$ and $\omega_2 $ are  purely imaginary. The system is stable for all $\tau\ge 0$ and a stationary solution to Eq. (\ref{EqLlinred}) always exists, as illustrated in Fig. \ref{Fig14}. The temperatures go to finite values as $\tau \rightarrow \infty$, 
\begin{align}
\label{Eqlim2}
\frac{T_x}{T}&\rightarrow  \frac{1}{Q_0\sqrt{1-g^2/Q_0^2}}\: \frac{1}{\mbox{Im}(\omega_1+\omega_2 )} \nonumber\\
\frac{T_v}{T}&\rightarrow  \frac{1}{Q_0}\: \frac{1}{\mbox{Im}(\omega_1+\omega_2)} \ .
\end{align}
Note that $T_v$ has a non-monotonic behavior as a function of $\tau$, in contrast with $T_x$.
\begin{figure}[hbt]
\begin{center}
\includegraphics[width=8cm]{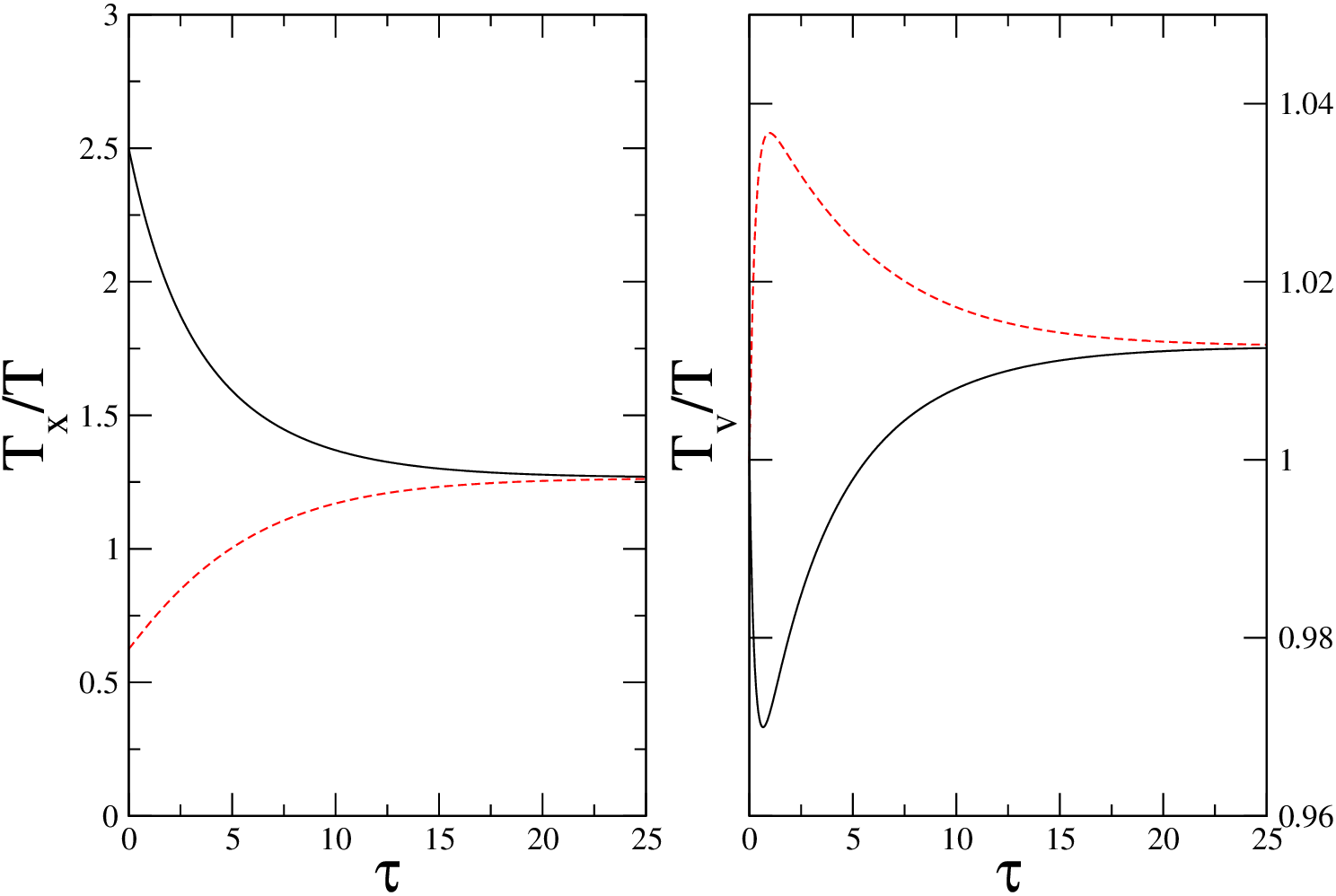}
 \caption{\label{Fig14} (Color on line) Same as Fig. \ref{Fig13} for $Q_0=0.25$, $g/Q_0=0.6$ (solid black line) and $g/Q_0=-0.6$ (dashed red line) (region 2).  $T_x/T$ and $T_v/T$ tend to $\approx 1.27$ and $\approx 1.01$, respectively, as $\tau \rightarrow \infty$, as predicted by Eqs. (\ref{Eqlim2}).}
\end{center}
\end{figure}

\vspace{0.25cm}

Region 3:  $Q_0>\frac{1}{2}$  and $\vert \frac{g}{Q_0}\vert <\frac{1}{Q_0}\sqrt{1-\frac{1}{4Q_0^2}}$. Then $\Delta <0$ and $\omega_1$ and $\omega_2 $ are complex conjugates. The system is stable for all $\tau\ge 0$ and a stationary solution to Eq. (\ref{EqLlinred}) always exists, as illustrated in Fig. \ref{Fig15}. The temperatures go to finite values as $\tau \rightarrow \infty$, 
\begin{align}
\label{Eqlim3}
\frac{T_x}{T}&\rightarrow  \frac{1}{2Q_0\sqrt{1-g^2/Q_0^2}}\: \frac{1}{\mbox{Im}(\omega_1)} \nonumber\\
\frac{T_v}{T}&\rightarrow \frac{1}{2Q_0} \frac{1}{\mbox{Im}(\omega_1)} \ .
\end{align}
The peaks in the effective temperatures  (or the variances $\langle x^2\rangle$ and $\langle v^2\rangle$) becomes more and more pronounced as $Q_0$ increases and one approaches the boundary with region 4: in this respect, these peaks (called  `destabilization  resonances' in \cite{PFFBT2006}) may be viewed as remnants of the oscillatory instability points (Hopf bifurcation points) that exist in region 4.
\begin{figure}[hbt]
\begin{center}
\includegraphics[width=8cm]{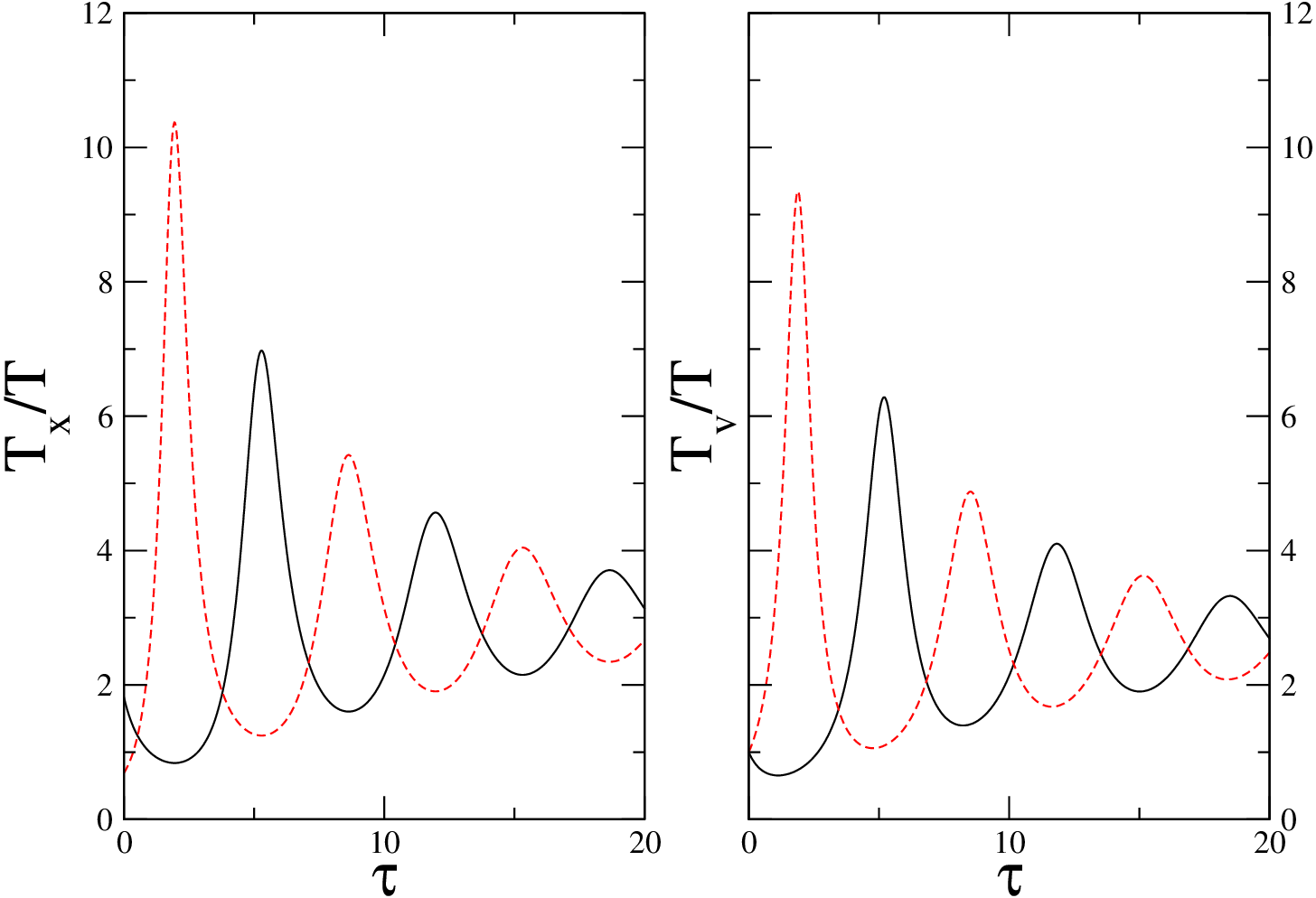}
 \caption{\label{Fig15} (Color on line) Same as Fig. \ref{Fig13} for $Q_0=2$, $g/Q_0=0.45$ (solid black line) and $g/Q_0=-0.45$ (dashed red line) (region 3). $T_x/T$ and $T_v/T$ tend to $\approx 2.95$ and $\approx 2.63$, respectively, as $\tau \rightarrow \infty$, as predicted by Eqs. (\ref{Eqlim3})}
\end{center}
\end{figure}

\vspace{0.25cm}

Region 4:  $Q_0>\frac{1}{\sqrt{2}}$ and $\frac{1}{Q_0}\sqrt{1-\frac{1}{4Q_0^2}}<\vert \frac{g}{Q_0}\vert <1$. Then $\Delta>0$ and both $\omega_1$ and $\omega_2 $ are real. This gives rise to multistability with a characteristic ``Christmas tree'' stability diagram (see Fig. \ref{FigChristmas_tree} in the main text), \textit{i.e.}, an increasing sequence of critical delays ordered as follows,
\begin{align}
\tau^*_{0,1}<\tau^*_{0,2}<\tau^*_{1,1}<\tau^*_{1,2}<...<\tau^*_{n-1,1}<\tau^*_{n-1,2}<\tau^*_{n,1}
\end{align}
up to a certain integer $n$.  As $\tau$  varies  from $0$ to $\tau^*_{n,1}$, the deterministic system switches from stability to instability and back to stability $2n+1$ times, and is unstable for  $\tau>\tau^*_{n,1}$.  A stationary solution to Eq. (\ref{EqLlinred}) only exists inside the stability domains, as illustrated in Fig. \ref{Fig16} (note that  $\tau^*_{0,1}$ and $\tau^*_{0,2}$ are negative for $g>0$ so that the series starts at $\tau^*_{1,1}$).
\begin{figure}[hbt]
\begin{center}
\includegraphics[width=8cm]{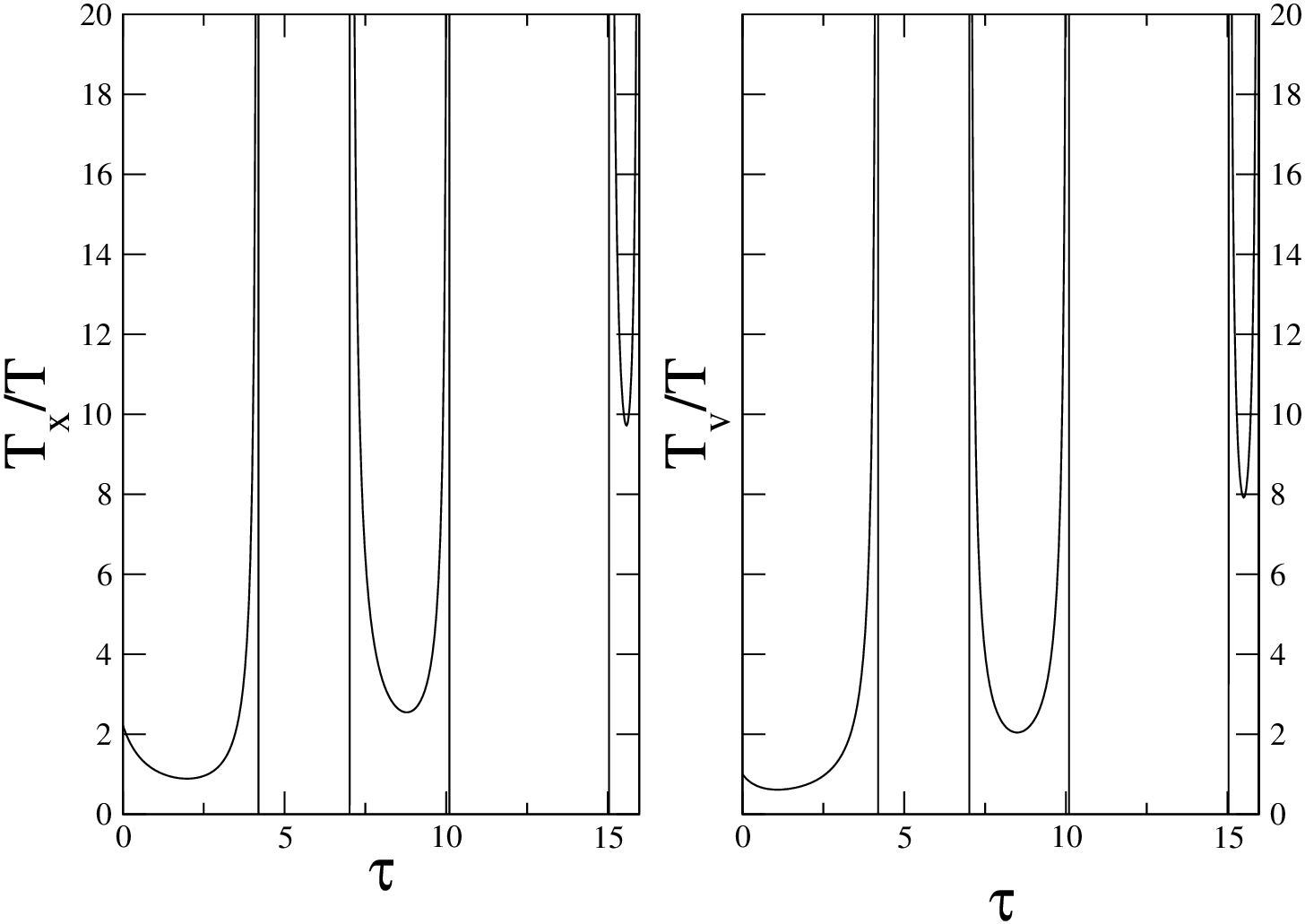}
 \caption{\label{Fig16} Same as Fig. \ref{Fig13} for $Q_0=2$ and $g/Q_0=0.55$ (region 4). The stationary solution disappears at the critical delays $\tau^*_{1,1}\approx 4.19, \tau^*_{2,1}\approx 10.08,\tau^*_{3,1}\approx 15.98$ and exists again at $\tau^*_{1,2}\approx 7.01, \tau^*_{2,2}\approx 15.03$. There is no stationary solution beyond $\tau^*_{3,1}$.}
\end{center}
\end{figure}

\section{Poles of $\widetilde \chi(s)$ and branch cuts for the function $\ln \widetilde \chi(s)/\chi_{g=0}(s)$}

In this Appendix we study the location of the poles of $\widetilde \chi(s)$ in the complex $s$-plane as a function of the system's parameters and we determine the corresponding branch cuts that define the single-valued function $\ln \widetilde \chi(s)/\chi_{g=0}(s)$  in Eq. (\ref{EqdefJs}). For brevity, the study is restricted to the case $g>0$.

From Eq. (\ref{Eqchis}), the poles  of $\widetilde \chi(s)$ (with $s=\sigma+i\omega$) are solutions of the  coupled equations
\begin{subequations}
\label{eq:subeqns}
\begin{align}
\frac{g}{Q_0}e^{\sigma\tau}\cos(\omega \tau)&=\sigma^2+\frac{\sigma}{Q_0}+1-\omega^2 \label{eq:subeq1}\\
\frac{g}{Q_0}e^{\sigma \tau}\sin(\omega \tau)&=2\omega(\sigma +\frac{1}{2Q_0}) \label{eq:subeq2} \ .
\end{align} 
\end{subequations} 

These equations have a (countably) infinite set of solutions which are either real ($\omega=0$) or complex conjugates ($\omega \ne 0$). In the first case, $\sigma$ is solution of 
\begin{align}
\label{Eqpolereel}
\frac{g}{Q_0}e^{\sigma \tau}=\sigma^2+\frac{\sigma}{Q_0}+1
\end{align} 
which has  $1,2$ or $3$ roots for $g>0$. In the second case, Eq. (\ref{eq:subeq2}) can be solved for $\sigma$ to obtain
\begin{align}
\label{Eqsigma}
\sigma=-\frac{1}{2Q_0}-\frac{1}{\tau}W(-g\frac{\tau\sin(\omega \tau )}{2Q_0\: \omega}e^{-\tau/(2Q_0)})
\end{align} 
where $W$ is the Lambert function of order $0$ or of order $-1$ (denoted respectively $W_0$  and $W_{-1}$) depending on whether $\sigma$ is smaller or larger than $1/\tau-1/(2Q_0)$\cite{note5}. Substituting Eq. (\ref{Eqsigma}) into Eq.  \eqref{eq:subeq1}  then yields a transcendental equation for $\omega$ that must be solved numerically.  Observe that the combination of Eqs.  \eqref{eq:subeq1}  and  \eqref{eq:subeq2}   yields a quadratic polynomial in $\omega^2$: as a result, there exist at most two pairs of complex conjugate poles with the same real part $\sigma$.

When $g/Q_0$ is small, one can expand Eqs. \eqref{eq:subeqns} in Taylor series in $g$, which gives
\begin{align}
\label{Eqpoleexp}
\sigma^{\pm}=&\sigma_0^{\pm}+g\frac{e^{\tau\sigma_0^{\pm}}}{1+2Q_0\sigma_0^{\pm}}+{\cal O}(g^2)\nonumber\\
\omega=&0
\end{align} 
for $Q_0\le 1/2$ (with $\sigma_0^{\pm}$ given by Eq. (\ref{spm})) and 
\begin{align}
\sigma &=-\frac{1}{2Q_0}+g\frac{e^{-\tau/(2Q_0)}}{\sqrt{4Q_0^2-1}}\sin(\frac{\tau}{2Q_0}\sqrt{4Q_0^2-1})+{\cal O}(g^2)\nonumber\\
\omega &=\pm \frac{1}{2Q_0}\sqrt{4Q_0^2-1} \mp g\frac{e^{-\tau/(2Q_0)}}{\sqrt{4Q_0^2-1}}\sin(\frac{\tau}{2Q_0}\sqrt{4Q_0^2-1})\nonumber\\
&+{\cal O}(g^2)
\end{align} 
for $Q_0\ge 1/2$. 
This category of poles is thus obtained  by continuously shifting the poles $s_0^+$ and $s_0^-$ of the causal response function $\chi_{g=0}(s)$. When they are complex conjugates (for $Q_0>1/2$), they are given by the solution of Eq. (\ref{Eqsigma})  involving the branch $W_{0}$ of the Lambert function.
In addition, there also exists a second category of poles  that appear nonperturbatively, with a real part of the order of $- \log(g)$ as $g\rightarrow 0$. These poles are either real or complex conjugates and then given by the solution of Eq. (\ref{Eqsigma}) involving the branch $W_{-1}$ of the Lambert function.
 
So long as $g/Q_0$ is small, there are only two poles of the first category on the left-hand side of the complex $s$-plane ($Re(s)<0$) and an infinity of  poles of the second category on the right-hand side ($Re(s)>0$). The situation becomes more complicated as $g$ increases and depends on the values of $Q_0$ and $\tau$: poles can move from the left-hand side to the right-hand side of the complex plane or vice-versa, new poles appear, and poles may cross each other. Such effects are nonperturbative and take also place if one varies $\tau$ at a fixed value of $g$. Complex conjugate poles given by the branch $W_{0}$, or their real counterparts, will be referred to as poles of ``type $0$": this generalizes the first category of poles encountered at small $g/Q_0$ in perturbative expansions (see above). On the other hand, complex conjugate poles obtained from the branch $W_{-1}$, or their real counterparts, will be referred to as poles of ``type $1$": they generalize the above second category of poles appearing at small $g/Q_0$ nonperturbatively.

The behavior may be rather intricate as illustrated by Fig. \ref{Fig17} which corresponds to an oscillator functioning in the second  stability lobe for $Q_0=34.2$ and $g/Q_0=0.25$ (see Fig. \ref{Fig_rates_vs_tau_34.2} in the main text). This figure shows the evolution as a function of $\tau$ of the real part of the two sets of poles that are closest to the imaginary axis. One  observes that for $7<\tau \lesssim 7.15$, there is one pair of (conjugate) poles of type $0$ and one pair of (conjugate) poles of type $1$. Then, for $\tau \gtrsim 7.15$,  all poles become of type 0 and the real parts cross each other for $\tau\approx 7.9$. Note also that one pair of poles moves from the left-hand side to the right-hand side of the complex plane  as $\tau$ increases whereas the other one goes in the opposite direction.
\begin{figure}[hbt]
\begin{center}
\includegraphics[width=7cm]{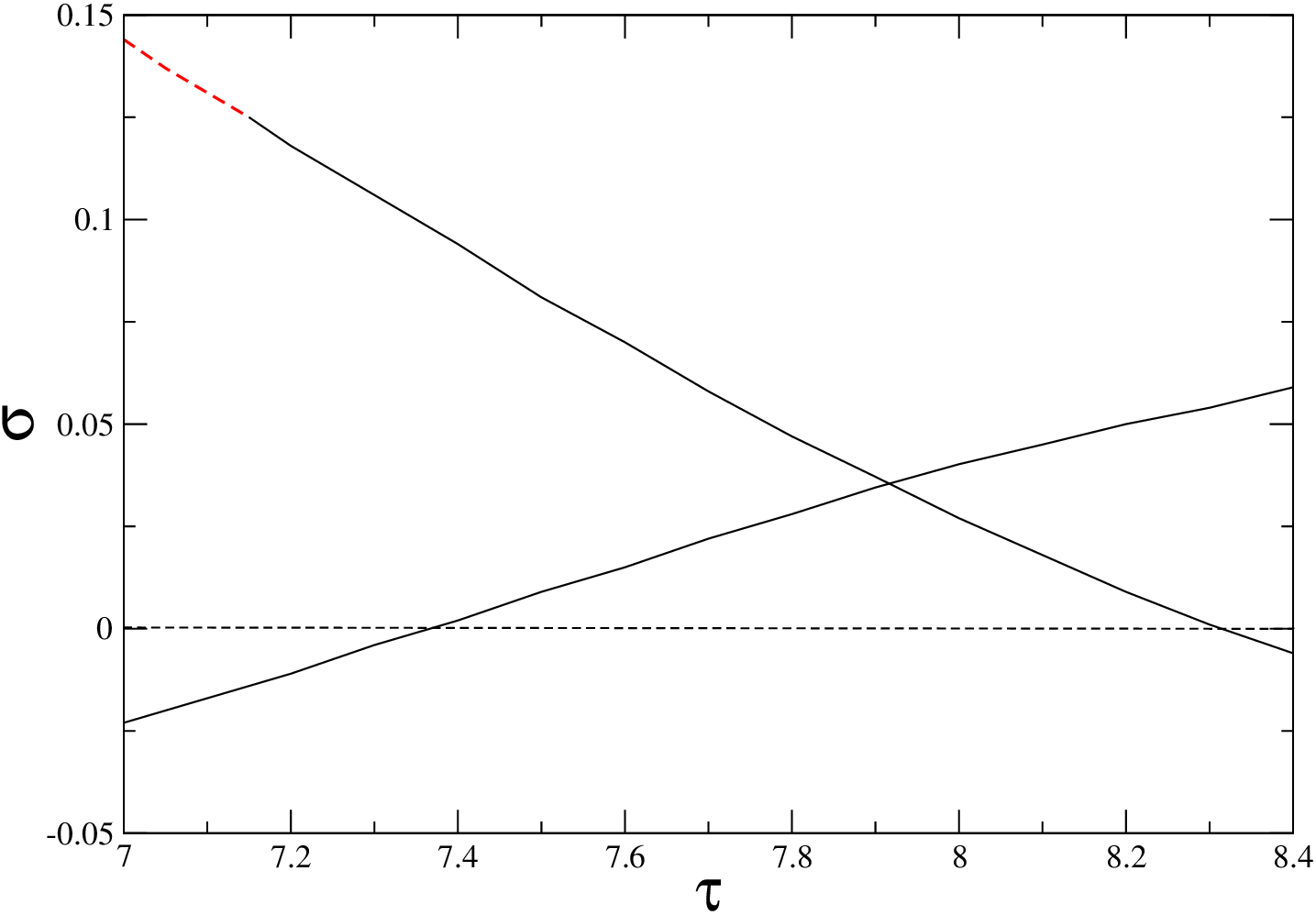}
 \caption{\label{Fig17} (Color on line) Evolution with $\tau$ of the real part of the two pairs of poles of $\tilde \chi(s)$ that are closest to the imaginary axis. Here,  $Q_0=34.2$ and $g/Q_0=0.25$ and only the second stability lobe is considered (see Fig. \ref{Fig_rates_vs_tau_34.2} of the main text). The poles are of type $0$ (solid black lines), except for a small interval  $7<\tau \lesssim 7.15$ where one pair of poles is of type $1$ (dashed red line).}
\end{center}
\end{figure}
 
Once the poles of $\tilde \chi(s)$ (and thus  the branch points of $\ln \widetilde \chi(s)/\chi_{g=0}(s)$) are determined, we need to introduce cuts that connect them pairwise. This amounts to studying the real and imaginary parts of the function $\widetilde \chi(s)/\chi_{g=0}(s)$, which from  Eqs. (\ref{Eqchis0}) and (\ref{Eqchis}) are given by 
\begin{align}
\label{Eqbranch}
&\Sigma(\sigma,\omega)=1+ge^{\sigma \tau}\times \nonumber\\
&\frac{\cos \omega \tau [Q_0(\omega^2-\sigma^2)-Q_0-\sigma]-\omega \sin(\omega \tau)(1+2Q_0\sigma)}{Q_0\omega^4+\frac{1}{2}[(1+2Q_0\sigma)^2+1-4Q_0^2]\omega^2+(Q_0\sigma^2+Q_0+\sigma)^2}
\end{align}
and 
\begin{align}
&\Omega(\sigma,\omega)=ge^{\sigma \tau}\nonumber\\
&\times \frac{\omega \cos \omega \tau (1+2Q_0\sigma)+\sin(\omega \tau)[Q_0(\omega^2-\sigma^2)-Q_0-\sigma]}{Q_0\omega^4+\frac{1}{2}[(1+2Q_0\sigma)^2+1-4Q_0^2]\omega^2+(Q_0\sigma^2+Q_0+\sigma)^2}\ ,
\end{align}
respectively. Since we are dealing with the principal value of the logarithm (with the argument in $(-\pi,\pi]$), the branch cuts are defined by $\Omega(\sigma,\omega)=0$ and $\Sigma(\sigma,\omega)<0$.  

Extensive numerical investigations for various values of $Q_0,g,\tau$ in regions $2,3,4$ of the parameter space (see Fig. \ref{Figdiag} in the main text) lead to the following  conclusions, illustrated schematically in Fig. \ref{FigCuts} in the main text:

\begin{itemize}

\item the poles of type $0$ have a smaller real part than any pole of type $1$,
\item the branch cuts originating from the poles of type 1 go all the way to $+\infty$,
\item when there are several poles of type 0, the branch cuts originating from the poles with the largest real part also go to $+\infty$,
\item the two branch cuts originating from the poles $s_0^{\pm}$ of $\chi_{g=0}(s)$ connect them to the leftmost poles of type 0 (this is also valid if the poles are real). 
\end{itemize}

From this analysis, we conclude that the Bromwich contour $Re(s)=c$ must be placed as illustrated in Fig. (\ref{FigCuts}) in the main text.  There is no alternative choice. In this regards, the third conclusion is crucial, since it tells us what to do in the situation depicted above in Fig. \ref{Fig17} when poles of type $0$ cross.  In this latter case the behavior of $\dot {\cal S}_{\cal J}$ is  nonanalytic as a function of $g$ or $\tau$: one must indeed consider only the leftmost pair of poles of type 0, whose identity changes after the crossing point. However, since Eq. (\ref{Eqfinal}) only involves the real part of the poles, this implies only a linear cusp 
in $\dot {\cal S}_{\cal J}$ and not a discontinuity.  This is the origin of the cusp found in the second instability lobe in Fig. \ref{Fig_rates_vs_tau_34.2} of the main text.

\section{``Acausal" response function $\widetilde \chi(t)$}
\label{App_acausal_response}

An acausal response function $\widetilde \chi(t)$ can tentatively be defined by the inverse Laplace transform of $\widetilde \chi(s)$: see Eq. (\ref{EqLaplace_response}) of the main text.

Since $\widetilde \chi(s)=(Q_0/g)e^{-s\tau}\psi(s)/[1-\psi(s)]$, $\widetilde \chi(t)$ is solution of the integral equation
\begin{align}
\label{EqVolt}
\widetilde \chi(t) -\int_{-\infty}^{t+\tau} dt'\:\widetilde \chi(t')\psi(t-t')=\frac{Q_0}{g}\psi(t-\tau)
\end{align}
which would be a Volterra integral equation of the second kind if the upper limit in the integration were $t$ instead of $t+\tau$ (as a consequence of  Eq. (\ref{Eqpsit})). $\widetilde \chi(t)$ is also solution of the second-order differential equation
\begin{align}
\ddot{\widetilde \chi}(t) +\frac{1}{Q_0} \dot{\widetilde \chi}(t)+\widetilde \chi(t) -\frac{1}{Q_0} \widetilde \chi(t+\tau)=\delta(t) \ ,
\end{align}
and $\widetilde \chi(t-t')=\langle x(t)\xi(t')\rangle$ with $x(t)$ solution of the acausal Langevin equation (\ref{EqLlcon}).

The acausal character of Eq. (\ref{EqVolt}) introduces unusual properties with respect to causal Volterra equations: i) it has an infinite number of solutions (or none at all), and ii)  these solutions may be unbounded as $t\rightarrow +\infty$ or $t\rightarrow -\infty$.   However, by fixing the Bromwich contour, which essentially amounts to fixing the ROC of $\widetilde \chi(s)$, we have selected a unique solution. In particular, by differentiating Eq. (\ref{EqdefJs}) with respect to $g$  and using Eq. (\ref{Eqfinal}), we derive that
\begin{align}
\label{EqdSJdg}
\widetilde \chi(\tau)=Q_0 \frac{\partial  \dot S_{{\cal J}}}{\partial g}=Q_0\frac{\partial(\widetilde s^{+}+\widetilde s^{-})}{\partial g}\ .
\end{align}

The behavior of $\widetilde \chi(t)$ for $t \ge 0$ is obtained from the property that there are only 2 poles, $\tilde s^{+}$ and $\tilde s^{-}$, to the left of the ROC of $\widetilde \chi(s)$. The inverse Laplace transform in Eq. (\ref{EqLaplace_response}) of the main text can thus be computed by closing the contour to the left with a large semi-circle (with radius taken to infinity). The only singularities inside the contour are the two poles $\tilde s^{\pm}$ and the residues at the poles then give $\widetilde \chi(t)$ in the form of a linear combination of $e^{\tilde s^{+}t}$ and $e^{\tilde s^{-}t}$. On the other hand, when $t \rightarrow - \infty$, the asymtotic behavior of $\widetilde \chi(t)$ is dominated by the pole(s), say $\tilde s_{\pm}^{(2)}$, that are the closest to the ROC on its right: this leads to an asymptotic dependence in $e^{\tilde s_{\pm}^{(2)}t}$. Some illustrative examples of the behavior of $\widetilde \chi(t)$ are shown in Figs. \ref{Fig18}-\ref{Fig20}. 

\begin{figure}[hbt]
\begin{center}
\includegraphics[width=7cm]{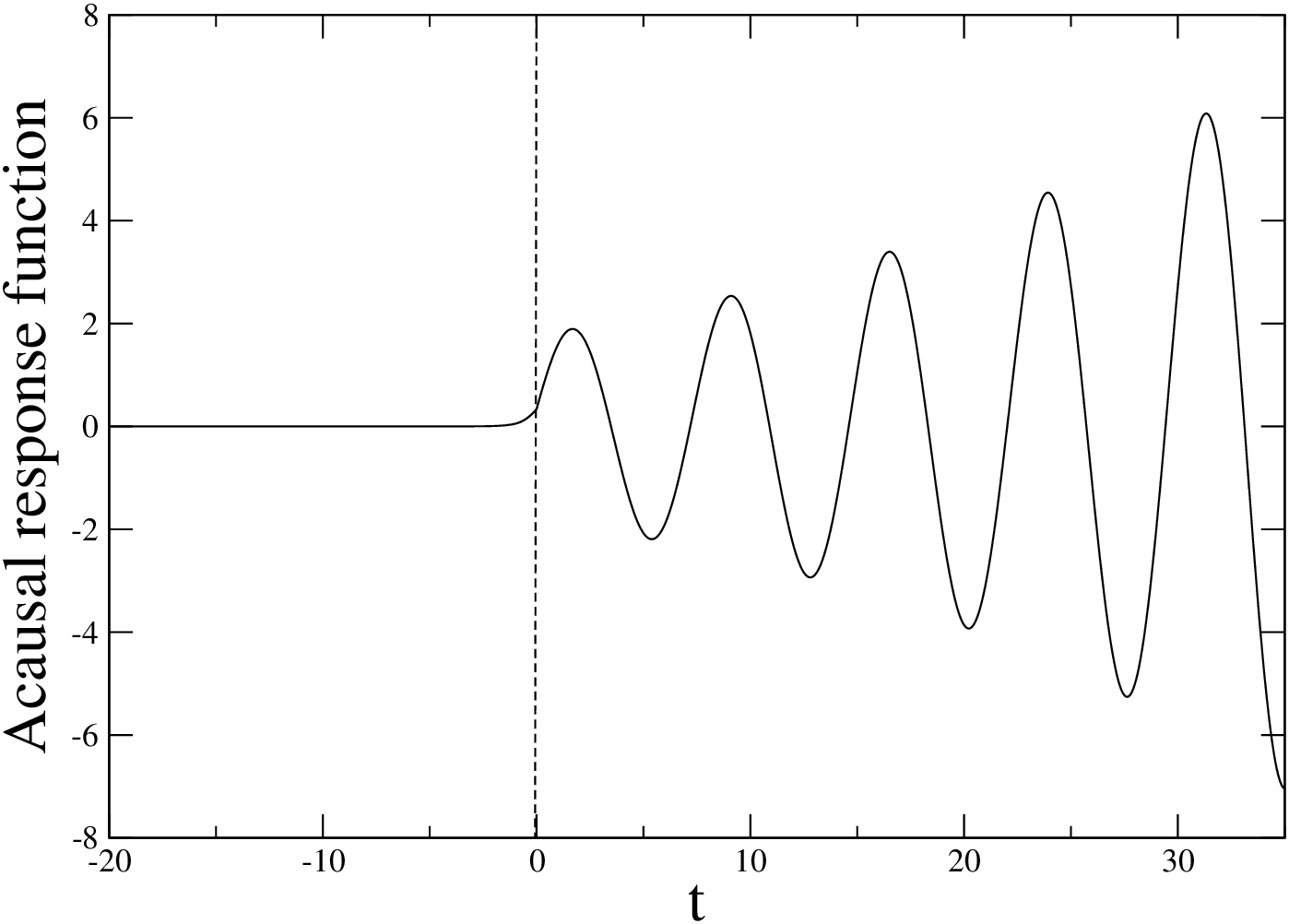}
 \caption{\label{Fig18}  Acausal response function $\widetilde \chi(t)$ versus time for $Q_0=2$, $g/Q_0=0.55$ and $\tau=1.2$. $\widetilde \chi(s)$ has no poles on the left-hand side of the complex plane. The poles  $\tilde s^{\pm}\approx 0.0394\pm 0.847\:i$ and $\tilde s^{(2)}\approx 1.977$ on the right-hand side control the behavior of $\widetilde \chi(t)$ for $t \ge 0$ (hence the oscillations and the diverging time dependence) and $t \rightarrow -\infty$ (hence the rapid decay to zero), respectively. The ROC of $\widetilde \chi(s)$ is defined by $\sigma_{min}=Re(\tilde s^{\pm})<Re(s)<\sigma_{max}=\tilde s^{(2)}$.}
\end{center}
\end{figure}
\begin{figure}[hbt]
\begin{center}
\includegraphics[width=7cm]{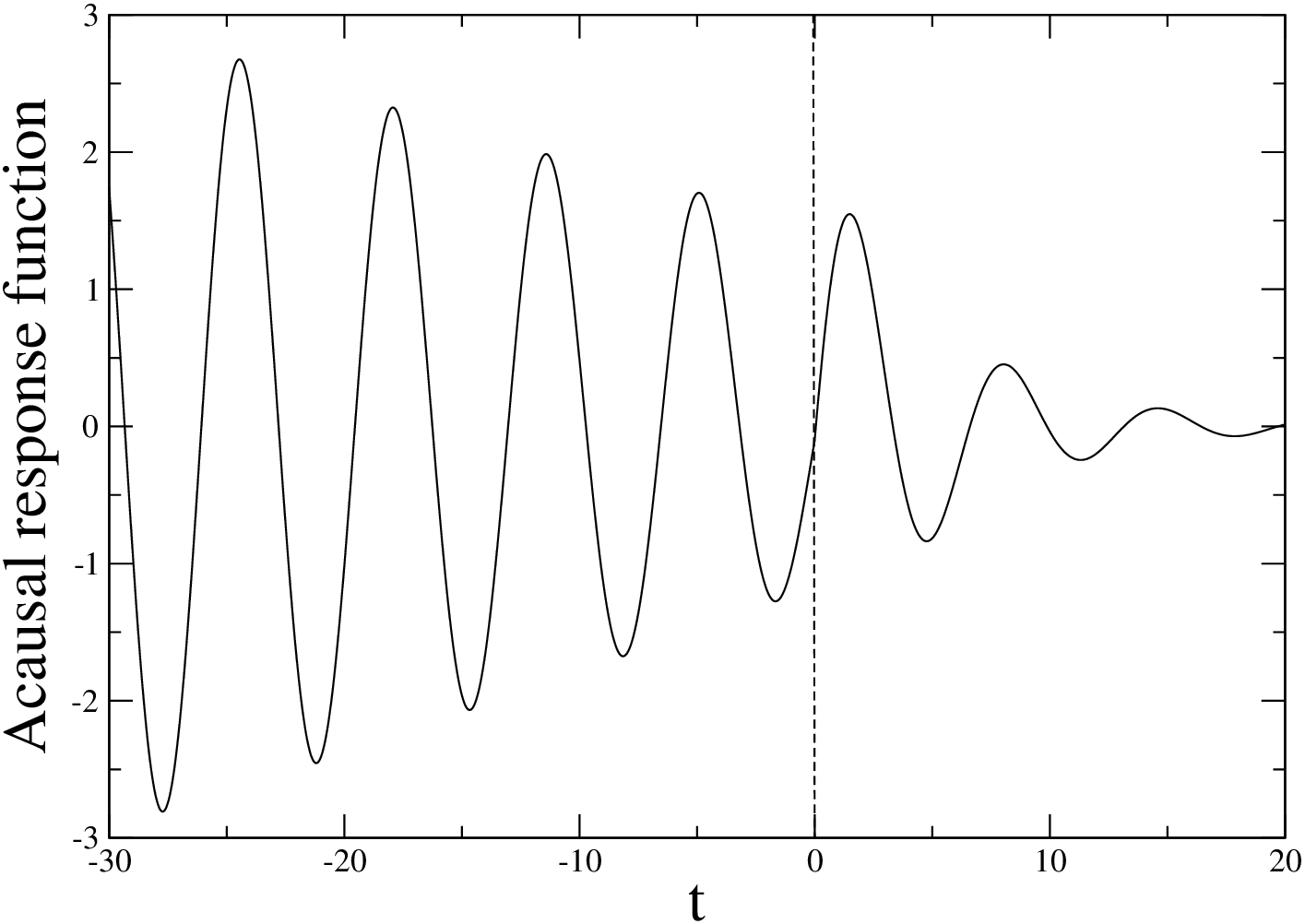}
 \caption{\label{Fig19}  Acausal response function $\widetilde \chi(t)$: Same as Fig. \ref{Fig18} for $\tau=8$. $\widetilde \chi(s)$ has $4$ poles on the left-hand side of the complex plane. The poles   $\tilde s^{\pm}\approx-0.187 \pm 0.958\:i$ and $\tilde s_{\pm}^{(2)}\approx -0.033\pm 0.967\:i$ control  the behavior of $\widetilde \chi(t)$ for $t\ge 0$ (hence the oscillation and the asymptotic decay to zero) and $t \rightarrow -\infty$ (hence the oscillations and the diverging asymptotic behavior), respectively. The ROC of $\widetilde \chi(s)$ is defined by $\sigma_{min}=Re(\tilde s^{\pm})<Re(s)<\sigma_{max}=Re(\tilde s_{\pm}^{(2)})$.} 
\end{center}
\end{figure}

\begin{figure}[hbt]
\begin{center}
\includegraphics[width=7cm]{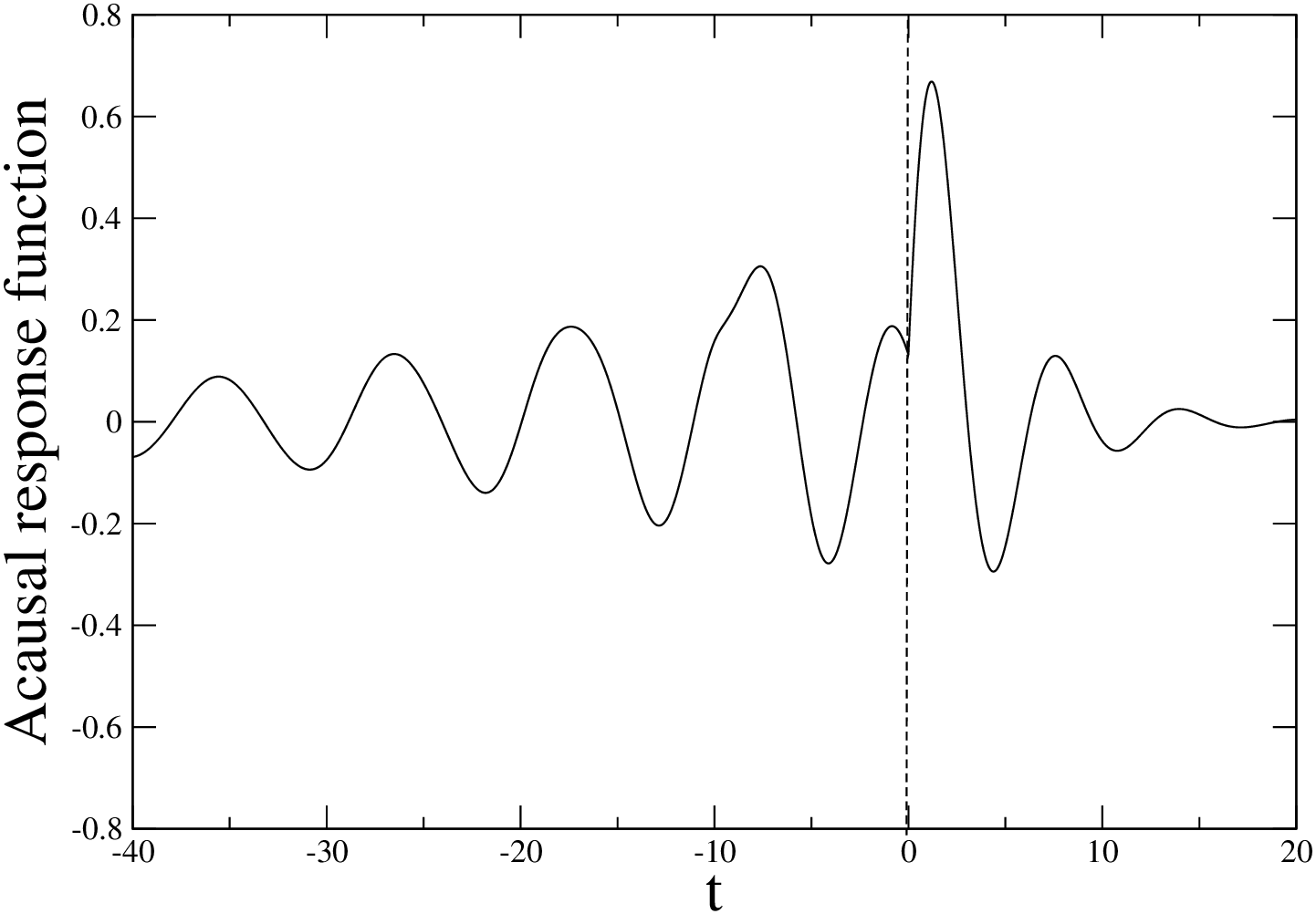}
 \caption{\label{Fig20}  Acausal response function $\widetilde \chi(t)$: Same as Fig. \ref{Fig18} for $g/Q_0=0.45$ and $\tau=10$. The poles   $\tilde s^{\pm}\approx-0.257 \pm 0.984\:i$ and $\tilde s_{\pm}^{(2)}\approx 0.0407\pm 0.692\:i$ control  the behavior of $\widetilde \chi(t)$ for $t\ge 0$ and $t \rightarrow -\infty$, respectively. Note the cusp behavior for $t=0$ and the weaker singularities for $ -\tau,-2\tau,$ etc.} 
\end{center}
\end{figure}

Note finally that $\widetilde \chi(t)$ is not $C^{\infty}$  at $t=0,-\tau, -2\tau,$ etc., as can be easily seen by differentiating Eq. (\ref{EqVolt}) twice: this is more clearly seen in Fig. \ref{Fig20}.


\begin{thebibliography}{10}
\bibitem{note4}  Eq. (\ref{EqdefJs}) differs from Eq. (18) in \cite{MR2014} which involves $G(s)=[s(s+1/Q_0)]^{-1}$ instead of $\chi_{g=0}(s)$. The two equations are equivalent except when the Bromwich contour has to be taken on the left-hand side of the complex plane. In this case, Eq. (18) in \cite{MR2014} does not yield the correct result for $\dot {\cal S}_{\cal J}$. Moreover, this equation is not suited for the small-$g$ expansion.
\bibitem{note9} This can be understood by rewriting the Langevin equation  (\ref{EqLlin}) as $m\dot v_t=-(k-k')x_t-k'\int_{t-\tau}^{t}v_s ds-\gamma v_t +\sqrt{2\gamma T}\xi_t$. Then, as $k'\rightarrow k$, the Brownian entity becomes unconfined and the Langevin equation only describes the stochastic dynamics of the velocity.
\end{thebibliography}
\end{document}